\documentclass[10pt]{iopart}
\usepackage{latexsym,epsfig,graphicx,bbm,bm,amssymb}
\usepackage[framemethod=tikz]{mdframed}

\usepackage{hyperref}
\hypersetup{
	colorlinks   = true
}

\newcommand{\ket}[1]{\left | \, #1 \right \rangle}
\newcommand{\kets}[1]{| \, #1 \rangle}

\newcommand{\eqr}[1]{Eq.~(\ref{#1})}
\newcommand{\fir}[1]{Fig.~\ref{#1}}
\newcommand{\secr}[1]{Sec.~\ref{#1}}

\newcommand{\bra}[1]{\left \langle #1 \, \right |}
\newcommand{\braket}[2]{\left\langle\, #1\,|\,#2\,\right\rangle}

\newcommand{\av}[1]{\langle #1\rangle}

\newcommand{\vac}{\ket{\textrm{vac}}}

\begin{document}

\title[Unifying NQS and CPS via Tensor Networks]{Unifying Neural-network Quantum States and Correlator Product States via Tensor Networks}

\author{Stephen~R~Clark\dag}
\address{\dag Department of Physics, University of Bath, Claverton Down, Bath BA2 7AY, U.K.}
\ead{s.r.clark@bath.ac.uk}

\begin{abstract}
Correlator product states (CPS) are a powerful and very broad class of states for quantum lattice systems whose amplitudes can be sampled exactly and efficiently. They work by gluing together states of overlapping clusters of sites on the lattice, called correlators. Recently Carleo and Troyer {\em Science} {\bf 355}, 602 (2017) introduced a new type sampleable ansatz called neural-network quantum states (NQS) that are inspired by the restricted Boltzmann model used in machine learning. By employing the formalism of tensor networks we show that NQS are a special form of CPS with novel properties. Diagramatically a number of simple observations become transparent. Namely, that NQS are CPS built from extensively sized GHZ-form correlators, which are related to a canonical polyadic decomposition of a tensor, making them uniquely unbiased geometrically. Another immediate implication of the equivalence to CPS is that we are able to formulate exact NQS representations for a wide range of paradigmatic states, including superposition of weighed-graph states, the Laughlin state, toric code states, and the resonating valence bond state. These examples reveal the potential of using higher dimensional hidden units and a second hidden layer in NQS. The major outlook of this study is the elevation of NQS to {\em correlator operators} allowing them to enhance conventional well-established variational Monte Carlo approaches for strongly correlated fermions. \\

\end{abstract}


\maketitle

\section{Introduction}
Quantum many-body systems represent a supreme `big data' challenge in modern physics. Formally an exponentially large amount of information is needed to fully describe a generic many-body quantum state, making a brute-force numerical approach intractable and limited to a few 10's of qubits even with the most advanced supercomputers~\cite{Lauchli2012}. Yet it is now becoming increasingly clear that physically relevant states typically occupy a very small corner of this vast Hilbert space~\cite{Eisert2010}. The many-body problem is then formulated as finding physically motivated and efficient schemes of capturing this corner of states and appealing to the variational principle to locate the best approximation within this class. 

In wider societal and commercial contexts increasing online connectivity has made big data problems ubiquitous~\cite{Snijders2012}. Currently deep learning is an increasingly popular technique for processing meaningful information from these problems, and is already having a transformative effect that finds far-reaching applications, ranging from self-driving cars to speech recognition to targeted online advertising~\cite{LeCun2015,Goodfellow2016}. Underlying this success are artificial neural networks, which are a powerful tool for compactly representing complex correlations in multi-variable functions and naturally allow patterns and abstractions in data to be revealed~\cite{Haykin2008}. Since their inception there has been a close connection between neural-networks and physics, specifically statistical mechanics, and this has provided insightful guidance on the `unreasonable success' of deep learning through its close links to renormalisation group methods~\cite{Beny2013,Mehta2014}. 

Over the past couple of years neural-networks and deep learning techniques have attracted significant attention in the communities working on many-body systems. This includes training a neural-network to identify symmetry-broken and topological quantum phases of interacting systems~\cite{Carrasquilla2017,vanNieuwenburg2017,Broecker2017,Wang2016,Chng2017}, to solve impurity problems in dynamical mean-field theory~\cite{Arsenault2015}, to model thermodynamics observables~\cite{Torlai2016}, to recommend cluster spin-flips that accelerate Monte Carlo simulations~\cite{Huang2017,Liu2017}, and to enhance wave-function tomography~\cite{Tubman2016,Torai2017}. Of particular importance here is the recent novel proposal by Carleo and Troyer~\cite{Carleo2017} to directly apply neural-network representations to the variational formulation of the quantum many-body problem. Their numerical evidence, and very recent extensions~\cite{Nomura2017}, on model systems suggests that this neural-network quantum state (NQS) approach is a promising and potentially disruptive concept for the field.   

Since then there have been a number of follow-up works aimed at understanding how expressive a neural-network inspired ansatz is. This has included the construction of exact NQS representations of several topological states~\cite{Deng2016}, a characterisation of states that can be described efficiently based on the depth of quantum circuits generating them~\cite{Gao2017}, and an analysis of the entanglement properties of NQS~\cite{Deng2017}. Importantly there has also been an effort~\cite{Chen2017} to connect NQS to another successful quantum many-body approach, namely tensor network theory (TNT)~\cite{Verstraete2008,Cirac2009,Orus2014}. Consequently a picture is now emerging about what kinds of quantum states are easy to capture in NQS and how it differs from other well-established ansatzes. 

This aim of this paper is to push these connections further by exposing in detail the connection between NQS and a very general class of {\em sampleable} many-body quantum states called correlator product states (CPS) introduced by Changlani {\em et al}~\cite{Changlani2009}. Although not formally required to understand or apply the NQS and CPS approaches, we will heavily exploit tensor network diagrammatics. In doing so we will demonstrate that TNT is a rather powerful and unifying form of visual calculus for revealing fundamental properties of these ansatzes and making clear pathways for extending them. The main results of this work are a collection of relatively simple observations, not necessarily all widely appreciated, that taken together provide key insights into NQS. Indeed we will show that NQS are a very interesting special case of CPS, a fact that allow us to construct a diverse set of non-trivial states with exact NQS representations, including weighted-graph states and resonating valence bond (RVB) states. Knowledge of such examples has proven invaluable in understanding the power and limitations of other ansatzes, like matrix product states (MPS), and so they are another useful contribution of this work. Moreover, these examples naturally suggest a number of extensions to NQS including (i) the use of higher-dimensional hidden units, (ii) using two hidden layers akin to deep neural networks, and (iii) elevating NQS to a form of projective Jastrow-type ansatz~\cite{Jastrow1955}. With this in mind we will argue that on practical level there are substantial advantages in using the tensor network framework to code these types of increasingly sophisticated ansatzes~\cite{Alassam2017}.    

\subsection{Summary of main results}
Correlator product states are built from by gluing together states, called {\em correlators}, for overlapping clusters of sites on the lattice, as formally defined in \secr{sec:cps}. In contrast NQS are based on restricted Boltzmann machines (RBMs) composed of $M$ binary hidden units (or neurons), as shown in \fir{fig_rbm} and formally defined in \secr{sec:nqs}. Here we briefly summarise the main results of this work: 

\begin{enumerate}
\item NQS can be viewed as a new and special form of CPS with each hidden unit associated to an extensively-sized correlators, analogous to string-bond states. 
\item NQS correlators are based on GHZ states and so are geometrically unbiased allowing them to address the system either globally or locally. 
\item In terms of tensor factorisations each hidden unit correlator has the structure of canonical polyadic decomposition (CPD).
\item NQS become more powerful with $r$-dimensional hidden units, corresponding to a CPD with rank $r$, and are equivalent to an NQS composed of two layers with $r$ and $\lceil \log_2(r) \rceil$ binary hidden units, respectively.
\item NQS with $M$ hidden units can be converted into MPS and projected entangled pair states (PEPS) with an internal dimension $\chi = 2^M$ and $\chi = 2^{M/2}$, respectively.
\item Graph states have an NQS representation with the number of hidden units equal to the minimum vertex cover of the graph, and this is generalised to a superposition of weighted graphs states by introducing a second layer of hidden units.
\item Uniform number states, such as the W state, can be described by an NQS with $M = \lceil N/2 \rceil$ hidden units, where $N$ is the number of sites in the system.  
\item The Laughlin state is encoded by an additional $M = N-1$ hidden units on top of a uniform number state with the required filling.
\item Toric code, fully-packed loop and dimer states share the same structure of NQS with $M = N/2$ hidden units, but the latter two states require $r=4$ dimensional hidden units.
\item The RVB state is shown to have a two-layered NQS representation both with $O(N)$ hidden units, rendering it an inefficient representation for exact sampling.
\item NQS can be generalised to correlator operators and applied to a wide class of references states such as fermionic wave functions.
\end{enumerate}

Details of these results are presented in the main text, which is structured into five sections. For completeness some foundational background is given in \secr{sec:background}, which begins by formally introducing the quantum many-body problem we consider in \secr{sec:manybody}, followed by giving a brief overview of TNT in \secr{sec:tensor_networks}, variational Monte Carlo (VMC) in \secr{sec:vmc} and CPS in \secr{sec:cps}. In \secr{sec:cps_tnt} we describe a key tensor network ingredient of this work, the COPY tensor in \secr{sec:copy_tensor}, and then formulate CPS as tensor networks using them in \secr{sec:cps_networks}. This section closes in \secr{sec:coherent_thermal} where we show that quantum states whose amplitudes follow from a partition function of a classical model with pairwise interactions, so called coherent thermal states, have a simple exact CPS representation. Building on this \secr{sec:nqs_overview} describes the essentials behind NQS, starting by introducing RBMs in \secr{sec:rbm}, which leads naturally to their formal definition in \secr{sec:nqs} and their identification as a special class of CPS. This section culminates with the corresponding tensor network for NQS being analysed in \secr{sec:tensor_network_nqs}. In \secr{sec:examples} we then introduce exact NQS representations for weighted-graph states in \secr{sec:graph_states}, uniform number states in \secr{sec:uniform_number}, the Laughlin state in \secr{sec:laughlin}, toric code states in \secr{sec:toric_code}, the fully-packed loop and dimer states in \secr{sec:fully_packed_dimer}, and the RVB state in \secr{sec:rvb_state}. In \secr{sec:extensions} we discuss extensions of NQS as correlator operators that can modify commonly used fermionic wave functions, before concluding the paper in \secr{sec:conclusions}.

\section{Background}\label{sec:background}
In this section we introduce the quantum many-body problem formally and give a brief overview of the approaches of TNT, VMC and CPS that will serve as useful background for later sections. 

\subsection{Quantum many-body problem}\label{sec:manybody}
For concreteness throughout this paper our considerations will be focused on a system composed of $N$ qubits (spin-1/2 subsystems) described by local basis states $\ket{v}$, with $v \in \{0,1\}$, being eigenstates of the $z$-Pauli operator $\hat{\sigma}^z\ket{v} = (-1)^v \ket{v}$. An arbitrary many-body state of this system can then be written as 
\begin{equation}
\ket{\Psi} = \sum_{\bf v} \Psi({\bf v})\ket{\bf v},
\end{equation}
where ${\bf v} = (v_1,v_2,\dots,v_N)^{\rm T} \in \{0,1\}^N$ is a bit string specifying a configuration basis state $\ket{\bf v}$ and $\Psi({\bf v})$ is its associated the complex amplitude. At zero temperature the quantum many-body problem commonly involves two tasks: (i) finding the ground state and/or low-lying excitations of a given Hamiltonian $\hat{H}$, and (ii) time-evolving a given initial state according to a (possibly time-dependent) Hamiltonian. Since a general quantum state possesses $2^N$ amplitudes $\Psi({\bf v})$, this represents an acute manifestation of the `curse of dimensionality'. We will now introduce the TNT and CPS approaches for sidestepping this issue.

\begin{figure}[ht]
\begin{center}
\includegraphics[scale=0.5]{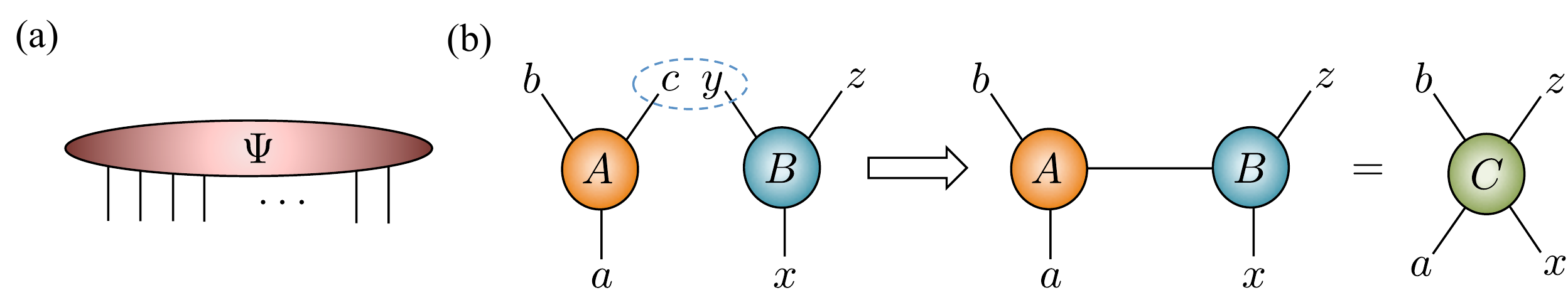}
\end{center}
\caption{(a) Tensors are represented graphically as shapes with legs protruding from them corresponding to an index of the tensor. Here a general order-$N$ tensor $\Psi$ of amplitudes for a quantum state is shown. (b) Two order-3 tensors $A_{abc}$ and $B_{xyz}$ are shown with the third index of $A$ and second index of $B$ circled. By contracting these indices, depicted as connected legs, we arrive at the order-4 tensor $C_{abxz}$.}
\label{fig_contract}
\end{figure}

\subsection{Tensor networks}\label{sec:tensor_networks}
The amplitudes $\Psi({\bf v})$ can be viewed as an order-$N$ tensor $\Psi_{v_1v_2\cdots v_N}$, represented diagrammatically as a shape with $N$ open legs as shown in \fir{fig_contract}(a). Tensor network theory is based on trying to decompose this structureless and monolithic object $\Psi_{v_1v_2\cdots v_N}$ into a network of lower order tensors. Such a network is defined by a graph $G$ where every vertex $\nu$ has associated to it a tensor $T^{(\nu)}$ possessing a small number of {\em internal} indices, each of dimension at most $\chi$, and may additionally possess physical indices $v_j$, of dimension $2$ here. The edges of $G$ then describe how the internal legs of each tensor are to be {\em contracted} together. Contraction is essentially the generalisation of matrix multiplication, e.g. for two order-3 tensors $A_{abc}$ and $B_{xyz}$ a contraction could form a new order-4 tensor as $C_{abxz} = \sum_{\alpha} A_{ab\alpha }B_{x\alpha z}$, so long as the dimension the third index of $A$ equals the that of the second index of $B$. This operation is represented graphically by joining legs together, as shown in \fir{fig_contract}(b). A tensor network decomposition therefore has the general form 
\begin{equation}
\Psi_{v_1v_2\cdots v_N} = {\rm tTr}\left[\otimes_{\nu \in G}\, T^{(\nu)}\right],
\end{equation}
where ${\rm tTr}$ is the tensor trace that performs all the contractions of the internal indices specified by $G$, leaving $N$ open physical indices $v_j$. Formally, if $\chi$ is allowed to scale exponentially with $N$ then any tensor network decomposition based on a connected graph can describe any state $\ket{\Psi}$. However, the practical utility of tensor networks relies on the broad observation that even with a bounded and small $\chi$ certain networks can provide extremely accurate and highly compressed descriptions of physically relevant states. 

\begin{figure}[ht]
\begin{center}
\includegraphics[scale=0.5]{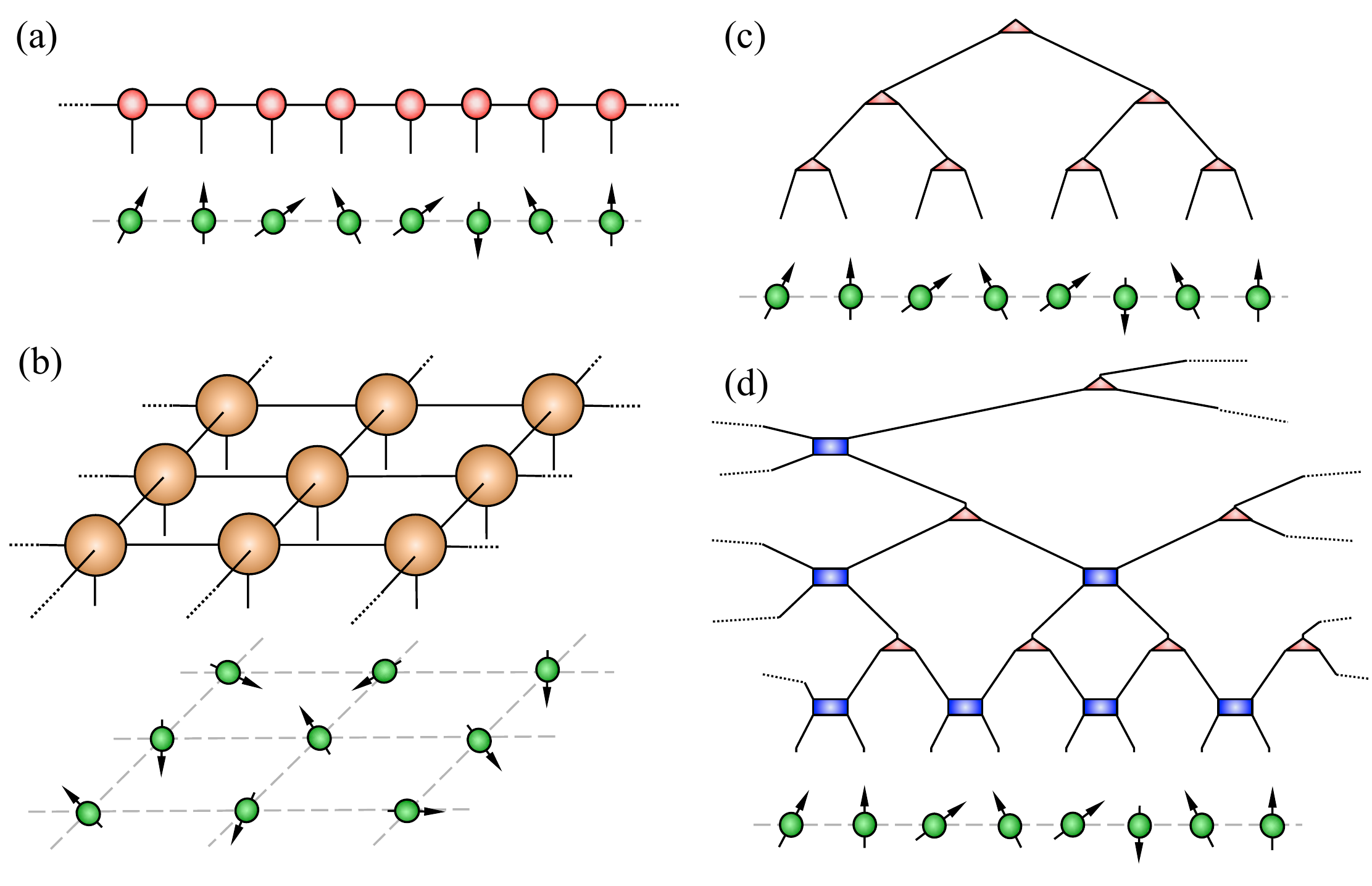}
\end{center}
\caption{Some common tensor networks. (a) Matrix product state (MPS), (b) projected entangled pair state (PEPS), (c) tree tensor network (TTN) and (d) multiscale entanglement renormalisation ansatz (MERA). In all cases the open legs of the network correspond to individual degrees of freedom (spins/qubits) in the physical system depicted beneath them.}
\label{fig_tns}
\end{figure}

Based on the entanglement area-law~\cite{Eisert2010} and ideas from renormalisation group there are a number of well established tensor networks suited to describing quantum many-body states arising as low-lying eigenstates of short-ranged Hamiltonians. This includes matrix product states (MPS)~\cite{Schollwock2011,Orus2014}, projected entangled pair states (PEPS)~\cite{Verstraete2008,Cirac2009}, tree tensor networks (TTN)~\cite{Shi2006,Murg2010} and multiscale entanglement renormalisation ansatz (MERA)~\cite{Evenbly2014}. For MPS and PEPS, shown in Figures~\ref{fig_tns}(a) and (b), the resulting network follows the geometry of the underlying physical system, e.g. a chain or lattice with coordination number $\mathcal{Z}$, and are built from tensors with $\mathcal{Z}$ internal indices and one physical index. The dimension $\chi$ of the internal indices is directly related to how much entanglement is captured by MPS and PEPS. A TTN has a hierarchical structure in which degrees of freedom are successively thinned down layer by layer to a dimension $\chi$ by order-3 isometric tensors, e.g. as in Kadanoff spin-blocking. This is shown in Figure~\ref{fig_tns}(c) for a 1D system. The MERA network in 1D is similar to a TTN, as seen in Figure~\ref{fig_tns}(d), but the layers of isometries are separated by layers of order-4 unitary tensors that `disentangle' prior to truncation. Both TTN and MERA can be generalised to 2D systems.

Finding a tensor network decomposition involves variationally minimising the tensor elements. A first step in this is computing expectation values $\bra{\psi}\hat{h}\ket{\psi}$, where $\hat{h}$ is some product operator, e.g. a term in the Hamiltonian $\hat{H}$. The contraction of the tensor network for $\bra{\psi}\hat{h}\ket{\psi}$ is therefore required. For MPS and TTN efficient and exact contractibility follows from the 1D chain or tree-like geometry, while for MERA it follows from its peculiar causal cone structure resulting from the unitary layers~\cite{Evenbly2014}. For PEPS, however, exact contraction is not efficient in general, but efficient approximate contraction can be performed~\cite{Verstraete2008}. Deterministic tensor network algorithms for computing ground state MPS, PEPS, TTN and MERA essentially boil down to performing a form of alternating least squares minimisation of the total energy with respect to given tensor(s) in the network~\cite{Ran2017}. Beyond stationary states MPS methods have proven particularly successful for simulating the dynamical time-evolution of 1D systems~\cite{Vidal2003}, with applications in cold-atoms~\cite{Clark2004,Bruderer2010}, periodically driven materials~\cite{Coulthard2017}, dissipative and disordered systems~\cite{Mendoza2015,Mendoza2016,Znidaric2017}, as well as classical stochastic problems~\cite{Johnson2010,Johnson2015}. Substantial efforts have been and continue to be made to mimic this success in higher dimensions with other tensor networks.

\subsection{Variational Monte-Carlo}\label{sec:vmc}
Contractibility of a tensor network is a rather constraining property and so we will instead focus on the weaker property sampleability. This means that the amplitudes $\Psi({\bf v})$ of a given ansatz in some fixed basis $\ket{{\bf v}}$ can be efficiently computed. Once armed with such a representation standard Monte Carlo methods allow the state to be variationally minimised~\cite{Foulkes2001,Gubernatis2016}. Specifically, the expectation value of an observable $\hat{A}$ can be written in a form suited to Monte Carlo sampling as
\begin{equation}
\av{\hat{A}} = \sum_{\bf v} p({\bf v}) A({\bf v}), \quad {\rm where} \quad p({\bf v}) =  \frac{|\Psi({\bf v})|^2}{\sum_{\bf v} |\Psi({\bf v})|^2},
\end{equation}
is the probability of a configuration $\bf v$ and
\begin{equation}
A({\bf v}) = \sum_{{\bf v}'} \frac{\Psi({\bf v}')}{\Psi({\bf v})}\bra{\bf v}\hat{A}\ket{{\bf v}'}, \label{eq:estimator}
\end{equation}
is the estimator of $\hat{A}$. The sum over ${\bf v}'$ in \eqr{eq:estimator} is restricted to only those configurations for which the matrix element $\bra{\bf v}\hat{A}\ket{{\bf v}'} \neq 0$. Thus, so long as $\hat{A}$ is sparse in the chosen fixed basis its expectation value can be efficiently estimated by flipping one or more qubits via a Markov-chain algorithm such as Metropolis-Hastings\footnote{Since it is not necessary to explicitly compute the $\braket{\Psi}{\Psi}$ in these methods throughout this paper we will not concern ourselves with normalisation constants.}. Typical terms comprising short-ranged Hamiltonians $\hat{H}$ fulfil this requirement. As a result variational minimisation of a sampleable ansatz can proceed by evaluating its energy $E = \av{\hat{H}}$ and its variance, along with their gradient vectors with respects to parameters of an ansatz, updating them by a small step along the direction of steepest descent, and iterating until convergence~\cite{Gubernatis2016}. More sophisticated approaches such as modified stochastic optimisation~\cite{Lou2007}, `linear method'~\cite{Nightingale2001,Toulouse2007} and stochastic reconfiguration~\cite{Sorella2001} are also commonly used. 

\begin{figure}[ht]
\begin{center}
\includegraphics[scale=0.5]{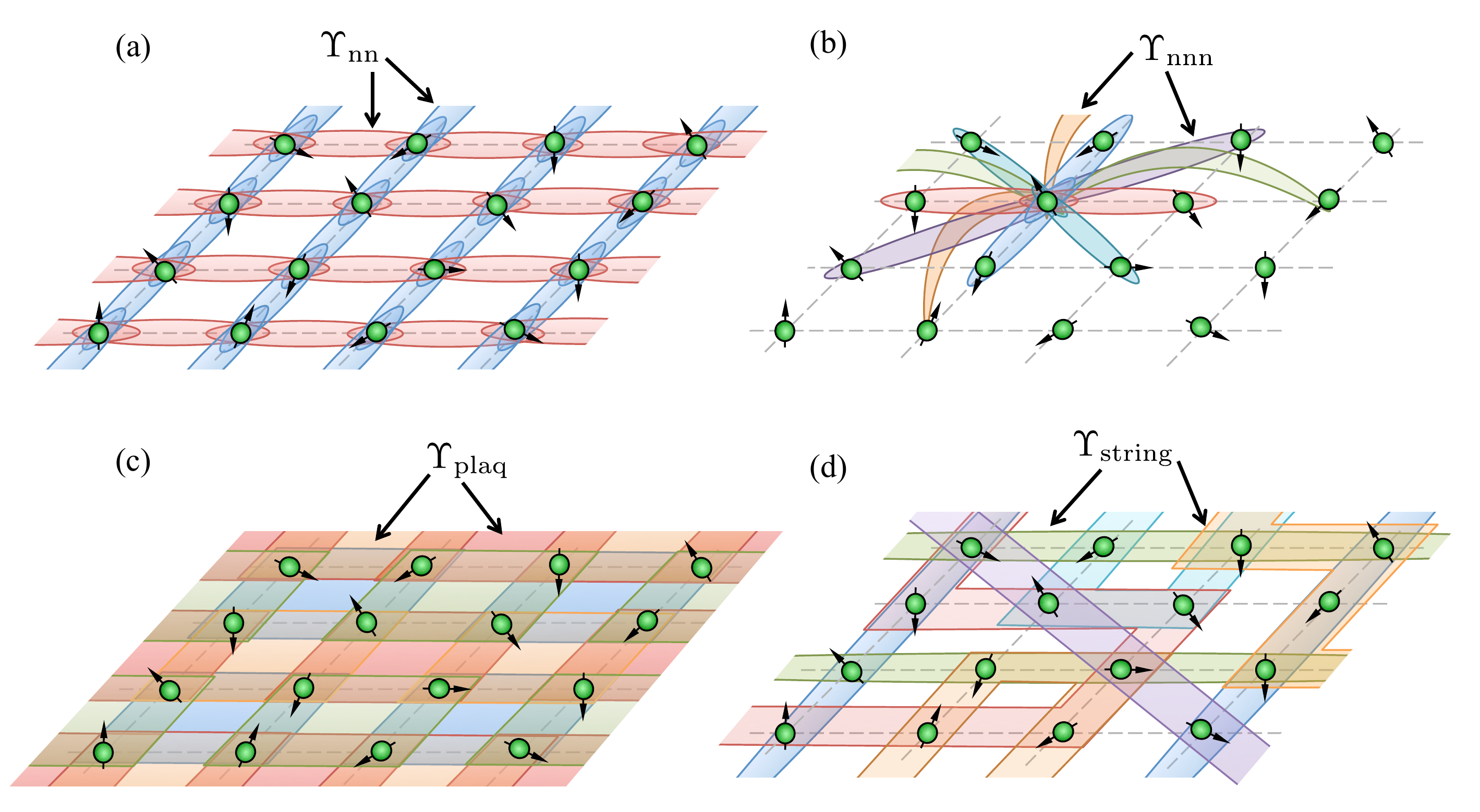}
\end{center}
\caption{(a) The covering pattern for nearest-neighbour pairwise correlators $\Upsilon_{\rm nn}$. (b) A depiction of a longer-range CPS possessing next-nearest-neighbour correlators $\Upsilon_{\rm nnn}$, shown for one site only for clarity. (c) Plaquette correlators $\Upsilon_{\rm plaq}$ spanning four sites. A given site is a member of four overlapping plaquettes giving the patchwork quilt pattern shown. (d) A depiction of a variety of extensive string correlators $\Upsilon_{\rm string}$ threading across the system vertically, horizontally, diagonally, and along other paths. In all cases the colour of correlators is only to guide the eye.}
\label{fig_cps}
\end{figure}

\subsection{Correlator product states}\label{sec:cps}
A very flexible approach for constructing ansatzes with sampleable amplitudes $\Psi({\bf v})$ are correlator product states~\cite{Changlani2009}. The essential idea is to use quantum states of subsets of sites as {\em correlators}, and construct a full state of the system by overlapping the states of many such subsets. Suppose we have a covering of our lattice $\mathcal{C}$ composed of $\ell$-site subsets, the $c$-th member being specified by sites $\{x_1(c),x_2(c),\dots,x_{\ell}(c)\}$. Each subset has an $\ell$-site correlator $\Upsilon^{(c)}_{v_1v_2\cdots v_\ell}$, comprising of $2^\ell$ complex numbers, associated to it and a CPS is formed as the product of overlapping amplitudes 
\begin{equation}
\kets{\Psi_{\rm CPS}} = \sum_{\bf v} \prod_{c \in \mathcal{C}} \Upsilon^{(c)}_{v_{x_1(c)}v_{x_2(c)}\cdots v_{x_{\ell}(c)}}\ket{\bf v}. \label{eq:correlator_product_state}
\end{equation}
Consequently the amplitudes $\braket{{\bf v}}{\Psi_{\rm CPS}}$ reduce to products of elements of the correlators $\Upsilon^{(c)}_{v_1v_2\cdots v_\ell}$. 

A common choice for CPS is to use completely general structureless correlators, which limits their size $\ell$ to a small number of sites. The simplest example is a two-site correlator $\Upsilon^{(ij)}_{v_iv_j}$ associated to all nearest neighbouring pairs $\langle ij\rangle$, as depicted in \fir{fig_cps}(a). This can be easily extended to longer-ranged two-site correlators by enlarging the covering set, as illustrated in \fir{fig_cps}(b) for one site. The extreme limit of this is where every pair of sites shares a two-site correlator, giving a special subclass of CPS called the {\em complete-graph tensor network} ansatz~\cite{Marti2010}. Increasing the size of correlators allows for the inclusion of plaquettes of the underlying lattice with an overlapping covering. This subclass of CPS, shown in \fir{fig_cps}(c), are also referred to as {\em entangled plaquette states}~\cite{Mezzacapo2009}. Other geometrical arrangements and covering are possible and potentially desirable~\cite{Neuscamman2011}. 

The CPS formalism equally applies to extensively-sized correlators, e.g. correlators that involve all or a fraction of the total system. To avoid the curse of dimensionality extensive correlators cannot be generic and must themselves posses internal structure. One possibility is to use a tensor network decomposition of the correlators. For instance, employing an MPS structure naturally allows correlators to thread across the whole system, as shown in \fir{fig_cps}(d). This subclass of CPS are also referred to as {\em string bond states}~\cite{Schuch2008}. A key observation of our work here is that NQS can be interpreted as a different tractable alternative for defining extensive correlators, as we shall see shortly in \secr{sec:tensor_network_nqs}.

\section{Expressing CPS with TNT}\label{sec:cps_tnt}
While contractibility is a key property for the conventional tensor network approach, the diagrammatic tensor formalism is both applicable and useful beyond this for describing sampleable ansatzes. Here we shall illustrate this by formulating CPS as sampleable tensor networks. 

\subsection{The COPY tensor} \label{sec:copy_tensor}
The crucial ingredient we shall exploit frequently here is the COPY tensor~\cite{Biamonte2011,Denny2012}, defined for three indices to have the diagonal elements
\begin{equation}
\delta_{ijk} = 
\left\{
\begin{array}{cc}
1, & i = j = k \\
0, & {\rm otherwise} 
\end{array}
\right., \label{eq:copy_tensor}
\end{equation}
and so is zero unless all its indices are equal. It generalises to any number of indices straightforwardly and is denoted graphically as a dot $\bullet$ with legs for each index as shown in \fir{fig_copy_tensor}(a). The name COPY tensor reflects that if we interpret any single leg as an input qubit and terminate it with one of the basis states $\ket{0}$ or $\ket{1}$, then these states are copied to all the legs representing the output qubits\footnote{Consistent with the no-cloning theorem copying only occurs for inputs in this fixed basis.}, as shown in \fir{fig_copy_tensor}(b). Irrespective of its number of indices the COPY tensor factorises for basis state inputs. This algebraic property is the cornerstone of expressing a large class of many-body quantum states as {\em sampleable} tensor networks. Terminating a leg with $\ket{+} = \ket{0} + \ket{1}$ deletes it giving a COPY tensor with an order reduced by one, as shown in \fir{fig_copy_tensor}(c).

\begin{figure}[ht]
\begin{center}
\includegraphics[scale=0.5]{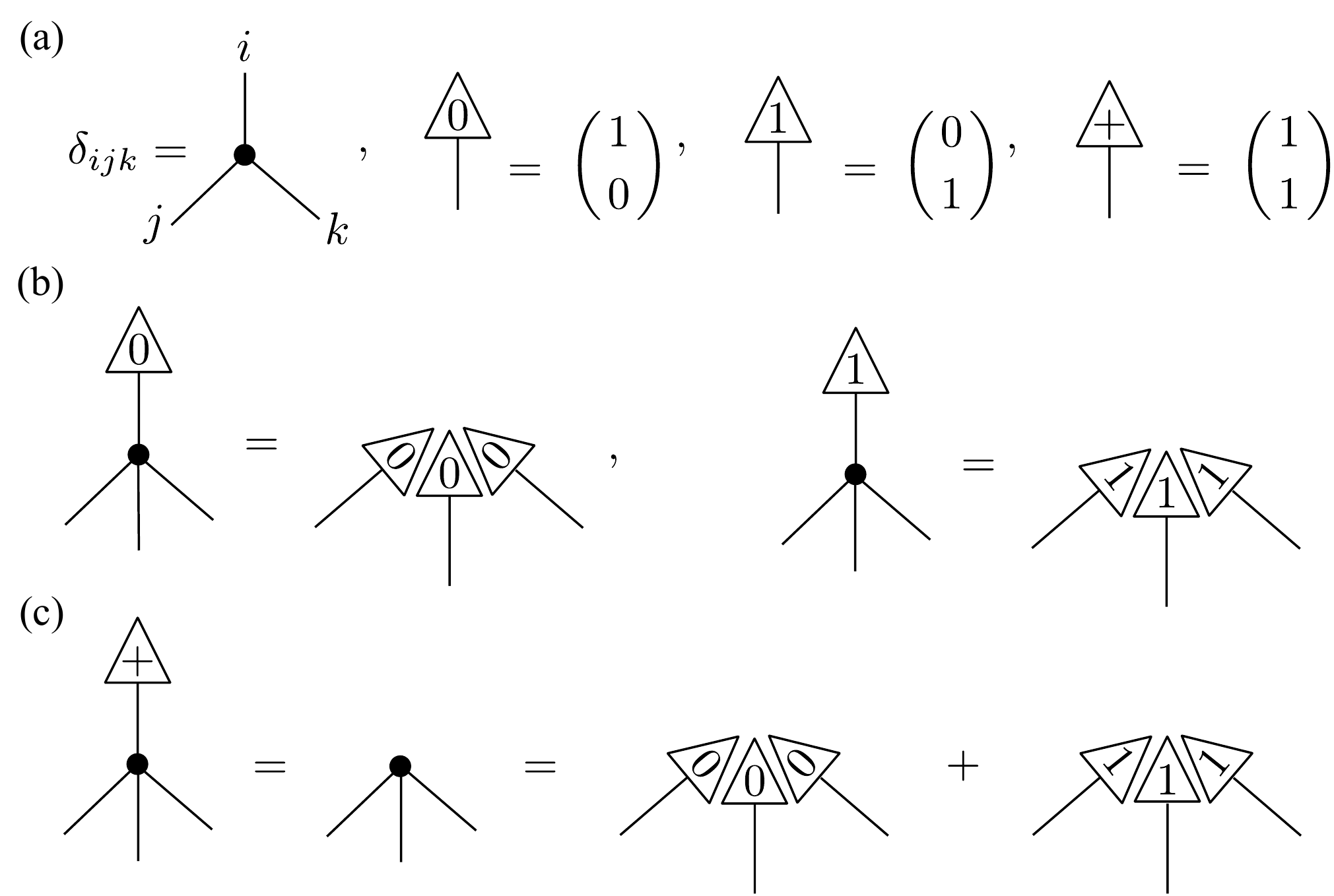}
\end{center}
\caption{(a) A diagram of the order-3 COPY tensor defined in \eqr{eq:copy_tensor}, along with diagrams for the order-1 tensors for the qubit states $\ket{0}$, $\ket{1}$ and $\ket{+}$. (b) The copy property in action for an order-4 COPY tensor when terminating any leg with the $\ket{0}$ and $\ket{1}$ state. Note that the COPY tensor factorises into a product of order-1 tensors. (c) Terminating any leg with a $\ket{+}$ deletes the leg leaving a COPY tensor one order less. Since $\ket{+}$ is a sum of $\ket{0}$ and $\ket{1}$ we can equally view this as showing the COPY tensor is a sum of products of order-1 tensors given in (b).}
\label{fig_copy_tensor}
\end{figure}

If two COPY tensors have one or more legs contracted together then they obey a ``fusion" rule allowing them to be amalgamated into one COPY tensor, as shown in \fir{fig_ghz}(a). The rule also applies in reverse so we can take a COPY tensor and split it up into an arbitrary network of connected COPY tensors with the same number of open legs. An immediate application of this is presented by an order-$N$ COPY tensor. We can interpret this tensor as amplitudes of an $N$ qubit GHZ state
\begin{equation}
\ket{\Psi_{\rm GHZ}} = \ket{0,0,0,\dots,0} + \ket{1,1,1,\dots,1},
\end{equation}
as depicted in \fir{fig_ghz}(b). Owing to its global correlations the GHZ state has no intrinsic geometry. This is reflected on a tensor level by using the fusion rule repeatedly to breakup the single COPY tensor into different networks that impose a given geometry. For example two splits isolate the central qubit, as shown in \fir{fig_ghz}(c). Further applications can then give an MPS network with a 1D geometry, as in \fir{fig_ghz}(d), or a  PEPS network with a 2D geometry, as in \fir{fig_ghz}(e). In both cases the internal dimension is the same as the physical dimension, so $\chi = 2$.

\begin{figure}[ht]
\begin{center}
\includegraphics[scale=0.5]{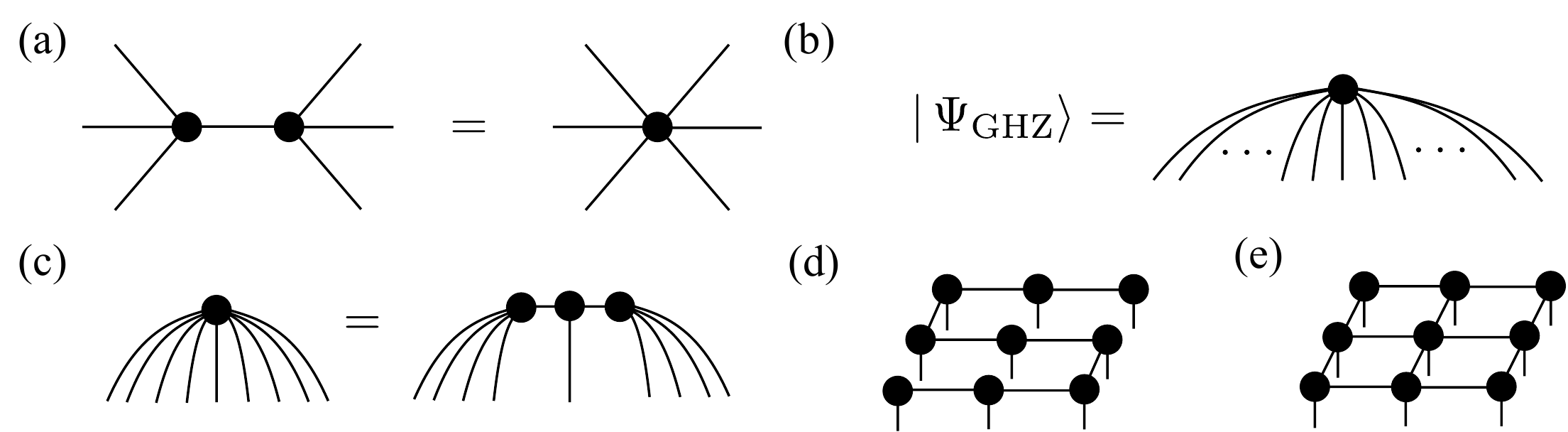}
\end{center}
\caption{(a) An order-$N$ COPY tensor is equivalent to the tensor of amplitudes for a GHZ state. (b) The COPY tensor has a {\em fusion rule} meaning that the contraction of one or more legs of two COPY tensors can be fused into a single COPY tensor. Likewise we can always breakup a COPY tensor into one or more contracted COPY tensors. (c) Applying the fusion rule two times to a 9 qubit GHZ state allows the central qubit to be isolated. (d) Applying the fusion rule 9 times decomposes the order-9 COPY tensor into a chain equivalent to an MPS tensor network. (e) Similarly the fusion rule allows the order-9 COPY tensor to be brought into a grid equivalent to a PEPS tensor network.}
\label{fig_ghz}
\end{figure} 

Naturally we can change the fixed basis of a COPY tensor by unitarily transforming its legs. For example by applying the $d=2$ Fourier matrix (Hadamard gate) $H_{jk} = (-1)^{jk}$ the COPY basis is transformed into the $\hat{\sigma}^x$ basis $\ket{\pm} = \ket{0} \pm \ket{1}$, as shown in \fir{fig_xor_tensor}(a). For an order-3 COPY tensor this gives a so-called XOR tensor defined as
\begin{equation}
\mathbbm{X}_{ijk} = 
\left\{
\begin{array}{cc}
1, & i \oplus j \oplus k = 0 \\
0, & {\rm otherwise} 
\end{array}
\right., \label{eq:xor_tensor}
\end{equation}
whose non-zero elements correspond to the truth table of a classical XOR gate. Its generalisation to higher order follows straightforwardly. The XOR tensor copies in the $\ket{\pm}$ basis, as shown in \fir{fig_xor_tensor}(b), and has legs deleted by termination with $\ket{0}$, as shown in \fir{fig_xor_tensor}(c). An order-$N$ XOR tensor is equivalent to the quantum state
\begin{equation}
\ket{\Psi_{\rm XOR}} = \sum_{\bf v} \Big(\mathcal{P}({\bf v}) \oplus 1\Big) \ket{\bf v},
\end{equation}
where $\mathcal{P}({\bf v}) = v_1 \oplus v_2 \oplus \cdots \oplus v_N$ is the parity function of the bit string $\bf v$. The state $\ket{\Psi_{\rm XOR}}$ is therefore an equal superposition of all configurations states with even parity. 
 
\begin{figure}[ht]
\begin{center}
\includegraphics[scale=0.5]{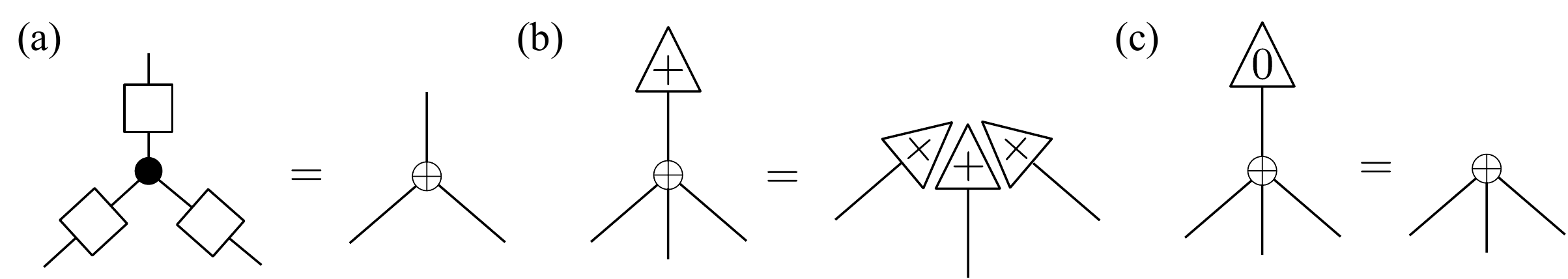}
\end{center}
\caption{(a) Changing the basis on each leg of a COPY tensor to the eigenstates $\ket{\pm}$ of $\hat{\sigma}^x$ via a Hadamard gate ($\square$) gives the XOR tensor, shown here for order-3. (b) The XOR generalises to any order and copies the $\ket{\pm}$ basis. (c) Terminating with the state $\ket{0}$ now deletes legs from the XOR tensor.}
\label{fig_xor_tensor}
\end{figure} 
 
\subsection{CPS as sampleable tensor networks}\label{sec:cps_networks}
Building many-body ansatzes as products of overlapping states for subsets of sites is easily expressed as a tensor network using the tools from \secr{sec:copy_tensor}. Each generic correlator $\Upsilon_{v_1v_2\cdots v_\ell}$ is an order-$\ell$ tensor whose indices are glued together by a COPY tensor of the physical index~\cite{Alassam2011}. In \fir{fig_cps_tensor}(a) the tensor network equivalent to the pairwise nearest-neighbour CPS in \fir{fig_cps}(a) is shown, along with the next-nearest-neighbour example from \fir{fig_cps}(b) in \fir{fig_cps_tensor}(b). Similarly the plaquette CPS from \fir{fig_cps}(c) results in the tensor network in \fir{fig_cps_tensor}(c). While the nearest-neighbour and plaquette CPS networks bare a strong resemblance to PEPS, and so could be approximately contracted, this is not at all guaranteed to be the case for more general CPS, such as those with long-ranged pairwise correlators. Crucially so long as a CPS has poly$(N)$ number of correlators, each with a small bounded size, irrespective of what pattern they decorate the lattice, or how many correlators a given site is encompassed by, the factorising properties of the COPY tensor in \fir{fig_copy_tensor}(b) guarantees the tensor network is efficiently sampleable in the fixed basis. This is shown in \fir{fig_cps_tensor}(d) for nearest-neighbour and \fir{fig_cps_tensor}(e) plaquette CPS. We can therefore view CPS as a very broad and flexible class of sampleable tensor networks.   
 
\begin{figure}[ht]
\begin{center}
\includegraphics[scale=0.5]{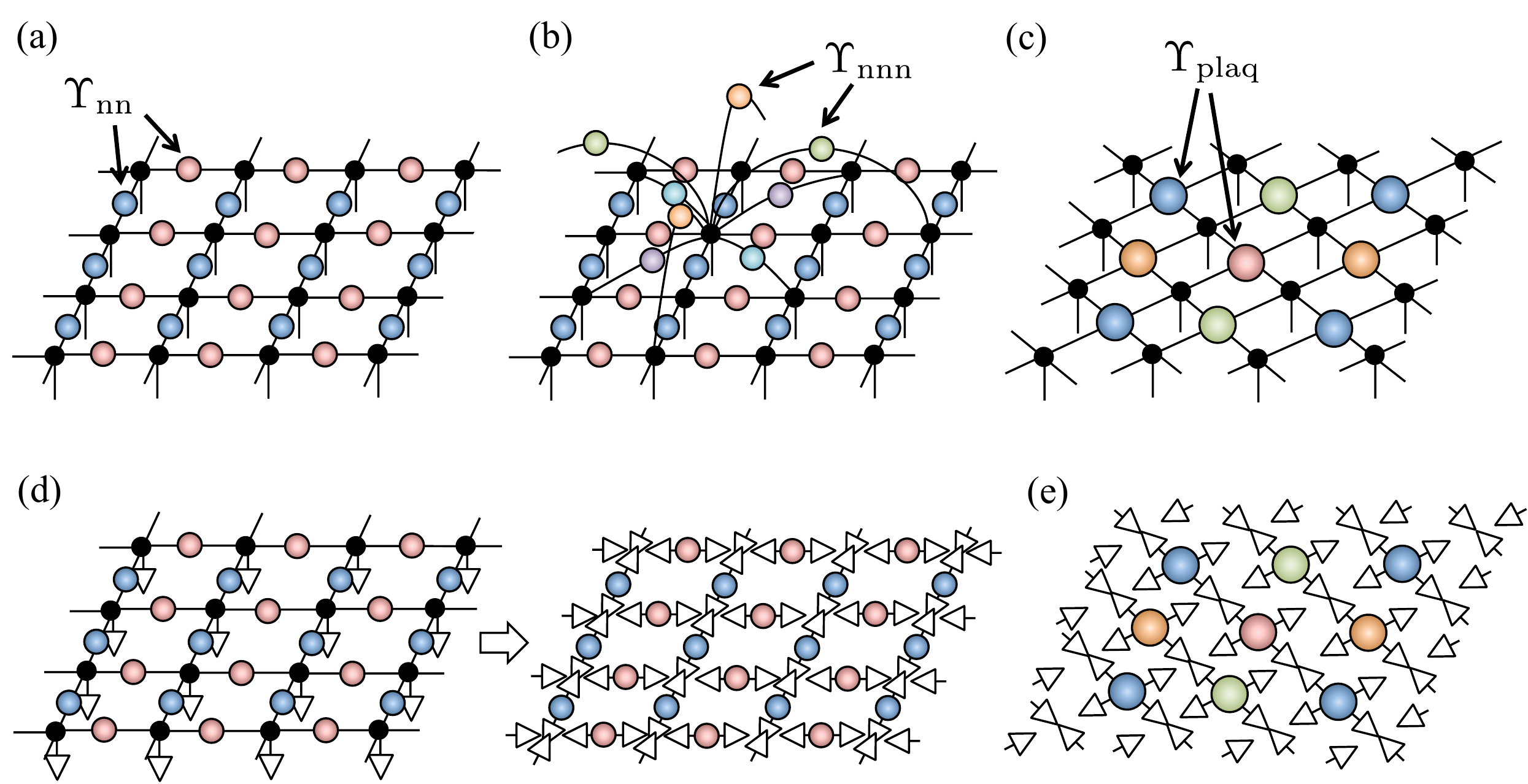}
\end{center}
\caption{(a) The tensor network of the nearest-neighbour CPS shown in \fir{fig_cps}(a) where each correlator $\Upsilon_{\rm nn}$ is a $2 \times 2$ order-2 tensor glued together by COPY tensors. (b) The generalisation to next-nearest-neighbour correlators is shown equivalent to \fir{fig_cps}(b) for a single site. (c) Similarly the plaquette CPS shown in \fir{fig_cps}(c) is built from $2 \times 2 \times 2 \times 2$ order-4 tensors $\Upsilon_{\rm plaq}$. (d) Sampling the tensor network involves terminating each physical index with an input state as depicted by $\triangle$ order-1 tensors. Here it is shown for the nearest-neighbour CPS tensor network from (a). If these are states from our chosen fixed basis set then the COPY tensor factorises leaving a product of the pairwise correlator elements in this case. (e) For the plaquette CPS from (c) sampling gives a product of the plaquette correlator elements.}
\label{fig_cps_tensor}
\end{figure}  

A similar construction applies to CPS with extensive correlators, but with additional constraints. For string bond states each correlator $\Upsilon_{\rm string}$, as in \fir{fig_cps}(d), is itself decomposed as MPS tensor network. In \fir{fig_cps_string}(a) the resulting tensor network for a CPS composed of overlapping horizontal and vertical strings is shown. Terminating the physical indices with basis states again factorises the COPY tensors, this time leaving a product of MPS similarly terminated, as illustrated in \fir{fig_cps_string}(b). The sampleability of a string bond type CPS is then inherited from COPY tensor factorability {\em and} the efficient contractibility of the MPS involved~\cite{Schuch2008}. 

\begin{figure}[ht]
\begin{center}
\includegraphics[scale=0.5]{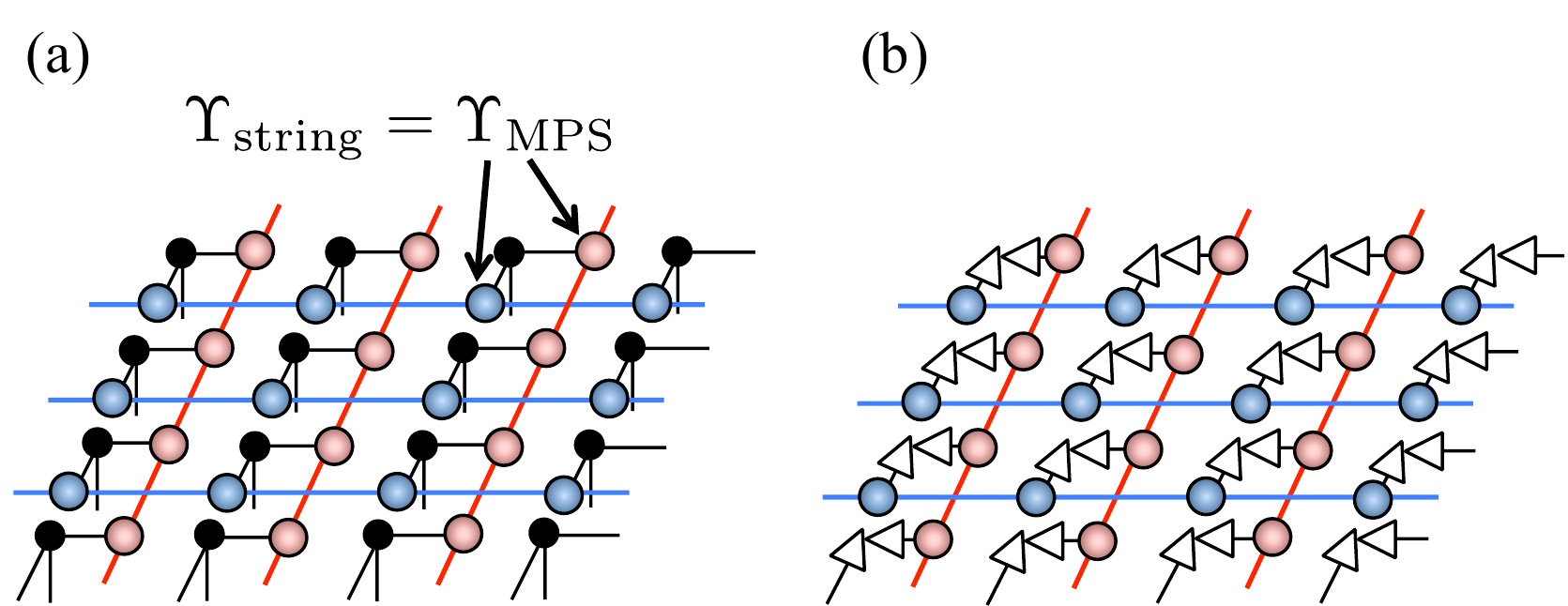}
\end{center}
\caption{(a) A string bond state like \fir{fig_cps}(d) has extensive correlators $\Upsilon_{\rm string}$ each of which is decomposed itself as MPS tensor networks. This gives a complete CPS tensor network shown here for a state composed of overlapping horizontal and vertical strings. (b) Sampling the tensor network with our fixed basis states causes the COPY tensors to factorise leaving a product of sampled MPS.}
\label{fig_cps_string}
\end{figure} 
 
\subsection{CPS for coherent thermal states} \label{sec:coherent_thermal}
A classical thermal probability distribution can be described exactly within the CPS and tensor network formalism~\cite{Changlani2009}. Given a lattice system composed of discrete classical binary units with configurations $\bf v$ the Boltzmann distribution follows as $p_{\bm \gamma}({\bf v}) = \exp[-E_{\bm \gamma}({\bf v})]/Z_{\bm \gamma}$ with partition function $Z_{\bm \gamma} = \sum_{{\bf v}} \exp[-E_{\bm \gamma}({\bf v})]$. Here ${\bm \gamma} = \{{\bf a}, {\bf K}\}$ denotes the parameters of the energy function $E_{\bm \gamma}({\bf v})$, which for this case are taken to be an $N \times N$ upper-triangular matrix $\bf K$ specifying pairwise Ising couplings $K_{ij}$, and a vector of local fields ${\bf a} = (a_1,a_2,\dots,a_N)^{\rm T}$. Together these define the energy function for the system as
\begin{equation}
E_{\bm \gamma}({\bf v}) = -{\bf v}^{\rm T}{\bf K}{\bf v} - {\bf a}^{\rm T}{\bf v}.
\end{equation}
The (unnormalised) probabilities $p_{\bm \gamma}({\bf v})$ have a CPS description built from two-site correlators 
\begin{equation}
\Upsilon^{(ij)}_{v_iv_j} = \exp\left(v_iK_{ij}v_j + \frac{a_i}{\mathcal{Z}_i}v_i + \frac{a_j}{\mathcal{Z}_j}v_j\right), \label{eq:thermal_correlator}
\end{equation}
defined between every pair of sites $i,j$ with a coupling $K_{ij} \neq 0$, and where $\mathcal{Z}_i$ is the coordination of site $i$. The connectivity of the CPS follows directly from $\bf K$ and so if only nearest-neighbour couplings are present on a 2D square lattice the thermal distribution has a tensor network of the form given in \fir{fig_cps_tensor}(a).    

In the context of quantum systems associating the Boltzmann distribution as amplitudes gives a {\em coherent thermal state}
\begin{equation}
\ket{\Psi_{\rm therm}} = e^{-\hat{H}_{\rm cl}}\ket{+}\ket{+}\cdots\ket{+} = \sum_{\bf v} e^{-E_{\bm \lambda}({\bf v})} \ket{\bf v},
\end{equation}
where $\hat{H}_{\rm cl} = -\sum_{i,j=1}^N \mathbbm{P}_i K_{ij} \mathbbm{P}_j - \sum_{i=1}^N a_i \mathbbm{P}_i$ and $\mathbbm{P}_i = \frac{1}{2}(\mathbbm{1} - \hat{\sigma}^z_i)$ are local projectors. With the addition of complex phases this form of quantum state has found application in describing frustrated spin systems~\cite{Huse1988}.  
 
\section{Constructing neural-network quantum states}\label{sec:nqs_overview} 
Having introduced tensor networks and CPS we now move on to define neural-network quantum states. Their origin is grounded in classical probabilistic models that generalise the coherent thermal states introduced in \secr{sec:coherent_thermal}. 
 
\subsection{Restricted Boltzmann machines}\label{sec:rbm}
A more general set of classical probabilistic models are {\em Boltzmann machines}~\cite{Goodfellow2016}. In addition to the $N$ physical or `visible' units these include $M$ hidden binary units whose configuration is specified by ${\bf h} = (h_1,h_2,\dots,h_M)^{\rm T} \in \{0,1\}^M$. This system of classical units is governed by a pairwise energy function 
\begin{equation*}
E_{\bm \lambda}({\bf v},{\bf h}) = -{\bf h}^{\rm T}{\bf W} {\bf v} - {\bf v}^{\rm T}{\bf K} {\bf v} - {\bf h}^{\rm T}{\bf Q} {\bf h} - {\bf a}^{\rm T} {\bf v} - {\bf b}^{\rm T} {\bf h}, 
\end{equation*}
whose parameters ${\bm \lambda} = \{{\bf a},{\bf b}, {\bf W},{\bf K}, {\bf Q}\}$ are extended to include Ising coupling between the visible and hidden units $\bf W$, the hidden units with themselves $\bf Q$, and local fields ${\bf b} = (b_1,b_2,\dots,b_M)^{\rm T}$ on the hidden units. The full joint probability distribution is then thermal as $p_{\bm \lambda}({\bf v},{\bf h}) = \exp[-E_{\bm \lambda}({\bf v},{\bf h})]/Z_{\bm \lambda}$ with $Z_{\bm \lambda} = \sum_{{\bf v},{\bf h}} \exp[-E_{\bm \lambda}({\bf v},{\bf h})]$. However, the marginal distribution for the visible units comprising our system $p_{\bm \lambda}({\bf v}) = \sum_{\bf h} p_{\bm \lambda}({\bf v},{\bf h})$ can be non-thermal. The inclusion of hidden units significantly broadens the probability distributions $p_{\bm \lambda}({\bf v})$ captured by the model.

In the context of neural networks {\em restricted Boltzmann machines} (RBMs) are a popular subclass of Boltzmann models in which ${\bf K} = {\bf Q} = 0$~\cite{Fischer2012}. They can therefore be viewed as a two-layer system in which only interlayer Ising couplings $\bf W$ are permitted between visible units in the lower layer and hidden units in the upper layer, as shown in \fir{fig_rbm}. The geometry of this bipartite graph has important implications. Namely that the visible and hidden variables are conditionally independent
\begin{eqnarray*}
p_{\bm \lambda}({\bf h}|{\bf v}) &=&  \prod_{i=1}^M p_{\bm \lambda}(h_i|{\bf v}), \quad {\rm and} \quad  p_{\bm \lambda}({\bf v}|{\bf h}) = \prod_{j=1}^N p_{\bm \lambda}(v_j|{\bf h}),
\end{eqnarray*}
with $p_{\bm \lambda}(h_i=1|{\bf v}) = \sigma\left(b_i + \sum_{j=1}^N W_{ij}v_j\right)$ and $p_{\bm \lambda}(v_j=1|{\bf h}) = \sigma\left(a_i + \sum_{i=1}^M h_iW_{ij}\right)$, where $\sigma(x) = 1/(1+e^{-x})$ is the sigmoid function. This enables efficient block Gibbs sampling of an RBM. Furthermore the absence of connections between hidden units makes the marginal distribution of the visible units straightforward to compute as~\cite{Goodfellow2016} 
\begin{eqnarray}
\fl \qquad 
p_{\bm \lambda}({\bf v}) &=& \frac{1}{Z_{\bm \lambda}}\sum_{\bf h} e^{-E_{\bm \lambda}({\bf v},{\bf h})} = \frac{1}{Z_{\bm \lambda}}\left\{\prod_{j=1}^N e^{a_jv_j}\right\}  \prod_{i=1}^M \left(1 + e^{b_i + \sum_{j=1}^N W_{ij}v_j}\right). \label{eq:rbm_marginal}
\end{eqnarray}
The goal of machine learning with RBMs is to obtain a set of parameters $\bm \lambda$ generating a distribution $p_{\bm \lambda}({\bf v})$ that is as close as possible to an unknown distribution $p_{\rm data}({\bf v})$ governing the data. This is achieved by minimising the Kullback-Leibler divergence $D_{\rm KL}(p_{\rm data}|| p_{\bm \lambda})$ between the two distributions. Although the partition function $Z_{\bm \lambda}$ of an RBM is intractable, both $p_{\bm \lambda}({\bf v})$ and its derivatives with respect to components of $\bm \lambda$ can be efficiently sampled using Markov-chain Monte Carlo via block Gibbs sampling~\cite{Goodfellow2016}. The optimisation problem can then be solved using a gradient descent algorithm. Alternatively, a computationally cheaper proxy for the KL divergence, like the contrastive divergence~\cite{Hinton2000}, can be used to find a good approximate model distribution prior to fine-tuning.  

\begin{figure}[ht]
\begin{center}
\includegraphics[scale=0.5]{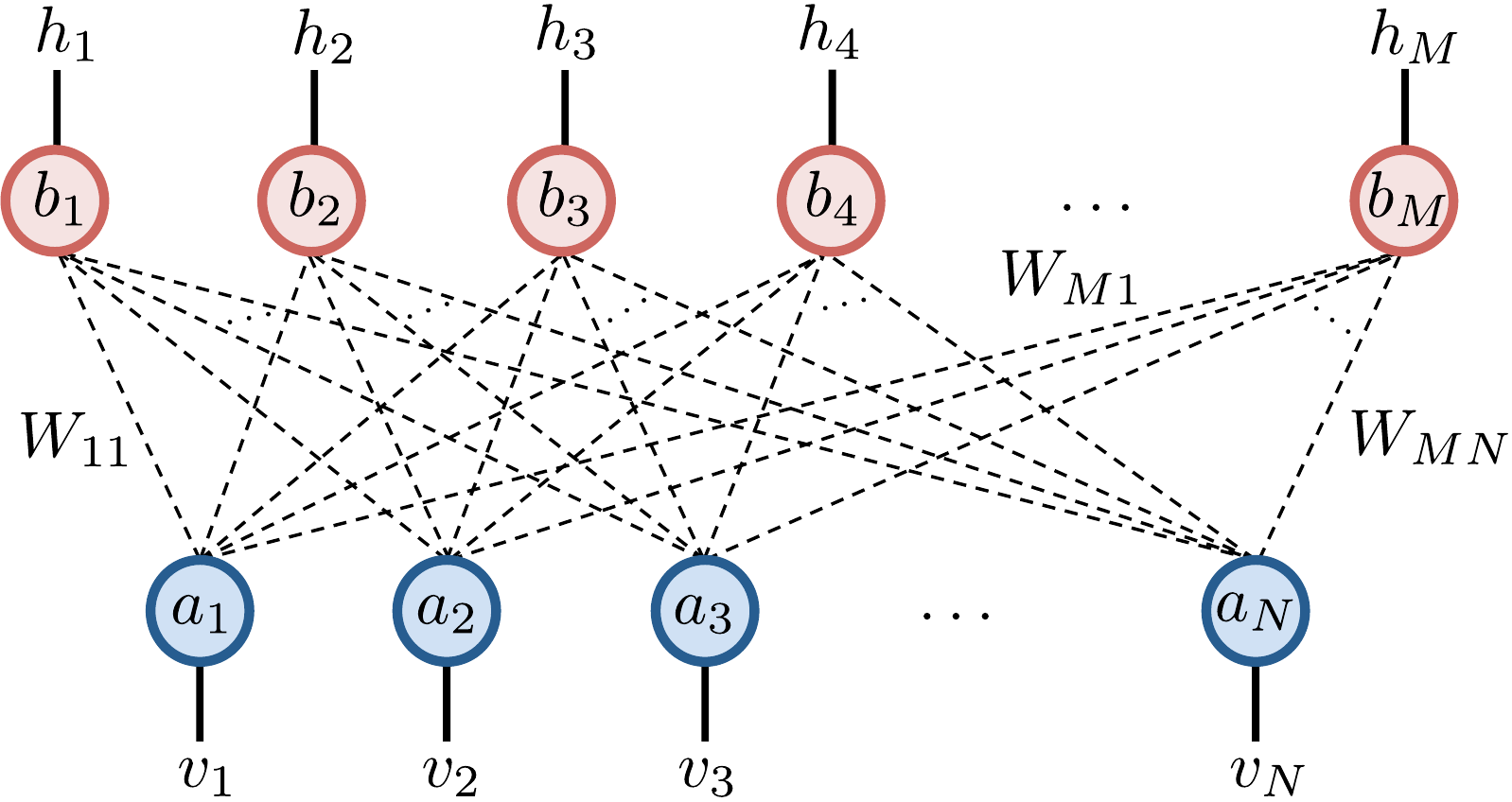}
\end{center}
\caption{The undirected bipartite graph of an RBM with $N$ visible and $M$ hidden spins. The elements of the binary vectors ${\bf v}$ and ${\bf h}$ are shown, along with the local fields ${\bf a}$ and ${\bf b}$. A few Ising couplings for the connections specified by $\bf W$ are also labelled.}
\label{fig_rbm}
\end{figure}

When applied to images, for example like the MNIST dataset of handwritten numerical digits~\cite{MNISTwebsite}, visible configurations $\bf v$ specify the `on' pixels. Optimisation of $p_{\bm \lambda}({\bf v})$ is performed by sampling $p_{\rm data}({\bf v})$ via a training set of example images. Once trained an RBM can reveal correlations and patterns in the data with intuitive interpretations. Typically the resulting couplings $\bf W$ correspond to features like pen strokes, and the activation of certain sets of hidden units by input images $\bf v$ are strongly linked a given digit. As such these extracted features can then be used for pattern recognition on new data, or can be used to generate new samples according to the optimised model distribution~\cite{Goodfellow2016}. 

\subsection{Neural-network Quantum States}\label{sec:nqs}
The NQS ansatz is an extended form of coherent thermal state in which the complex amplitudes $\Psi({\bf v})$ are taken to be the `marginal distribution' of an RBM, as given in \eqr{eq:rbm_marginal}. Further generality is included by allowing complex parameters ${\bm \lambda} = \{{\bf a},{\bf b},{\bf W}\}$. After rewriting \eqr{eq:rbm_marginal} in the following form~\cite{Carleo2017}
\begin{eqnarray}
\fl \qquad \Psi_{\bm \lambda}({\bf v}) &=& \prod_{i=1}^M\left\{\prod_{j=1}^N e^{\frac{a_j}{M}v_j} + \prod_{j=1}^Ne^{W_{ij}v_j + \frac{a_j}{M}v_j + \frac{b_i}{N}}\right\} = \prod_{i=1}^M \Upsilon^{(i)}_{v_1v_2\cdots v_N}, \label{eq:nqs_form}
\end{eqnarray}
it becomes clear that NQS are CPS constructed from the product of $M$ extensive $N$-site correlators $\Upsilon^{(i)}_{v_1v_2\cdots v_N}$ associated to each hidden unit. From a tensor-network perspective these correlators are built from a sum of products of hidden-visible $2 \times 2$ coupling matrices ${\bf C}^{(ij)}$ as 
\begin{equation}
\fl \qquad \Upsilon^{(i)}_{v_1v_2\cdots v_N} = \sum_{h_i=0}^1 \prod_{j=1}^N C^{(ij)}_{h_iv_j} = C^{i1}_{0v_1}C^{(i2)}_{0v_2}\cdots C^{(iN)}_{0v_N} + C^{(i1)}_{1v_1}C^{(i2)}_{1v_2}\cdots C^{(iN)}_{1v_N}, \label{eq:nqs_correlators}
\end{equation}
and therefore have a very special tractable structure. We reproduce \eqr{eq:nqs_form} by choosing the coupling matrices to have elements
\begin{equation}
C^{(ij)}_{h_iv_j} = \exp\left(h_iW_{ij}v_j + \frac{b_i}{N}h_i + \frac{a_j}{M}v_j\right), \label{eq:rbm_coupling_mat}
\end{equation}
We will find that the coupling matrices ${\bf C}^{(ij)}$ between the $i$-th hidden and $j$-th visible unit, discussed further in \ref{app:coupling_mats}, are a more transparent way to parameterise an NQS than $\bm \lambda$.

As with the modelling of probability distributions $p_{\bm \lambda}({\bf v})$ the RBM ansatz becomes increasingly expressive as more hidden units are added \cite{LeRoux2008}. Indeed if $\Psi({\bf v})$ possesses $k$ non-zero complex amplitudes then at most $k$ hidden units are needed for $\Psi_{\bm \lambda}({\bf v})$ to be an exact description, as shown in \ref{app:approximator}. An arbitrary state $\Psi({\bf v})$ can therefore be captured by NQS, but only at the price of using exponentially many hidden units. In contrast, once the number of hidden units scales at most as poly$(N)$ for a state then its NQS description is efficient. 

\subsection{Tensor network for NQS} \label{sec:tensor_network_nqs}
Since the full RBM distribution $p_{\bm \lambda}({\bf v},{\bf h})$ is thermal with pairwise interactions over the bipartite graph structure in \fir{fig_rbm}, it follows from \secr{sec:coherent_thermal} that it is described by CPS with pair correlators equal to the ${\bf C}^{(ij)}$ matrices. The tensor network for the NQS is then obtained by tracing out the classical hidden units, equivalent to quantum mechanically projecting, by terminating their legs with a $\ket{+}$, as shown in \fir{fig_nqs_tensor}(a). After using the results from \secr{sec:copy_tensor} the diagrammatic version of \eqr{eq:nqs_form} and \eqr{eq:nqs_correlators} emerge transparently as \fir{fig_nqs_tensor}(b) and \fir{fig_nqs_tensor}(c), respectively. 

\begin{figure}[ht]
\begin{center}
\includegraphics[scale=0.5]{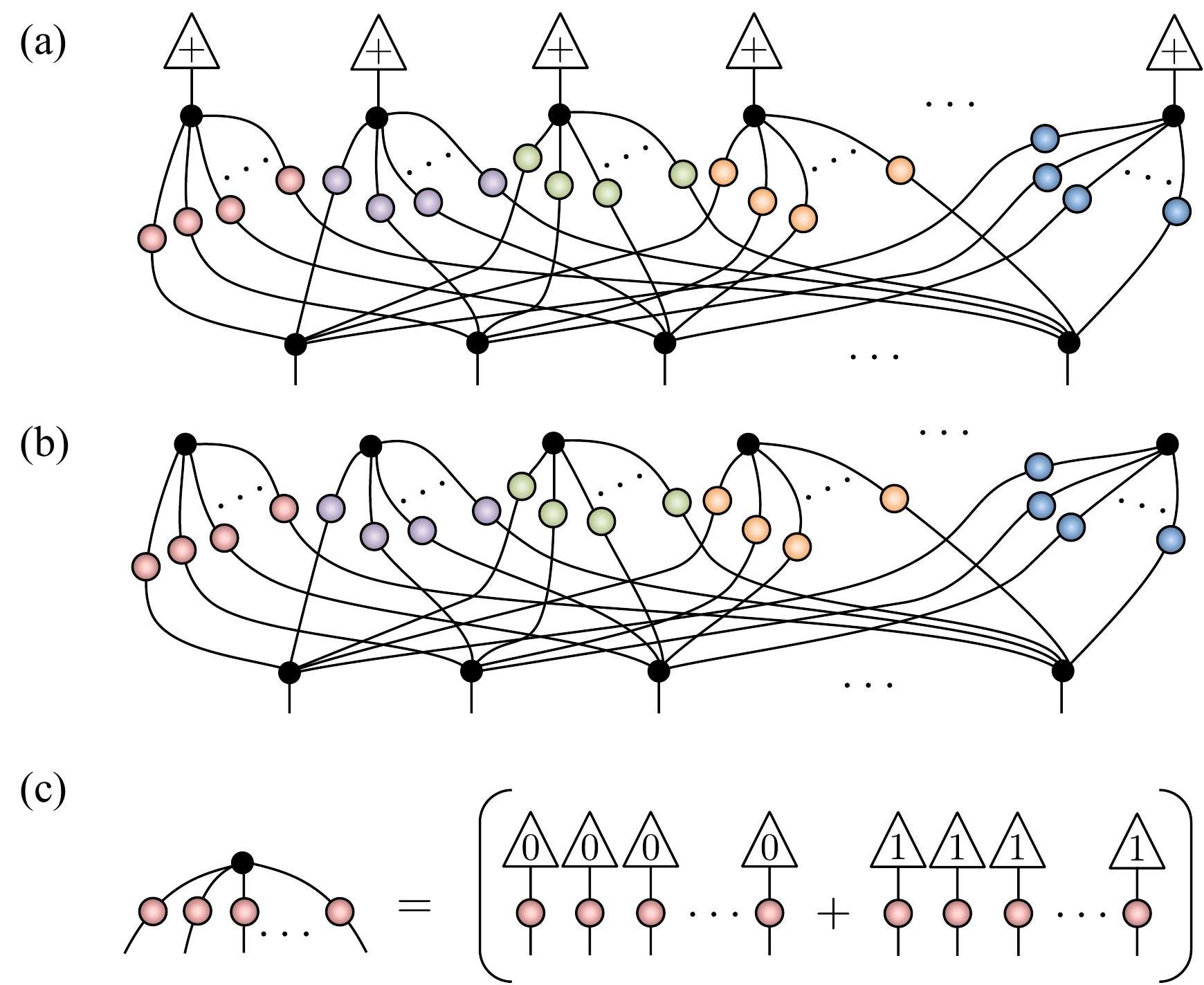}
\end{center}
\caption{(a) The tensor network built from coupling matrices ${\bf C}^{(ij)}$ according to the RBM geometry. The classical tracing out of the hidden spins corresponds to terminating their legs with $\ket{+}$. (b) Tracing out leaves a CPS with extensive correlators, as in \eqr{eq:nqs_form}. (c) Each correlator has the form of a GHZ-state and so can be broken into a sum of two terms that are the product of the coupling matrices, as in  \eqr{eq:nqs_correlators}.}
\label{fig_nqs_tensor}
\end{figure}
\noindent Viewed as a tensor network a number of useful observations about NQS emerge:

{\em Underlying GHZ correlators} -- For string bond states an MPS decomposition of an extensive correlator is used to ensure tractable sampling. For NQS an even simpler primitive is exploited that can also be sampled efficiently. Specifically, \fir{fig_nqs_tensor}(c) reveals that the underlying structure of $N$-site correlators $\Upsilon^{(i)}_{v_1v_2\cdots v_N}$ within an NQS is simply a GHZ-state. This basic unit of correlation is then modified locally for each visible unit via the coupling matrices ${\bf C}^{(ij)}$. Sampling an NQS amounts to \fir{fig_nqs_sampling}(a), which then breaks up into a product of GHZ form extensive correlators that can also be exactly and efficiently sampled as shown in \fir{fig_nqs_sampling}(b).

\begin{figure}[ht]
\begin{center}
\includegraphics[scale=0.5]{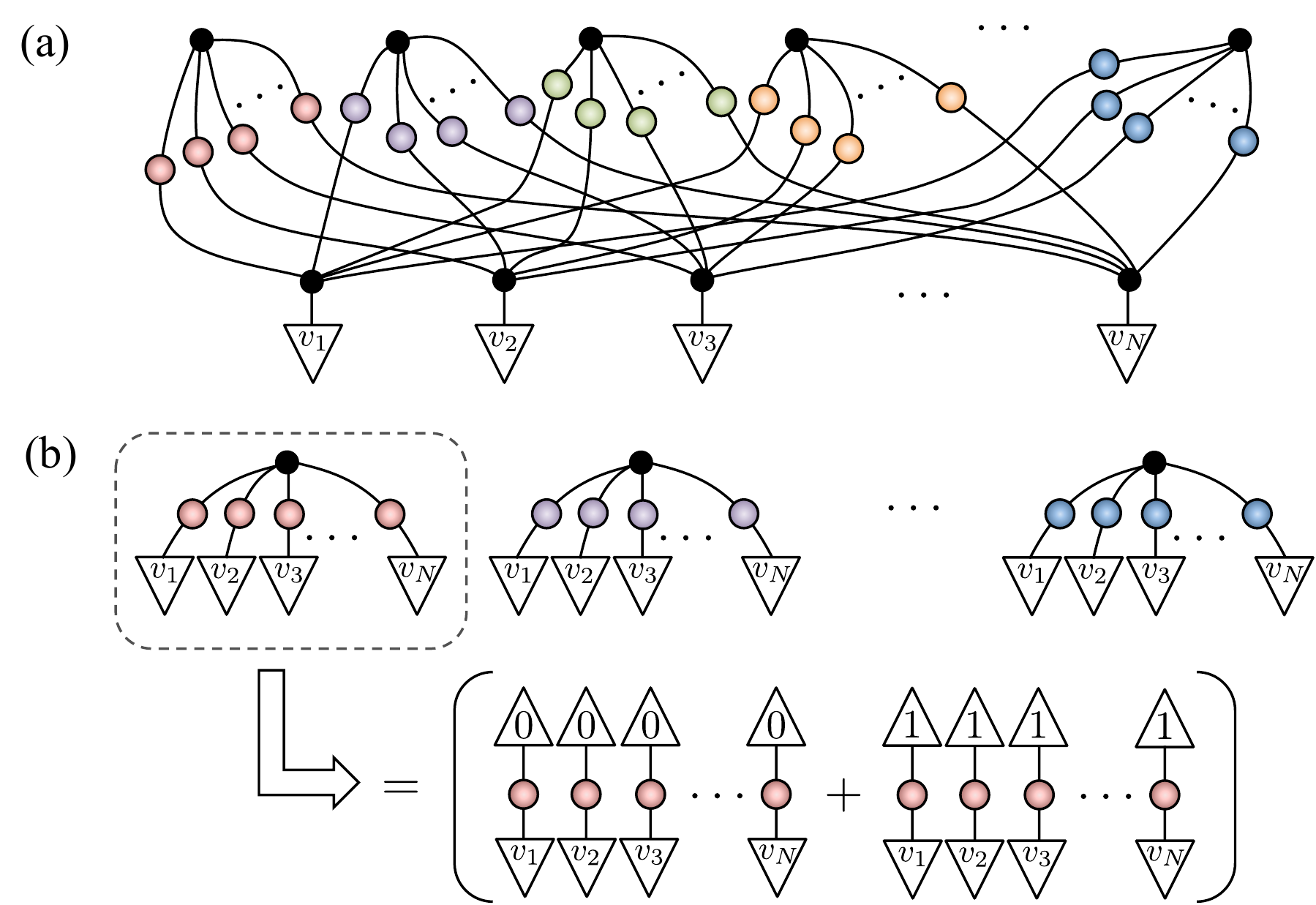}
\end{center}
\caption{(a) Sampling an NQS network involves terminating the physical. (b) Following \fir{fig_nqs_tensor}(c) this factorises into the product of GHZ form correlators, each of which can be individually sampled as shown.}
\label{fig_nqs_sampling}
\end{figure}

\begin{figure}[ht]
\begin{center}
\includegraphics[scale=0.5]{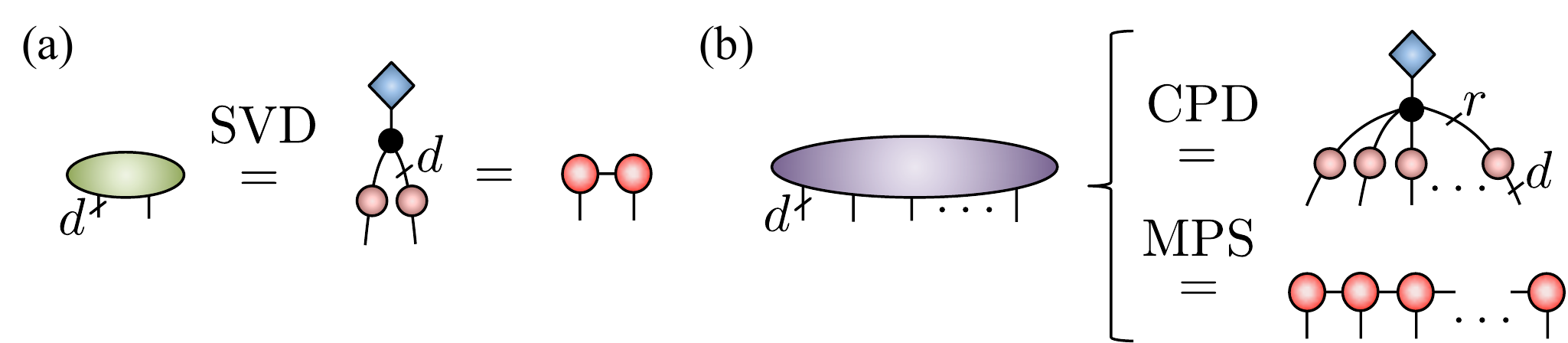}
\end{center}
\caption{(a) The SVD of a matrix (order-2 tensor) can be written in a GHZ form (the diagonal $\Diamond$ tensor of singular values can be absorbed into either or both of the unitaries), or MPS form. The slash notation through a leg denotes the dimension of that leg. The original matrix is considered to be $d \times d$, and so the GHZ form is made from a $d$-dimensional COPY tensor. (b) For an order-$N$ tensor, taken to be $d \times d \times \dots \times d$, a CPD decomposes it into a GHZ form built from an $r$ dimensional COPY tensor (where $r$ is the tensor rank), and the matrices on each leg are $r \times d$. Sequential application of the SVD can similarly decompose an order-$N$ tensor into an MPS.}
\label{fig_tensor_cpd}
\end{figure}

{\em Link to tensor decompositions} -- In terms of tensor networks there is a pleasing synergy between the MPS and GHZ correlators. For an order-2 tensor $\bf T$ the singular value decomposition (SVD) decomposes it into a product ${\bf U}{\bf S}{\bf V}^\dagger$ of unitary matrices $\bf U$ and $\bf V$ and a diagonal matrix of non-negative real numbers $\bf S$~\cite{Trefethen1997}. Diagrammatically we can view the SVD as bringing the $\bf T$ into equivalent GHZ or MPS forms, as shown in \fir{fig_tensor_cpd}(a). However, generalising the SVD for an order-$N$ tensor $T$ gives at least two inequivalent alternatives, shown in \fir{fig_tensor_cpd}(b). First, we could apply the SVD sequentially to $T$ bringing it into MPS form~\cite{Schollwock2011}. Second, we could use the direct multi-linear generalisation of the SVD called a canonical polyadic decomposition (CPD)~\cite{Hackbusch2016,Kolda2009}. This is where $T$ is factorised as
\begin{equation}
T_{v_1v_2\dots v_N} = \sum_{\alpha = 1}^r \lambda_\alpha \, A^{(1)}_{\alpha v_1}A^{(2)}_{\alpha v_2}\cdots A^{(N)}_{\alpha v_N},
\end{equation}
where ${\bf A}^{(j)}$ are component matrices for each index, $\lambda_\alpha$ are non-negative coefficients, and $r$ is the rank of $T$, i.e. the minimum number of terms for the decomposition to be exact. The COPY tensor in this case is the $r$ dimensional generalisation of that introduced in \secr{sec:copy_tensor}. Examining \eqr{eq:nqs_correlators} we see that hidden unit correlators are therefore equivalent to a CPD with its rank restricted to $r=2$ owing to them being binary. 

{\em Higher dimensional hidden units} -- This raises an interesting question about whether it is desirable to allow the hidden units to be higher dimensional degrees of freedom, analogous to MPS and PEPS bond dimension $\chi$. The answer appears to be in the affirmative. First, higher dimension hidden units can be easily handled within Monte Carlo so long as $r$ is not intractably large. Second, in \ref{app:cpd_to_nqs} it is shown that a general rank $r$ CPD, which involves the superposition of $r$ states, is equivalent to a more complex two-layer RBM network\footnote{Special cases exist where a tensor with a rank $r>2$ can still be described by a single hidden layer exist, as we shall see.} composed of $r$ and $\lceil \log_2(r) \rceil$ binary hidden units, respectively. This is depicted in \fir{fig_cpd_to_nqs}. Sampleability of this two-layer network relies on exhaustively summing the $r$ configurations of the second hidden layer. We will naturally encounter higher dimension hidden units in several NQS examples described in the \secr{sec:examples}.

\begin{figure}[ht]
\begin{center}
\includegraphics[scale=0.5]{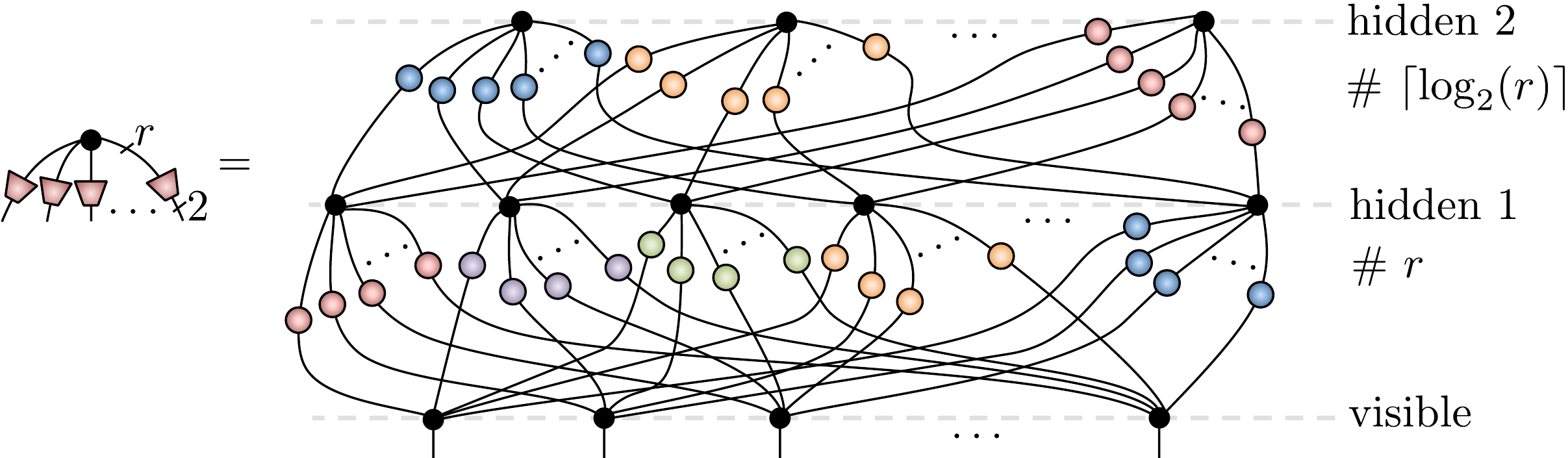}
\end{center}
\caption{A CPD already has an RBM-like geometry based around a higher dimensional hidden unit. It can be put into an NQS form based on only binary hidden units by using a two-hidden-layer geometry composed of $r$ and $\lceil \log_2(r) \rceil$ hidden units, respectively. See \ref{app:cpd_to_nqs} for details.}
\label{fig_cpd_to_nqs}
\end{figure}

{\em Adaptable receptive fields} -- While MPS correlators in string bond states can be considerably more complex than GHZ ones, they do suffer from having intrinsically one dimensional short-ranged correlations. A highly desirable feature of NQS is that GHZ states have no underlying geometry and possess infinite-ranged correlations. This makes them a useful building block for strongly correlated states. Specifically, the one-to-all connections of a hidden unit to the visible units is entirely adaptable. By fixing the coupling matrix to the projector $\ket{+}\bra{+}$ as
\begin{equation}
{\bf C}_{\rm dis}^{(ij)} = \left[
\begin{array}{cc}
1 & 1  \\
1 & 1  
\end{array}
\right],
\end{equation} 
connections are deleted and a hidden unit will only `talk' to a relevant subset of visible units. In the parlance of machine learning this would define a {\em local receptive field} of the hidden neuron~\cite{Goodfellow2016}. Variational minimisation of the NQS therefore has the capability to localise the receptive field of hidden units to capture local correlations and constraints of a state, provided enough hidden units are available. This is nicely illustrated by some examples of non-trivial states, e.g. toric code states, with exact NQS representations in \secr{sec:examples}.

\begin{figure}[ht]
\begin{center}
\includegraphics[scale=0.5]{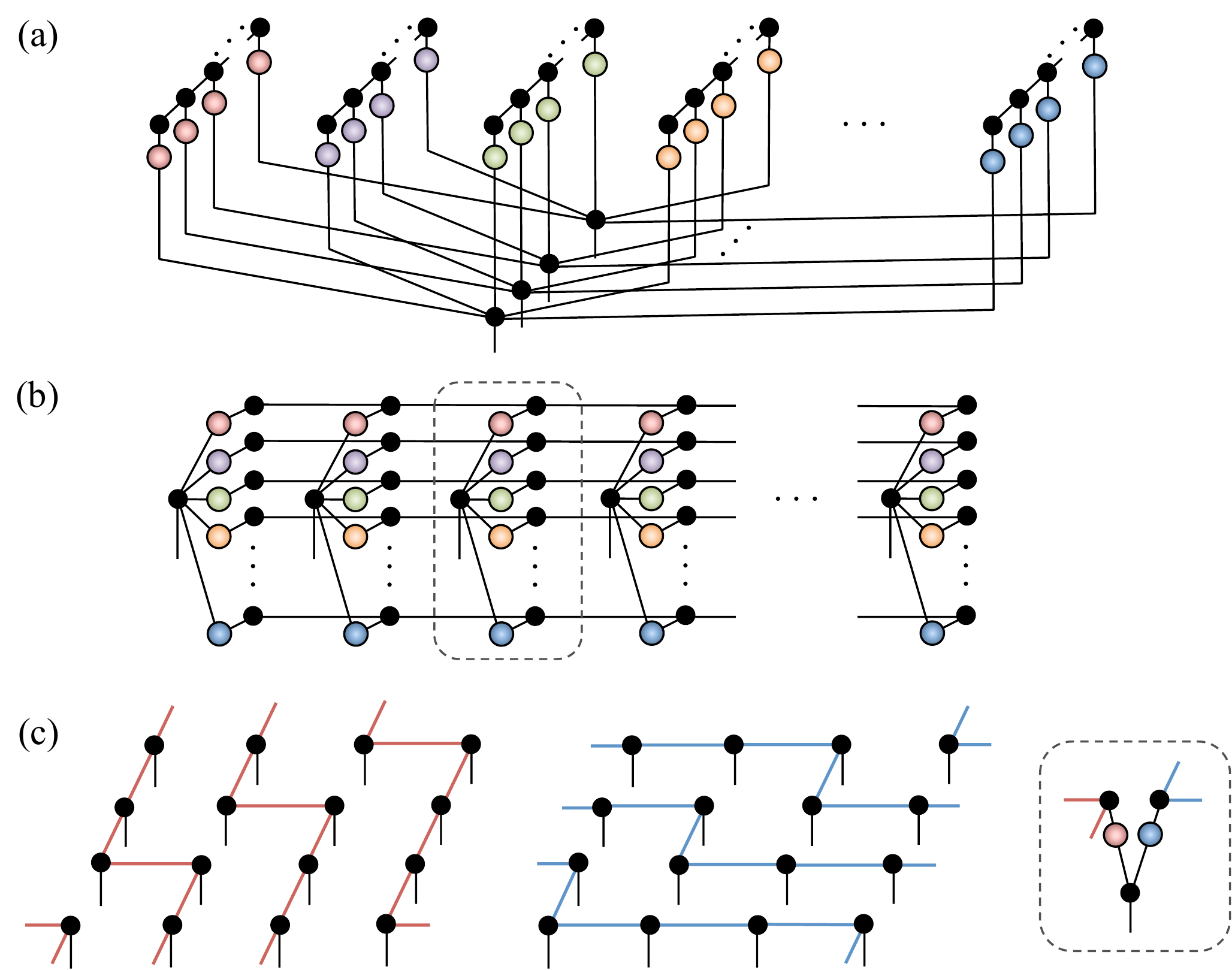}
\end{center}
\caption{(a) The fusion rule is used to split the COPY tensor into a chain of $N$ COPY tensors, like in \fir{fig_ghz}(d), each with one of the coupling matrices attached to it. Compared \fir{fig_nqs_tensor}(b) the visible spins have also been organised in a linear arrangement. Here we assume the most general case where every hidden unit correlator is fully connected, i.e. there are no ${\bf C}_{\rm dis}$ coupling matrices. (b) We can then bundle together all the COPY tensors and coupling matrices involving a given site. For each site these tensors could in principle be merged by contraction to form an MPS tensor (as highlighted in the dashed box) with a `fat' internal leg made of $M$ legs merged with a dimension $2^M$. An open boundary MPS is shown, but it is equally valid to have legs looping around and connecting the first and last tensors to give a periodic boundary MPS. (c) For a $4 \times 4$ square lattice two non-overlapping spanning periodic MPS arrangements for the COPY tensors are shown. This generalises straightforwardly for bigger lattices. Merging two hidden unit correlators with these MPS patterns will give a periodic boundary PEPS network with $\chi = 2$ (as highlighted by dashed box).}
\label{fig_nqs_to_mps}
\end{figure}

{\em Connections to MPS and PEPS} -- The geometric freedom of the GHZ state, reflected by the fusion rule for the COPY tensor shown in \fir{fig_ghz}(a), means that we can deform a NQS correlator into other tensor networks. By following \fir{fig_ghz}(d) an NQS is converted into an MPS via the steps shown in \fir{fig_nqs_to_mps}. The end result of this construction is an MPS with tensors possessing an internal dimension $\chi = 2^M$, as already pointed out recently in Ref.~\cite{Chen2017}. This is a rather loose upper-bound and in practice the tensors may be highly compressible owing to their internal structure, e.g. as apparent from \fir{fig_nqs_to_mps}(b). A completely analogous construction can be applied to give a PEPS by using \fir{fig_ghz}(e) to arrange each correlator into a grid network of COPY tensors. Similarly the resulting PEPS tensors will have an internal dimension $\chi = 2^M$. However, a slightly more efficient scheme instead exploits the fact that for a 2D square lattice we can always weave two non-overlapping spanning loops across the lattice, as pictured in \fir{fig_nqs_to_mps}(c). Thus, by converting half the correlators to one type of periodic MPS and half into the other type, making the NQS a string-bond state, and then merging the networks we obtain a periodic boundary PEPS with $\chi = 2^{M/2}$.

\section{Examples of NQS} \label{sec:examples}
While we know that formally any state can have an NQS representation we do not currently have a comprehensive understanding of what states have an efficient description. As a starting point, in this section we exploit the tools developed to express a variety of nontrivial many-body quantum states exactly and efficiently as an NQS. This provides important clues as to what physics can be captured with only a small number of hidden units. The examples are quite diverse. We begin by highlighting the straightforward conversion of several well known exact CPS examples~\cite{Changlani2009} into the NQS formalism, and then introduce some new examples including the dimer state and resonating valence bond state. 

\begin{figure}[ht]
\begin{center}
\includegraphics[scale=0.5]{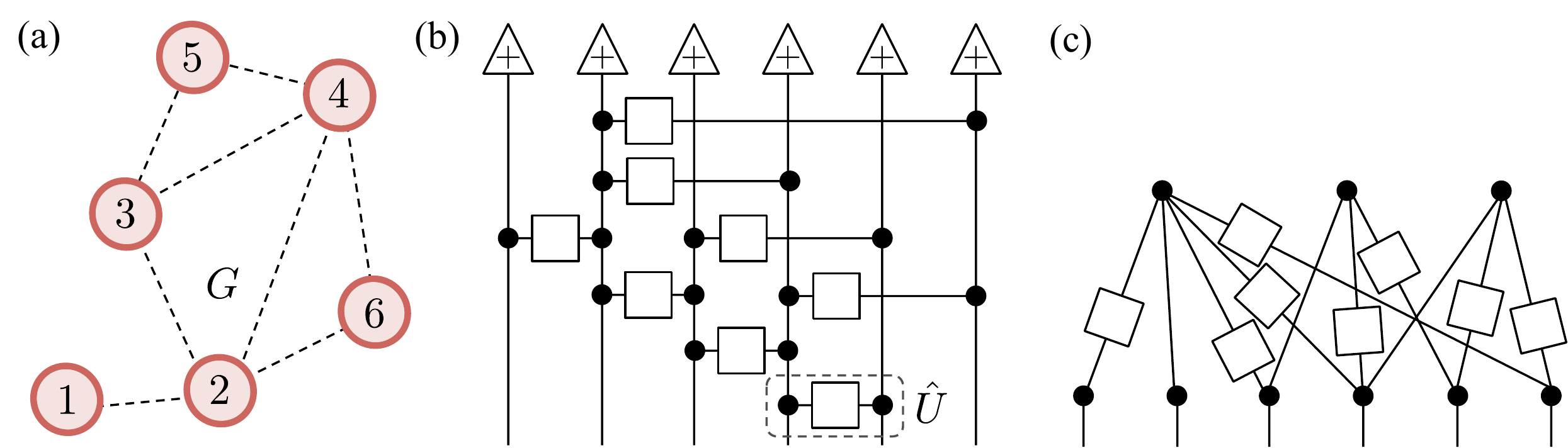}
\end{center}
\caption{(a) A graph $G$ whose edges (dashed lines) define a graph state of 6 qubits. (b) The quantum circuit defining $\ket{\Psi_G}$ from (a) is equivalent to a tensor network. The controlled-Z gates (one of which is highlighted by a dashed box) have been expressed as COPY tensors connected with a Hadamard gate (box) between them. (c) Using the COPY tensor rules this tensor network can be rearranged into an RBM geometry, in this case with 3 hidden units.}
\label{fig_nqs_graph_state}
\end{figure}

\subsection{Graph states} \label{sec:graph_states}
Graph states $\ket{\Psi_{G}}$ form an important resource for measurement based quantum computation~\cite{Raussendorf2001,Clark2005} and are defined for a set of qubits as 
\begin{equation}
\ket{\Psi_G} = \prod_{\langle i,j \rangle \in G} \hat{U}_{ij} \ket{+}\ket{+}\cdots\ket{+}, \label{eq:graph_state}
\end{equation}
where $G$ is a graph with qubits associated to the vertices and the edge $\langle i,j \rangle$ connecting the $i$-th and $j$-th qubits represents the application of a controlled-Z gate $\hat{U}_{ij} = \mathbbm{1}_i\otimes\mathbbm{1}_j + \mathbbm{1}_i\otimes\hat{\sigma}^z_i$ between them. Evidently graph states contain non-zero amplitudes for every configuration $\ket{\bf v}$ equal in magnitude, but with a sign structure imposed by the $\hat{U}$ gates. Nonetheless the amount of entanglement between a subsystem of $N_A$ qubits can scale as $O(N_A)$, and hence with its volume~\cite{Hein2004}. An example graph is shown in \fir{fig_nqs_graph_state}(a). Since all the $\hat{U}$ gates commute the amplitudes of $\ket{\Psi_G}$ have a pairwise correlator product form
\begin{equation}
\ket{\Psi_G} = \sum_{\bf v} \prod_{\langle i,j \rangle \in G} \Upsilon^{\rm H}_{v_iv_j} \ket{\bf v},
\end{equation}
where $\Upsilon^{\rm H}_{v_iv_j} = (-1)^{v_i v_j}$ and with the pattern of long-range correlators in the CPS following directly from the graph $G$. 

This CPS can be directly converted into an NQS by associating a hidden unit with each edge~\cite{Gao2017}. However, in this case each hidden unit only has a local receptive field composed of a pair of qubits. A more efficient NQS can be found by examining the quantum circuit underlying \eqr{eq:graph_state}, an example of which is shown in \fir{fig_nqs_graph_state}(b). A quantum circuit is equivalent to a tensor network. Since the $\hat{U}$ gate is built from a Hadamard gate sandwiched between COPY tensors the rules introduced in \secr{sec:copy_tensor} allow this network to be rearranged into an RBM geometry, as shown in \fir{fig_nqs_graph_state}(c). 

The minimum number of hidden units $M_G$ needed is found by rank ordering vertices by their coordination in $G$ and descending the list until the edges attached to the top ranked vertices includes all edges in the graph. In graph theory this is called the {\em minimal vertex covering} of the graph $G$~\cite{Bondy2011}. We denote the vertex from this rank list from which $j$-th hidden unit emerged from as $r_j$. Coupling matrices for the NQS then have elements $C_{v_ih_j} = (-1)^{v_i h_j}$, for all vertices $i$ attached to $r_j$ in $G$, along with $C_{v_{r_j}h_j} = \delta_{v_{r_j}h_j}$, while all others are disconnected with ${\bf C}_{\rm dis}$. That graph states have an efficient NQS representation already illustrates that the NQS formalism is not limited to describing area-law entanglement scaling states like MPS and PEPS are~\cite{Eisert2010}. They can capture massively entangled states, as has been greatly elaborated on in Ref.~\cite{Deng2017}. 

The above result is also easily generalised to so-called {\em weighted graph states}~\cite{Anders2006}. The first step in their construction is to replace the controlled-Z gate between qubits $i$ and $j$ in the graph by a more general controlled-phase gate $\hat{U}_{ij}(\varphi_{ij}) = \mathbbm{1}_i\otimes\mathbbm{1}_j + \mathbbm{1}_i\otimes\hat{P}_{j}(\varphi_{ij})$, where $\hat{P}(\varphi_{ij}) = {\rm diag}(1,e^{{\rm i} \varphi_{ij}})$ and $\varphi_{ij}$ is a pair-dependent phase. This gives
\begin{equation}
\ket{\Psi_{G}({\bm \varphi})} = \prod_{\langle i,j \rangle \in G} \hat{U}_{ij}(\varphi_{ij})\ket{+}\ket{+}\cdots\ket{+}, \label{eq:phased_graph_state}
\end{equation}
where $\bm \varphi$ is an adjacency matrix of pair phases. Consequently $\ket{\Psi_{G}({\bm \varphi})}$ retains the same structure of NQS as $\ket{\Psi_{G}}$ with $M_G$ hidden units, but with coupling matrix elements for all vertices $i$ attached to $r_j$ in $G$ changed to $C_{v_ih_j} = \exp({\rm i}\varphi_{i r_j}v_i h_j)$.

To weight the graph state additional diagonal deformation operators $\hat{D}_i = {\rm diag}(1,e^{d_i})$ with $d_i \in \mathbbm{C}$ are applied to each qubit\footnote{The definition of weighted graphs states also applies single qubit unitaries $\hat{u}_1\otimes\hat{u}_2\otimes\cdots\otimes\hat{u}_N$ after the deformations~\cite{Anders2006}. We will ignore this in the NQS representation since a fixed local change of basis can be accounted for by rotating the observables we compute within variational Monte Carlo.}, giving a state
\begin{equation}
\ket{\Psi_{WG}({\bf d},{\bm \varphi})} = \prod_{i=1}^N \hat{D}_i \ket{\Psi_{G}({\bm \varphi})}\label{eq:weighted_graph_state}
\end{equation}
parameterised by ${\bf d} = (d_1,d_2,\dots,d_N)^{\rm T}$. Such weighted graph states have been proposed as a useful class of variational states for interacting qubit systems, owing to them being an (over-complete) basis irrespective of $G$, but also a potentially highly entangled set of states depending on $G$~\cite{Anders2006}. Specifically the ansatz is composed of $k$ weighted graph states with identical $\bm \varphi$ but differing qubit deformations ${\bf d}^{(m)}$ superposed as
\begin{equation}
\ket{\Psi} = \sum_{m=1}^k \alpha_m \kets{\Psi_{WG}({\bf d}^{(m)},{\bm \varphi})}, \label{eq:sup_weighted_graph_states}
\end{equation} 
with amplitudes $\alpha_m$. The state $\ket{\Psi}$ has an NQS representation with $M_G$ binary hidden units to encode the amplitudes of the state $\ket{\Psi_{G}({\bm \varphi})}$ in \eqr{eq:phased_graph_state}, and one $k$ dimensional hidden unit to encode a superposition of $k$ deformations ${\bf d}^{(m)}$ and amplitudes $\alpha_m$ applied to this state. Following the discussion in earlier \secr{sec:tensor_network_nqs} and \ref{app:cpd_to_nqs} this higher dimensional hidden unit can be converted into a two-layer RBM with binary hidden units.

\subsection{Uniform number states} \label{sec:uniform_number}
Uniform number states are a basic set of states comprising of an equal superposition of all configuration states $\ket{\bf v}$ where $\bf v$ possesses a fixed number $n$ of 1's, as
\begin{equation}
\ket{\Psi_n} = \sum_{{\bf v} | \mathcal{S}({\bf v}) = n} \ket{\bf v},
\end{equation}
where $\mathcal{S}({\bf v}) = v_1 + v_2 + \cdots v_N$ is the sum function. Depending on how we interpret our physical system and the configuration basis $\ket{\bf v}$, the state $\ket{\Psi_n}$ could be a spin state with fixed total $z$-magnetisation $\hbar(\frac{1}{2}N - n)/2$, or a state of hard-core bosons/spinless fermions with fixed total particle number $n$. 

For any $0 < n < N$ the states $\ket{\Psi_n}$ have an NQS representation with $M = \lceil N/2\rceil$ hidden units forming fully-connected extensive correlators. The explicit construction is described in detail in \ref{app:uniform_number}. In short it works by each hidden unit correlator giving zero for $\bf v$'s in one or more of the number sectors $\neq n$. A product of several of such correlators is built so that overall the amplitude vanishes for all configurations except those in the desired number sector $n$.

A special case of a uniform number state with $n=1$ is commonly known as a W-state. In general the W-state is a superposition of all translates of the configuration state $\ket{1,0,0,\dots,0}$ as
\begin{eqnarray}
\fl \qquad \ket{\Psi_{\rm W}} &=& \alpha_1\ket{1,0,0,\cdots,0,0} + \alpha_2\ket{0,1,0,\dots,0,0} + \alpha_N\ket{0,0,0,\cdots,0,1}.
\end{eqnarray}
where $\alpha_j$'s are arbitrary complex amplitudes. Owing to its single-particle nature the W-state NQS follows directly from that of $\ket{\Psi_{n=1}}$ by taking any one of its hidden units $i$ and simply right multiply each of its coupling matrices ${\bf C}^{(ij)}$ by the matrix ${\bf D}_j = {\rm diag}(1,\alpha_j)$.

\begin{figure}[ht]
\begin{center}
\includegraphics[scale=0.5]{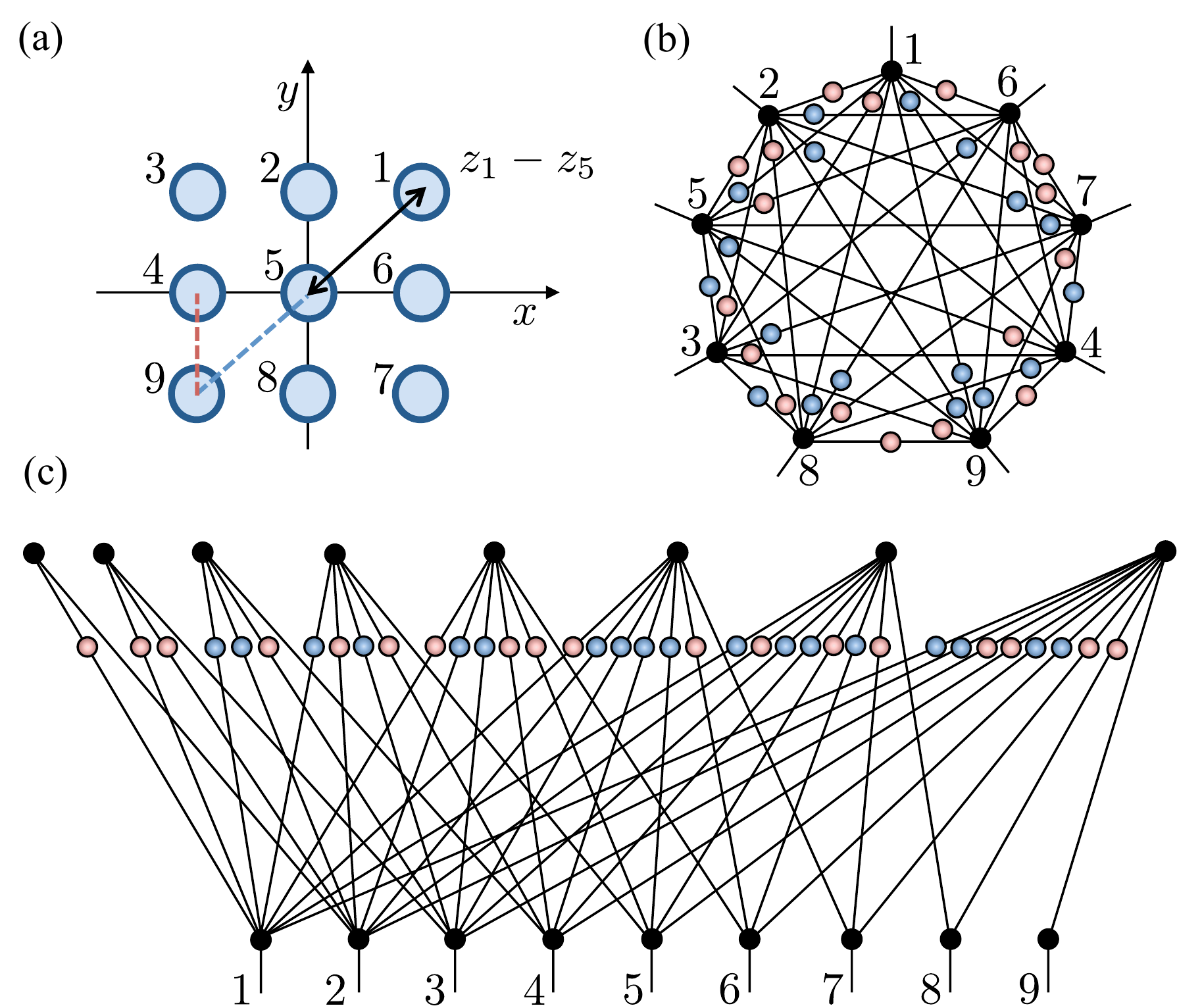}
\end{center}
\caption{(a) Take a $ 3 \times 3$ lattice in the $x$-$y$ plane with periodic boundary conditions and label the sites as shown. The two-site correlators depend on the complex coordinates $z_j$ of each site. (b) The CPS for the Laughlin amplitudes (on top of $\ket{\Psi_n}$) is a completely connected graph with the geometry of the lattice encoded in the two-site correlators. For illustration the colours of the correlators relates the whether it came from a nearest-neighbour or diagonal bond, as highlighted in (a). (c) The CPS reconfigured into an RBM geometry composed of 8 hidden units. Note the decreasing coordination of each hidden unit.}
\label{fig_nqs_laughlin}
\end{figure}

\subsection{Laughlin state} \label{sec:laughlin}
A less trivial modification of $\ket{\Psi_n}$ is the Laughlin quantum Hall state. In first quantisation this is a wave function for $n$ particles that fill a fraction $1/\nu$ of the lowest Landau level given as
\begin{equation}
\Psi_{\rm LS}({\bf r}_1,{\bf r}_2,\dots,{\bf r}_n) = \prod_{i =1}^n \prod_{j =i+1}^n (z_i - z_j)^\nu \prod_{k=1}^n e^{-|z_k|^2}, \label{eq:laughlin}
\end{equation}
where $z_k$ is the dimensionless complex coordinate of the $k$-th particle in the plane, normalised to the magnetic length scale~\cite{Laughlin1983}. To connect to a lattice system we map this state on to a set of coordinates $z_1,z_2,\dots,z_N$ for $N$ sites, as shown in \fir{fig_nqs_laughlin}(a), with $\ket{\bf v}$ being the occupation Fock basis as~\cite{Changlani2009}
\begin{equation}
\ket{\Psi_{\rm LS}} = \sum_{{\bf v}|\mathcal{S}({\bf v}) = n}\, \prod_{i=1}^N \prod_{j =i+1}^N \Upsilon^{{\rm LS}(ij)}_{v_iv_j} \ket{\bf v}, \label{eq:laughlin_lattice}
\end{equation} 
where
\begin{equation}
\Upsilon^{{\rm LS}(ij)}_{v_iv_j} = (z_i - z_j)^{\nu v_i v_j}\exp\left(-\frac{|z_j|^2v_j}{j-1}\right).
\end{equation}
Consequently the lattice Laughlin state in \eqr{eq:laughlin_lattice} is a CPS with two-site correlators $\Upsilon^{{\rm LS}(ij)}_{v_iv_j}$ between every pair of sites $i,j$, very similar to a fully-connected graph state, but with amplitudes restricted to configurations in the number sector $n$. 

To construct an NQS for the $\ket{\Psi_{\rm LS}}$ we therefore use the $\lceil N/2\rceil$ hidden unit correlators of $\ket{\Psi_n}$, and add to them the hidden unit correlators that imprint the lattice Laughlin amplitudes on to the $n$ particle configurations. The fully-connected geometry of the CPS~\cite{Marti2010}, shown in \fir{fig_nqs_laughlin}(b), can be unravelled into an RBM geometry giving $M = N-1$ additional hidden units, each with increasing coordination, as depicted in \fir{fig_nqs_laughlin}(c).

\subsection{Toric code state} \label{sec:toric_code}
NQS and commonly used CPS share an even closer relationship when the receptive field of the hidden units is limited to a small geometric motif. A simple but highly non-trivial example of this are the toric code states $\ket{\Psi_{\rm TS}}$ ~\cite{Kitaev2003}. If we take the physical qubits as being located on the bonds of a square lattice (i.e. on the dual lattice) with periodic boundary conditions, then $\ket{\Psi_{\rm TS}}$ arises as the ground state of the Hamiltonian
\begin{equation}
\hat{H}_{\rm toric} = -\sum_{v \in +} \hat{\mathcal{A}}_v- \sum_{p \in \square} \hat{\mathcal{B}}_p.
\end{equation}
where 
\begin{equation*}
\hat{\mathcal{A}}_v = \prod_{i \in v} \hat{\sigma}^z_i, \quad {\rm and} \quad \hat{\mathcal{B}}_p = \prod_{i \in p} \hat{\sigma}^x_i.
\end{equation*}
Here $+$ denotes the set of vertices with $\hat{\mathcal{A}}_v$ being the product of four $\hat{\sigma}^z_i$'s on qubits surrounding a vertex $v$, while $\square$ denotes the set of plaquettes with $\hat{\mathcal{B}}_p$ composed the product of four $\hat{\sigma}^x_i$'s on the qubits lying on the perimeter of a plaquette $p$. These are depicted in \fir{fig_toric_code_lattice}(a). 

\begin{figure}[ht]
\begin{center}
\includegraphics[scale=0.5]{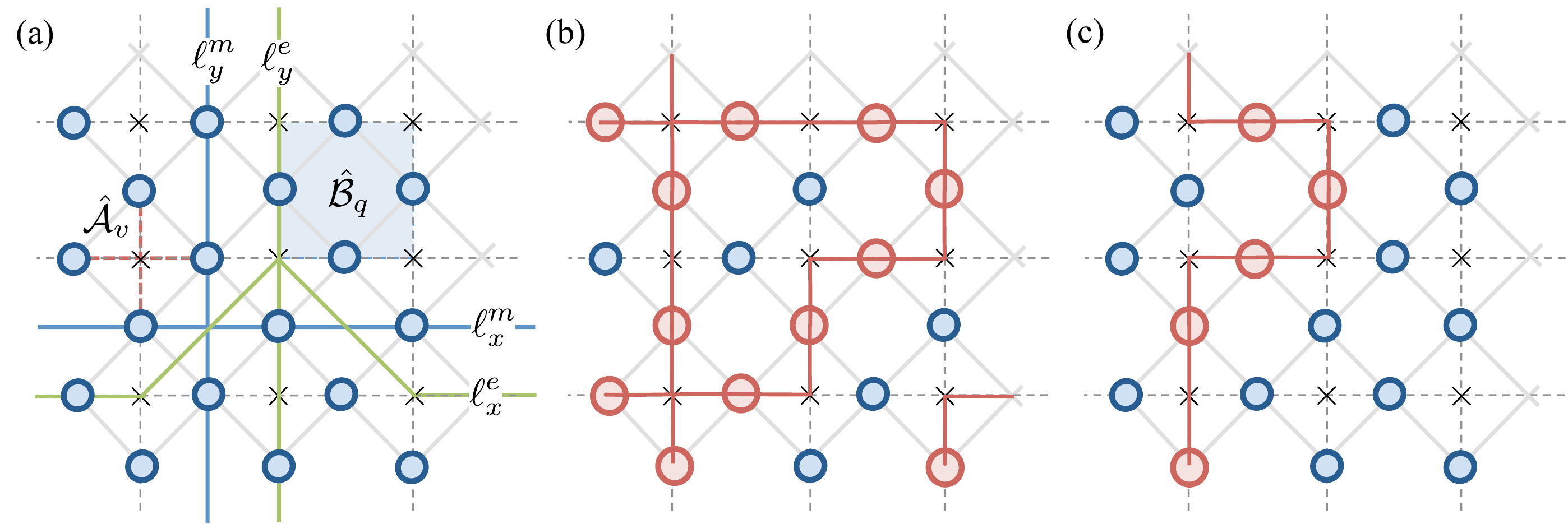}
\end{center}
\caption{(a) A $3 \times 3$ square lattice with vertices $\times$. Qubits (blue circles) are located at the centre of bonds and form a $6 \times 3$ dual lattice. A vertex operator $\hat{\mathcal{A}}_v$ and a plaquette operator $\hat{\mathcal{B}}_p$ are shown. Also possible paths $\ell^{e,m}_{x,y}$ for the four Wilson loop operators $\hat{\mathcal{W}}^{e,m}_{x,y}$ are shown. These closed loops need not be straight, as illustrated by $\ell^e_x$. (b) A closed loop state $\ket{\bf c}$ within the $(1,1)$ sector. Qubits in $\ket{0}$ are smaller blue circles, while the larger red circles are those in $\ket{1}$. The path between them follows the square lattice edges to form a closed loop. A closed loop state within the $(-1,1)$ topological sector.}
\label{fig_toric_code_lattice}
\end{figure}

Since overlapping vertex and plaquette terms share two qubits $\hat{H}_{\rm toric}$ is a sum of commuting terms. A ground state $\ket{\Psi_{\rm TS}}$ of $\hat{H}_{\rm toric}$ is therefore a simultaneous +1 eigenstate of all $\hat{\mathcal{A}}_v$ and $\hat{\mathcal{B}}_p$ operators. The former is satisfied by configuration states $\ket{\bf v}$ in which qubits in the basis state $\ket{1}$ form closed loops around the lattice, denoted as $\ket{\bf c}$, examples of which are shown in \fir{fig_toric_code_lattice}(b) and (c). To satisfy the latter we construct
\begin{equation}
\ket{\Psi_{\rm TS}} = \prod_{p \in \square} (\mathbbm{1} + \hat{\mathcal{B}}_p)\ket{\bf c}, \label{eq:toric_state}
\end{equation}
which is an equal superposition of all closed loops within the topological sector containing $\ket{\bf c}$. An example is shown in \fir{fig_toric_code_state}(a).   

The classes of closed loop states for a lattice with periodic boundary conditions are distinguished by so-called Wilson loop operators. These are two pairs of highly non-local operators of the form 
\begin{equation}
\fl \qquad
\hat{\mathcal{W}}^e_x = \prod_{i \in \ell^e_x} \hat{\sigma}^x_i, \quad \hat{\mathcal{W}}^e_y = \prod_{i \in \ell^e_y} \hat{\sigma}^x_i, \quad \hat{\mathcal{W}}^m_x = \prod_{i \in \ell^m_x} \hat{\sigma}^z_i, \quad \hat{\mathcal{W}}^m_y = \prod_{i \in \ell^m_y} \hat{\sigma}^z_i, \label{eq:wilson_loops}
\end{equation}
where $\ell^e_{x,y}$ is a set of qubits forming an $e=$ {\em electric} loop around the $x$- or $y$-axis of the lattice that cuts through vertices, while $\ell^m_{x,y}$ is an $m=$ {\em magnetic} loop which instead goes cuts through qubits via the centre of the plaquettes. Possible choices of these loops, which encircle the axes of the torus, are shown in \fir{fig_toric_code_lattice}(a). 

Closed loop states $\ket{\bf c}$ are eigenstates of the magnetic loop operators and their eigenvalues are the parity of the winding number $\pi_{x,y} = (-1)^{n_{y,x}}$, where $n_{x,y}$ is number of loops wrapping around the $x$- or $y$-axis, respectively. Together $(\pi_x,\pi_y)$ define four distinct topological sectors, containing loop states $\ket{\bf c}$ like those shown in \fir{fig_toric_code_state}(b)-(e), each of which generates unique ground states \eqr{eq:toric_state}. The application of electric loop operators transform ground states between topological sectors~\cite{Kitaev2003}. 

\begin{figure}[ht]
\begin{center}
\includegraphics[scale=0.5]{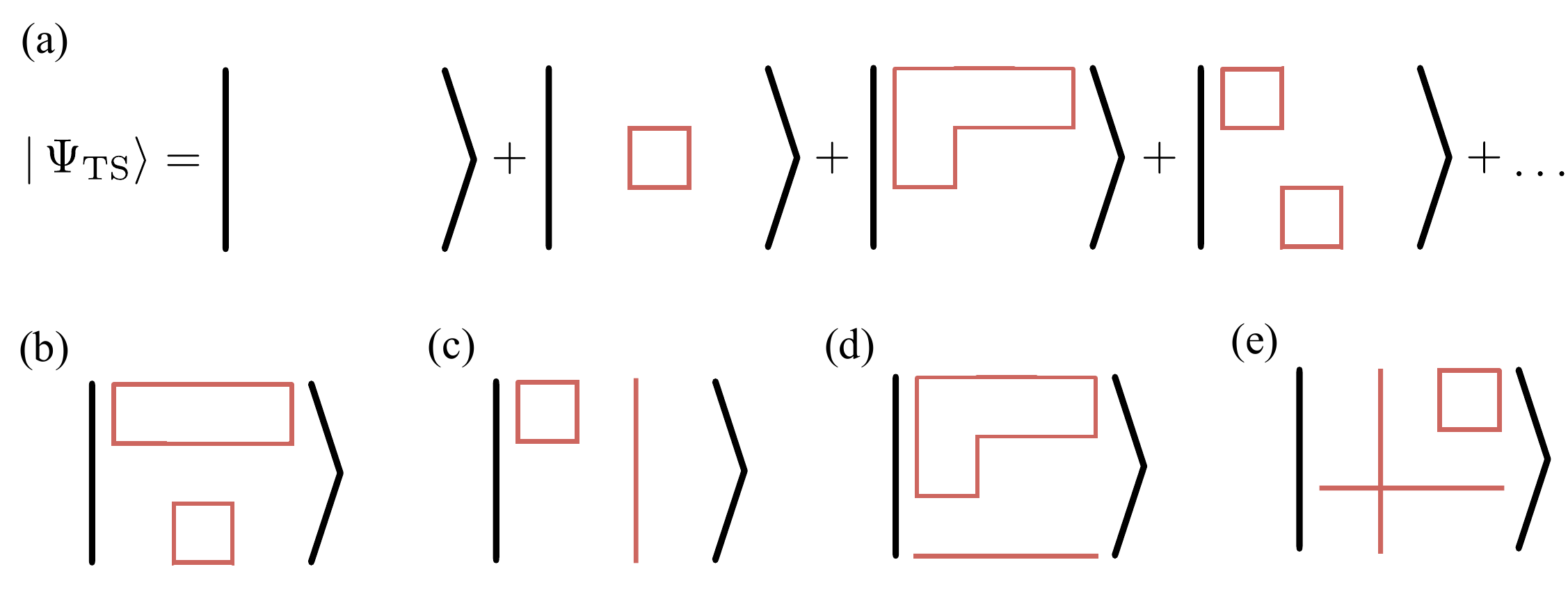}
\end{center}
\caption{(a) A toric code ground state $\ket{\Psi_{\rm TS}}$ is composed of an equal superposition of all closed loop states in a topological sector, as depicted. The loops inside $\ket{\quad}$ correspond to configurations like that in \fir{fig_toric_code_lattice}(b). (b) A closed loop state in the topological sector $(1,1)$, like those in (a). Similarly we have closed loop states in the topological sectors (c) $(-1,1)$, (d) $(1,-1)$ and (e) $(-1,-1)$.}
\label{fig_toric_code_state}
\end{figure}

The ground states $\ket{\Psi_{\rm TS}}$ have a remarkably simple and well studied representation in terms of overlapping plaquette CPS~\cite{Changlani2009} and related PEPS tensor network constructions~\cite{Verstraete2006}. To enforce closed loops an order-4 XOR tensor is placed at each vertex with connectivity mirroring the corresponding $\hat{\mathcal{A}}_v$ term~\cite{Denny2012}. The XOR tensor guarantees that only an even number of the four qubits around a vertex are in the basis state $\ket{1}$. By gluing together all the overlapping 4 qubit XOR state correlators with COPY tensors we obtain the tensor network shown in \fir{fig_nqs_toric_state}(a), which represents a superposition of all states that simultaneously satisfy all the vertex constraints for the sector $(1,1)$.  

\begin{figure}[ht]
\begin{center}
\includegraphics[scale=0.5]{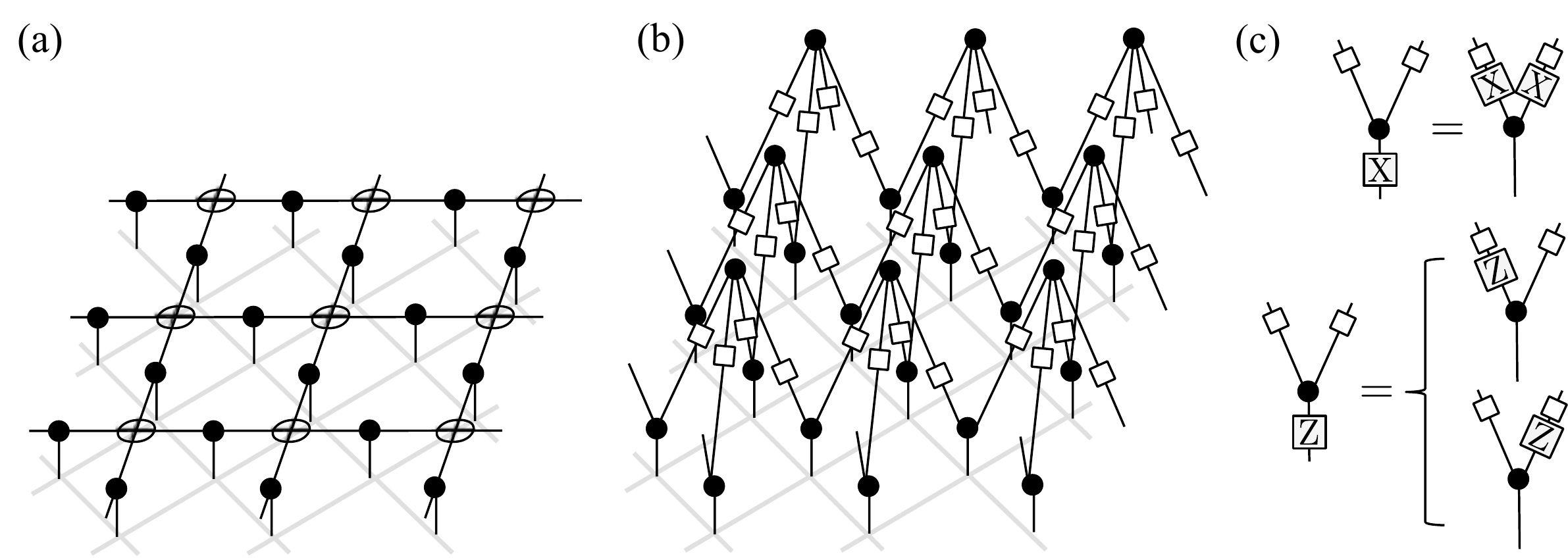}
\end{center}
\caption{(a) The tensor network representation of the toric code ground state $\ket{\Psi_{\rm TS}}$ built from XOR and COPY tensors. The light grey lines denote the dual lattice for the qubits. (b) The NQS representation of $\ket{\Psi_{\rm TS}}$ found by simple applying \fir{fig_xor_tensor}(a). The COPY tensors associated to hidden units have been raised out of the plane to show the RBM geometry. (c) Mapping to other topological sectors or creating excitations of $\hat{H}_{\rm toric}$ involve applying $\hat{\sigma}^x$ operators to strings of qubits. A $\hat{\sigma}^x$ operator (X box) gets copied across the COPY tensor and alters all the coupling matrices of the hidden units attached to that qubit, as shown. A $\hat{\sigma}^z$ operator (Z box) is diagonal in the configuration basis and so commutes with the COPY tensor, as shown.}
\label{fig_nqs_toric_state}
\end{figure}

The conversion of this tensor network into a NQS~\cite{Deng2016,Chen2017} is seen here to be a formality of performing a CPD on the plaquette correlator. This is trivial for the XOR tensor which has a rank $r=2$ decomposition, as shown earlier in \fir{fig_xor_tensor}(a), and so leaves a RBM with Hadamard gate coupling matrices on $N/2$ hidden units whose local receptive fields are illustrated in \fir{fig_nqs_toric_state}(b). The conclusion holds for the toric states in the other topological sectors\footnote{In fact all eigenstates of $\hat{H}_{\rm toric}$ can be expressed as a NQS of this form.}. Specifically, the application of $\hat{\mathcal{W}}^{e}_{x,y}$ loop operators can be pulled through the COPY tensors modifying the coupling matrices of hidden units attached to qubits along the path by multiplication with $\hat{\sigma}^x$, as shown in \fir{fig_nqs_toric_state}(c). 

\subsection{Fully-packed loop and dimer state}\label{sec:fully_packed_dimer}
The construction of NQS via overlapping plaquette correlators that enforce local constraints can be easily extended to build other non-trivial many-body states inspired from classical statistical mechanics~\cite{Batchelor1996,Fisher1961}. For instance instead of 4 qubit XOR state we could use the 4 qubit uniform number state $\ket{\Psi_2}$ as the correlator. The tensor network remains as \fir{fig_nqs_toric_state}(a) only with the XOR tensor replaced by an order-4 tensor $F$ with six non-zero elements $F_{1100} = F_{1010} = F_{1001} = F_{0110} = F_{0101} = F_{0011} = 1$, as depicted in \fir{fig_fully_packed_state}(a). Consequently $F$ enforces exactly two qubits around every vertex to be in the basis state $\ket{1}$. Gluing together these constraints means that configurations $\ket{\bf v}$ with non-zero amplitude are now so-called {\em fully packed} loops where the lattice is totally filled with non-touching loops~\cite{Batchelor1996}, an example of which is shown in \fir{fig_fully_packed_state}(b). The resulting tensor network gives the state $\ket{\Psi_{\rm FS}}$ that is an equal superposition of all such fully packed loops.

\begin{figure}[ht]
\begin{center}
\includegraphics[scale=0.5]{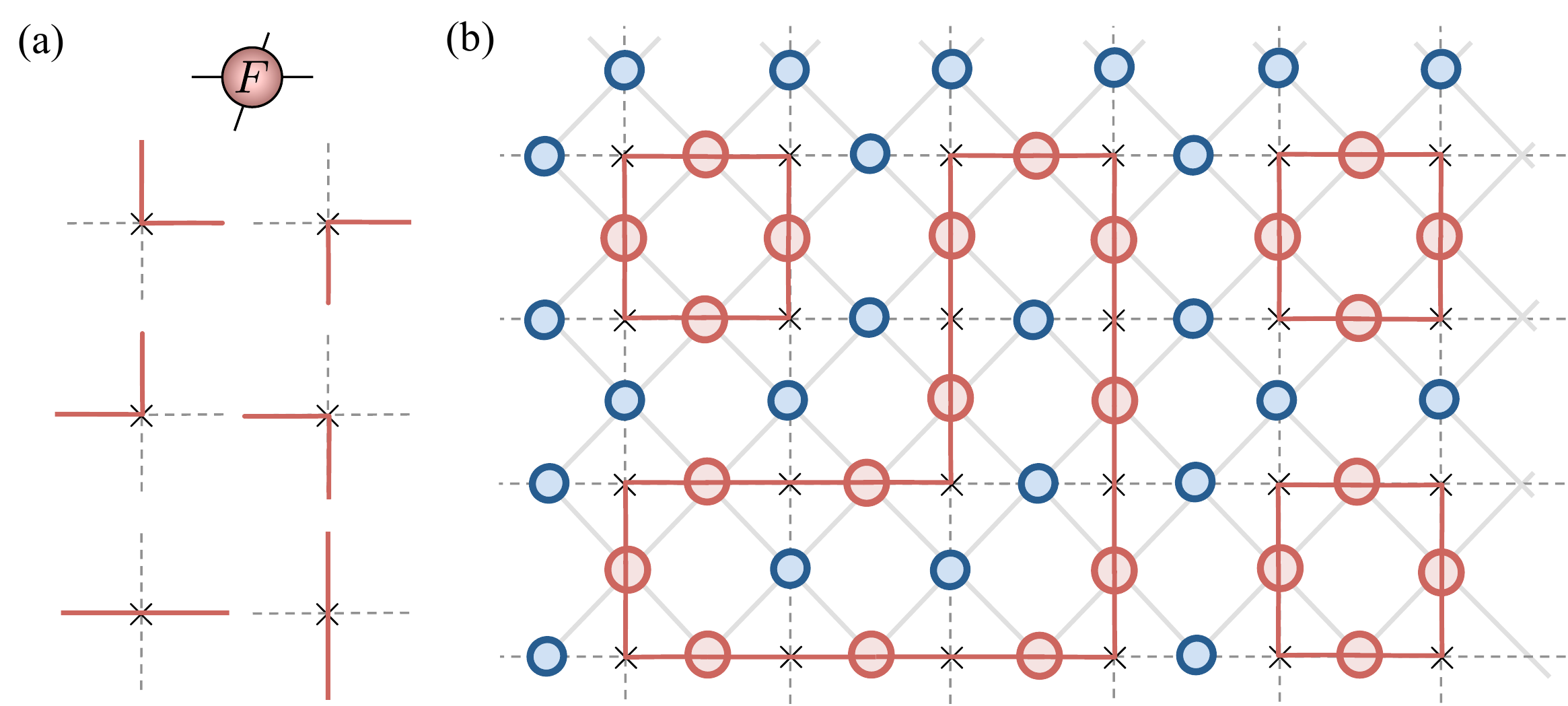}
\end{center}
\caption{(a) The order-4 tensor $F$ describing a $\ket{\Psi_2}$ correlator has only six non-zero elements. These can be interpreted as the ways in which a single line can enter and exit a vertex. (b) The constraints imposed by $F$ on each vertex means that allowable configurations of the system are now composed of fully packed loop states like that depicted. Since $F$ does not have a non-zero element for all four lines the loops are non-touching.}
\label{fig_fully_packed_state}
\end{figure}

Another example uses the 4 qubit uniform number state $\ket{\Psi_1}$, or the W-state, as the correlator. This the order-4 tensor $W$ then has non-zero elements $W_{1000} = W_{0100} = W_{0010} = W_{0001} = 1$ enforcing precisely one qubit around every vertex to be in the basis state $\ket{1}$. Applying this constraint to the results in a state $\ket{\Psi_{\rm DS}}$ that is an equal superposition of configurations of qubits $\ket{\bf v}$, each of which can be interpreted as a complete dimer covering of the underlying square lattice of vertices~\cite{Fisher1961}, as shown in \fir{fig_dimer_state}. 

In fact $\ket{\Psi_{\rm DS}}$ is closely related to the ground state of the well-known quantum dimer model~\cite{Rokhsar1988}
\begin{equation}
\fl\qquad\qquad
\hat{H}_{\rm RK} = \sum_{p \in \square} \Big\{-J\big(\ket{\bm \|}\bra{\bm =} + \ket{\bm =}\bra{\bm \|} \big) + V\big(\ket{\bm \|}\bra{\bm \|} + \ket{\bm =}\bra{\bm =} \big)\Big\}, 
\end{equation}
where the first term describes the kinetic energy of dimers which flips pairs of parallel nearest-neighbour dimers with energy $J$, and the second term describes a repulsion between such pairs with energy $V$. At the Rokhsar-Kivelson point $J=V$ the dimer state $\ket{\Psi_{\rm DS}}$ is a ground state of $\hat{H}_{\rm RK}$ with zero energy~\cite{Rokhsar1988}.

\begin{figure}[ht]
\begin{center}
\includegraphics[scale=0.5]{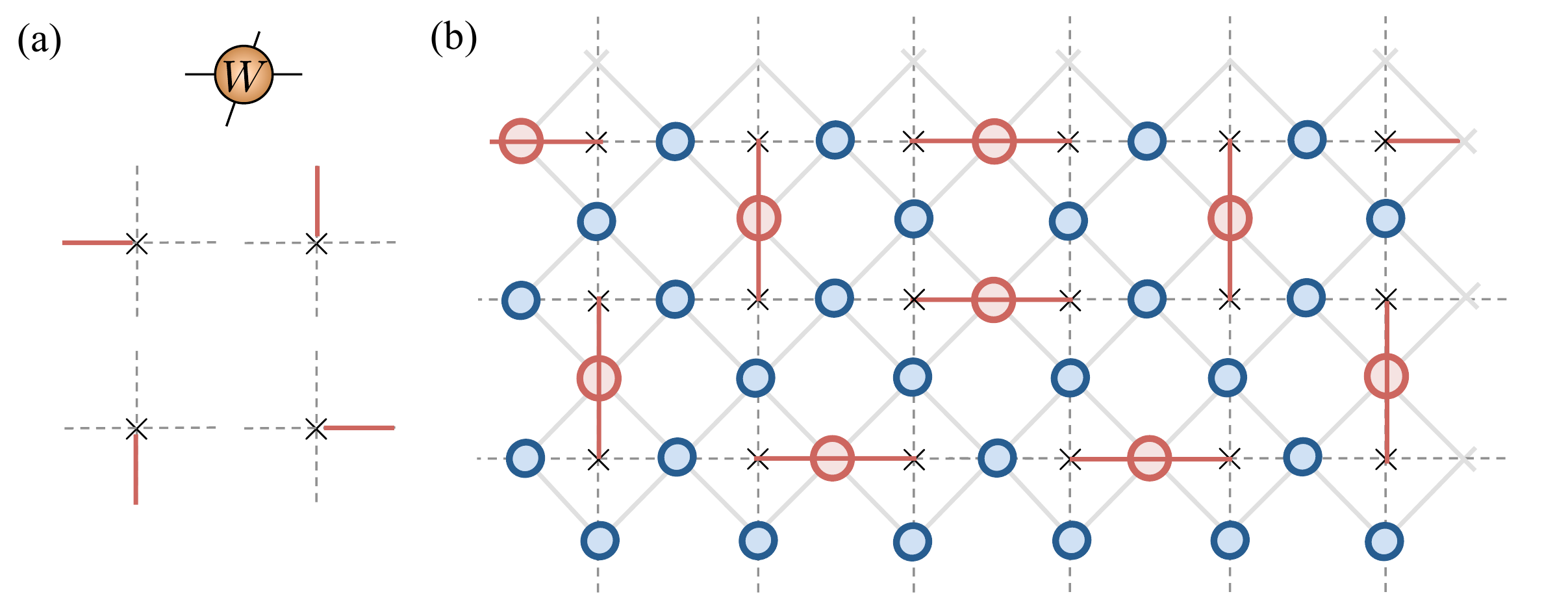}
\end{center}
\caption{(a) The order-4 tensor $W$ describing $\ket{\Psi_1}$ correlator has only four non-zero elements. These can be interpreted as fixing a line on a particular edge between vertices. (b) One possible qubit configuration compatible with $W$-state correlators at each vertex. The resulting state can be interpreted as a complete dimer covering of the underlying square lattice of vertices $\times$, as shown.}
\label{fig_dimer_state}
\end{figure}

Conversion of the tensor networks for both $\ket{\Psi_{\rm FS}}$ and $\ket{\Psi_{\rm DS}}$ into an NQS proceeds identically to the toric code ground state. However, an important difference is that the CPD of both the $F$ and $W$ tensor is now rank $r=4$. In the case of the $W$ tensor we have two options depicted in \fir{fig_w_singlet}(a), use a single 4 dimensional hidden unit in CPD form, or use two binary hidden units per plaquette correlator following the result from \secr{sec:uniform_number}. For clarity we will adopt the first approach.

\begin{figure}[ht]
\begin{center}
\includegraphics[scale=0.5]{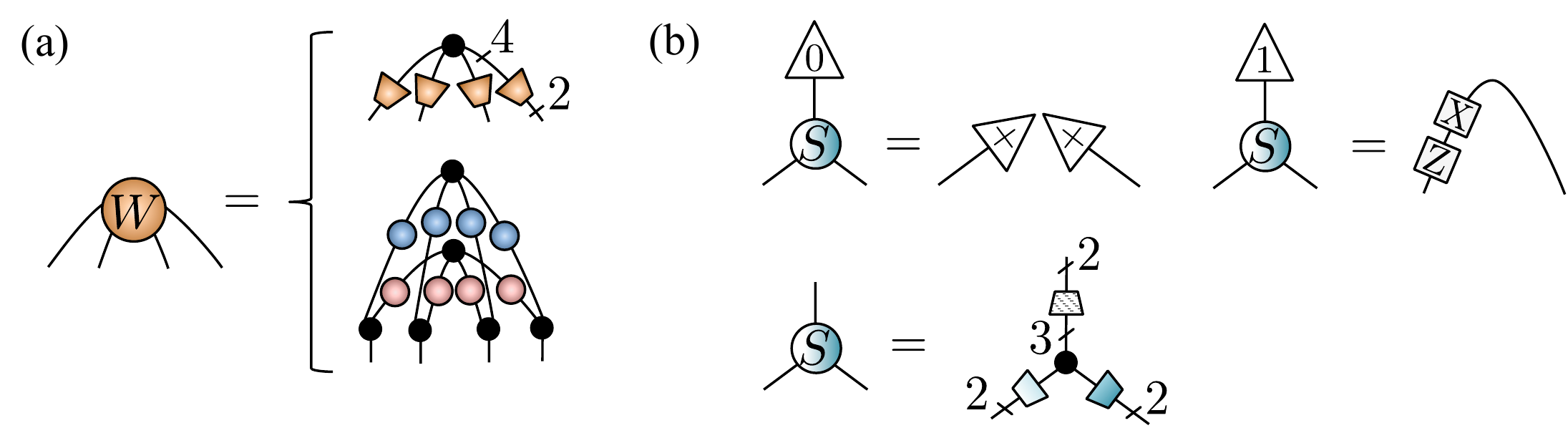}
\end{center}
\caption{(a) The order-4 tensor $W$ has a CPD with a rank of $r=4$ and therefore involves a 4 dimensional COPY tensor and $4 \times 2$ coupling matrices on each leg. The trapezoid shape denotes the rectangular nature of these coupling matrices. Alternatively it can be decomposed into a network involving two binary COPY tensors and $2 \times 2$ coupling matrices. (b) The singlet bond tensor $S$ is defined so that terminating its top leg with basis states results in $\ket{0} \mapsto \ket{+}\ket{+}$ and $\ket{1} \mapsto \ket{0,1} - \ket{1,0}$, as shown. The shading of $S$ indicates its orientation, and this is also reflected in its CPD which has a rank 3, as shown.}
\label{fig_w_singlet}
\end{figure}

\subsection{Resonating valence bond state} \label{sec:rvb_state}
The resonating valence bond (RVB) state was introduced by Anderson as a new form of insulating state~\cite{Anderson1973} and later as a trial wave function in the context of high temperature superconductivity~\cite{Anderson1987}. It continues to receive significant attention for describing quantum spin liquids~\cite{Savary2017,Zhou2017}. Generally for a bipartite lattice composed of $N$ qubits the RVB state is defined as~\cite{Liang1988}
\begin{equation}
\fl \qquad
\ket{\Psi_{\rm RVB}} = \sum_{i_\alpha \in A, j_\beta \in B} h(i_1 - j_1) \cdots h(i_{N/2} - j_{N/2}) \ket{(i_1,j_1)\cdots(i_{N/2},j_{N/2})},
\end{equation}
where $h(r)$ is a positive definite function of the bond length $r$ and $(i,j)$ denotes the spin singlet state $\ket{s} = \ket{0,1} - \ket{1,0}$ for qubits $i$ and $j$, with $i$ in sublattice $A$ and $j$ in sublattice $B$. The RVB state was originally proposed as a good variational trial for the ground state of spin-1/2 Heisenberg antiferromagnet 
\begin{equation}
\hat{H}_{\rm AFM} = J\sum_{\langle i,j\rangle}\left(\hat{\sigma}^x_i\hat{\sigma}^x_j + \hat{\sigma}^y_i\hat{\sigma}^y_j + \hat{\sigma}^z_i\hat{\sigma}^z_j\right),
\end{equation}
where $J>0$ is the exchange coupling and $\langle i,j\rangle$ denotes nearest-neighbour sites on the lattice. The rationale was that since each term individually has a ``valence bond" ground state $\ket{s}$, the state $\ket{\Psi_{\rm RVB}}$ composed of a superposition that ``resonates" between all possible singlet coverings connecting the two sublattices may be close to the ground state.   

As a final example we take the dimer state and transform it into the nearest-neighbour RVB state. To accomplish this we introduce an order-3 tensor $S$ that takes a qubit and maps its state to a pair of qubits as $\ket{0} \mapsto \ket{+}\ket{+}$ and $\ket{1} \mapsto \ket{s}$, and is illustrated in \fir{fig_w_singlet}(b). Thus, given a dimer state of qubits located on the bonds applying the $S$ transformation simultaneously to all nearest-neighbour pairs maps it to $\ket{\Psi_{\rm RVB}}$ over the underlying square lattice of qubits.

\begin{figure}[ht]
\begin{center}
\includegraphics[scale=0.5]{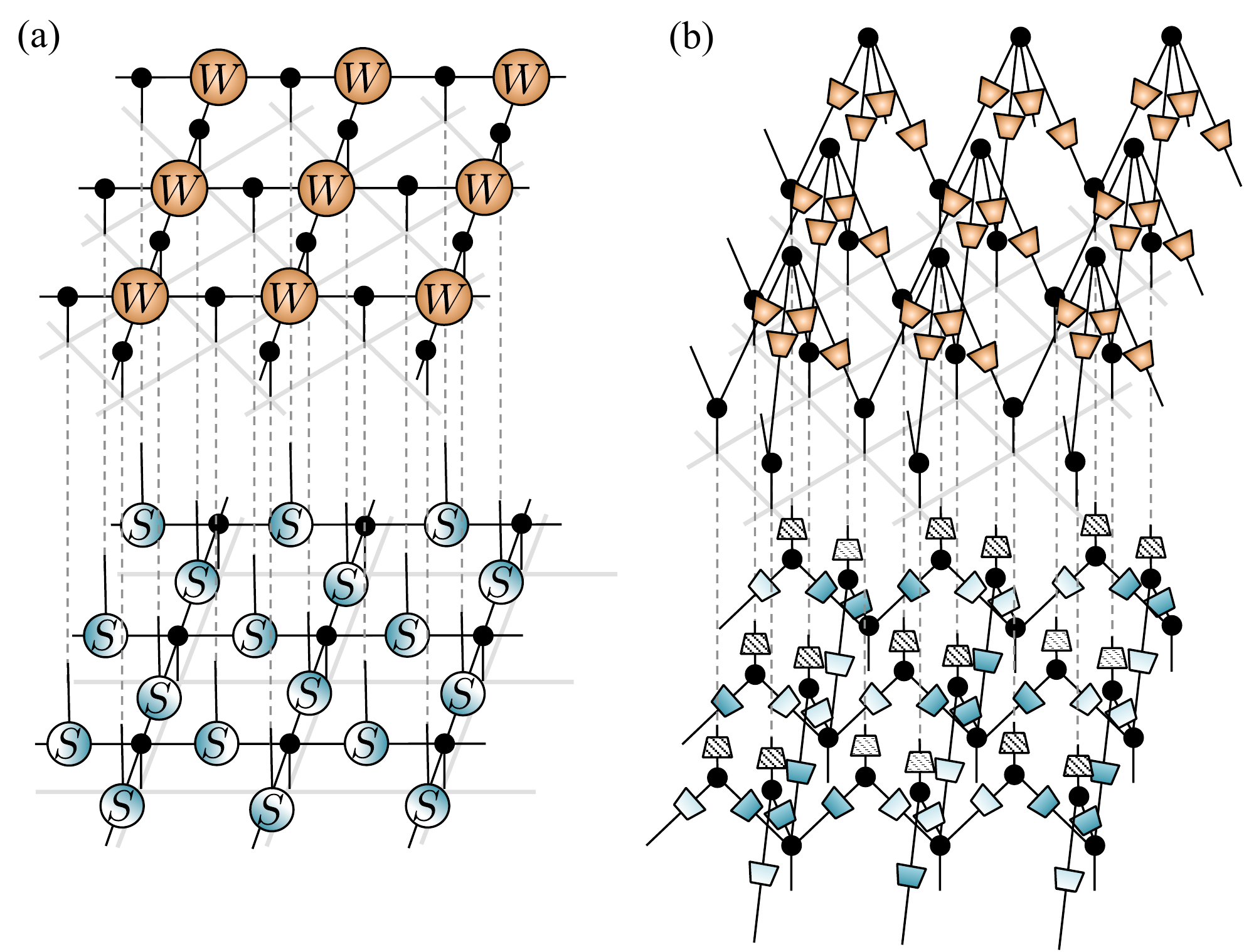}
\end{center}
\caption{(a) The tensor network built from $W$ tensors for the dimer state is shown. Displaced below it is the tensor network of $S$ transformations for each nearest-neighbour pair for a $3 \times 3$ patch of a larger system. Since the singlet state is antisymmetric the $S$ tensor orientation is important and is indicated by its shading. Here we use the convention that the first spin in any pair $\ket{0,1} - \ket{1,0}$ is located on sublattice $A$. The dashed lines indicate legs which are contracted together. (b) The same network as (a) but with CPD forms for the $W$ and $S$ tensors inserted.}
\label{fig_rvb_network}
\end{figure}

The tensor network describing the transformation is given by gluing together all the overlapping $S$ tensors with COPY tensors. The corresponding two-layer tensor network\footnote{It should be noted that a more efficient tensor network for the RVB state comprising of a PEPS geometry with bond dimension $\chi=3$ is known~\cite{Verstraete2006,Schuch2012}.} for $\ket{\Psi_{\rm RVB}}$ then follows as the product of the dimer state network with this transformation, as shown in \fir{fig_rvb_network}(a). 

Conversion to an NQS begins by inserting the CPD forms for the $W$ and $S$ tensors given in \fir{fig_w_singlet}(a) and (b). The resulting network is shown in \fir{fig_rvb_network}(b). This can rewired using the identity depicted in \fir{fig_rvb_state}(a), which is discussed further in \ref{sec_singlet_tensor}. The network is then composed of a visible layer of physical qubits along with two hidden layers with 3 and 4 dimensional hidden units, respectively, as shown in \fir{fig_rvb_state}(b). The result is thus an NQS with a ``deep" RBM geometry. In contrast to previous single layer NQS examples, including those with a single higher dimensional hidden unit, this representation cannot be exactly and efficiently sampled in the configuration basis since there $O(N)$ hidden units in the second layer\footnote{Forming a single layer NQS for the nearest-neighbour RVB state appears to be non-trivial. However, the solution is not really necessary for practical calculations as we are about to discuss in \secr{sec:extensions}.}. 
     
\begin{figure}[ht]
\begin{center}
\includegraphics[scale=0.5]{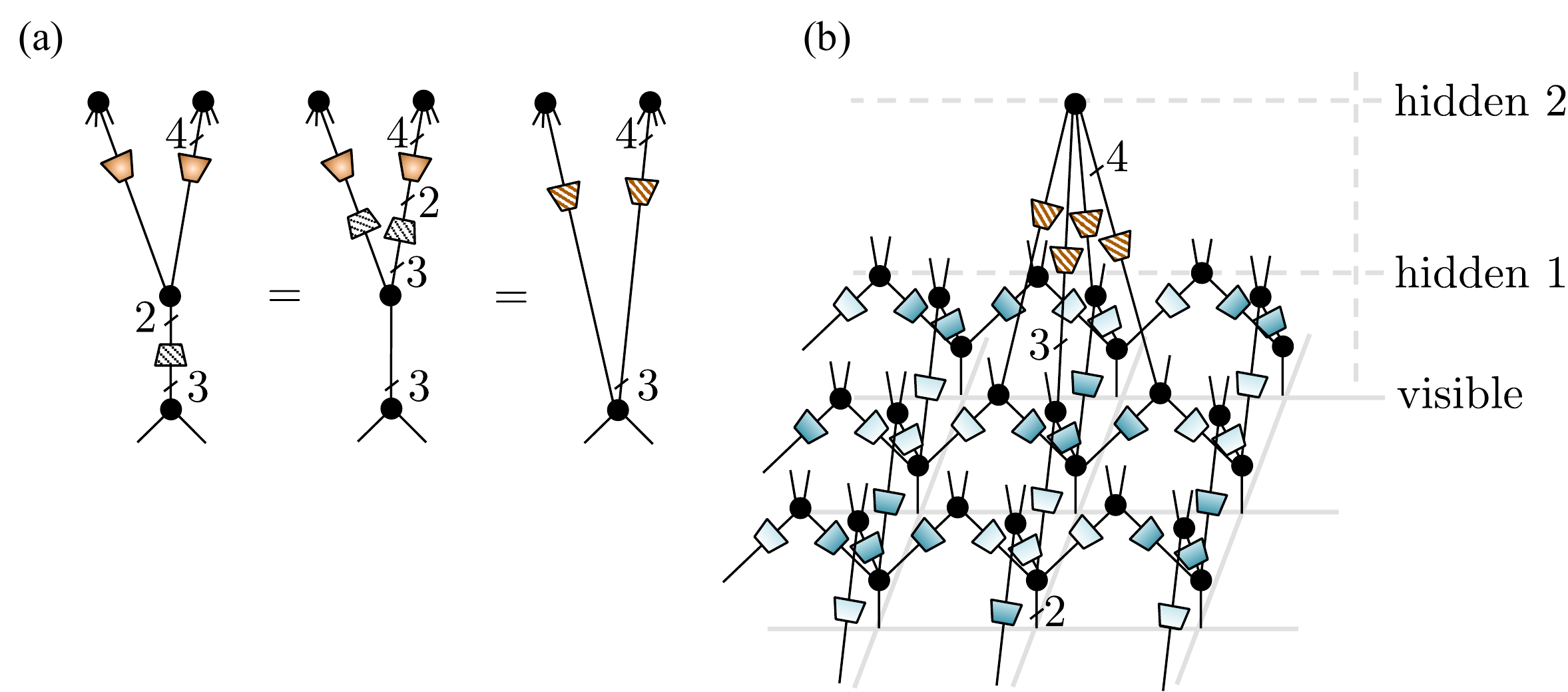}
\end{center}
\caption{(a) A manipulation of a segment of the network connecting the $S$ to two $W$ tensors. The top $3 \times 2$ component matrix for $S$ can be copied across the 2 dimensional COPY tensor converting it into a 3 dimensional COPY tensor (see \ref{sec_singlet_tensor}). This component matrix and the $2 \times 4$ component matrix from $W$ can then be combined into one $3 \times 4$ matrix. The two 3 dimensional COPY tensor can be fused together. (b) This transforms \fir{fig_rvb_network}(b) into a deep-RBM geometry with two hidden layers. For clarity only one hidden unit for the top layer is shown.}
\label{fig_rvb_state}
\end{figure}

\section{Extensions to correlator operator approach} \label{sec:extensions}
While the examples considered in \secr{sec:examples} shed light on the expressiveness of NQS to finish we outline a powerful extension of the approach made evident by the link to correlator product states. For any CPS, as in \eqr{eq:correlator_product_state}, each correlator $\Upsilon^{(c)}_{v_{x_1(c)}v_{x_2(c)}\cdots v_{x_{\ell}(c)}}$ can be elevated to a correlator operator $\hat{\Upsilon}^{(c)}$ defined to be diagonal in the fixed configuration basis~\cite{Changlani2009} as
\begin{equation}
\hat{\Upsilon}^{(c)} = \sum_{\bf v} \Upsilon^{(c)}_{v_{x_1(c)}v_{x_2(c)}\cdots v_{x_{\ell}(c)}}\ket{\bf v}\bra{\bf v}. \label{eq:correlator_operator}
\end{equation}
The general process of elevating any correlator (or state) to an operator by contracting order-3 COPY tensors to it is depicted in \fir{fig_nqs_projectors}(a). A CPS is then written as the product of these commuting correlator operators acting on a reference state $\ket{\Phi}$ as
\begin{equation}
\kets{\Psi_{\rm CPS}} = \prod_{c \in \mathcal{C}} \hat{\Upsilon}^{(c)}\ket{\Phi}, \label{eq:correlator_operator_product_state}
\end{equation}
allowing them to adjust the amplitudes of configurations already present in $\ket{\Phi}$. The generic CPS introduced in \eqr{eq:correlator_product_state} has $\ket{\Phi} = \ket{\Psi_+} = \ket{+}\ket{+}\cdots\ket{+} = \sum_{\bf v} \ket{\bf v}$ where all configurations appear uniformly. Importantly, we have the freedom to use other reference states $\ket{\Phi}$ that may be more physically relevant as well as computationally more convenient~\cite{Changlani2009}. A fundamental restriction in our choice is that $\braket{{\bf v}}{\Phi}$ must be efficiently computable if variational Monte Carlo is to be applicable.

A useful alternative $\ket{\Phi}$ is the uniform number state $\ket{\Psi_n} = \sum_{{\bf v}|\mathcal{S}({\bf v}) = n} \ket{\bf v}$, equivalent to applying a projection into the $n$-particle subspace to $\ket{\Psi_+}$. Since the sum function $\mathcal{S}({\bf v})$ is trivial to evaluate when sampling the $\ket{\bf v}$ basis this projection can be done on-the-fly within the Monte Carlo, thus removing the need for the CPS/NQS ansatz to explicitly enforce the constraint as we did earlier in \secr{sec:uniform_number}. 

\begin{figure}[ht]
\begin{center}
\includegraphics[scale=0.5]{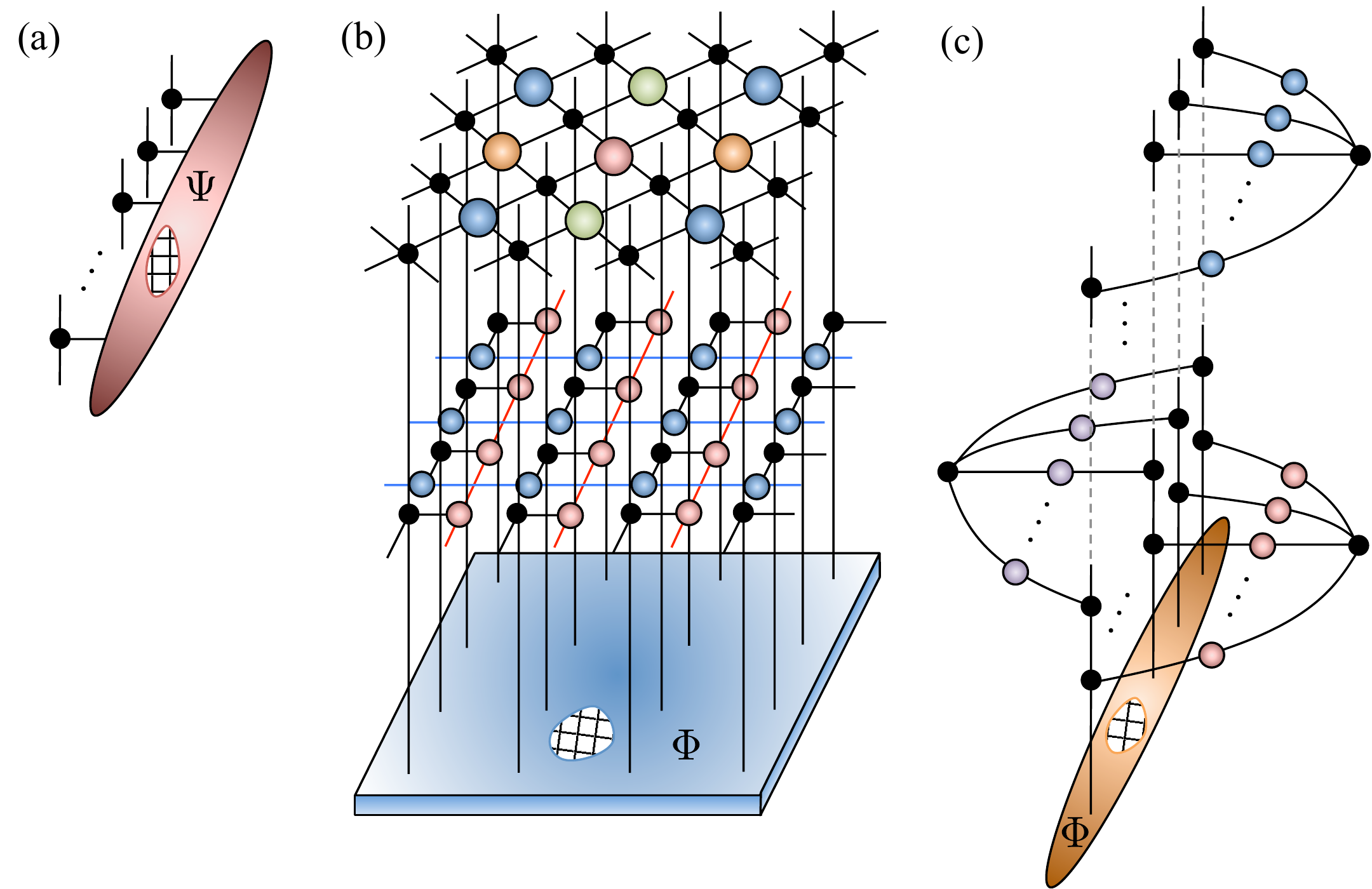}
\end{center}
\caption{(a) Any order-$N$ tensor $\Psi$ can be turned into a diagonal correlator operator by simply contracting an order-3 COPY tensor on to each physical leg. If $\Psi$ is samplable, as indicated by a hashed region, then the correlator operator can be used within variational Monte Carlo. (b) A reference state $\ket{\Phi}$ has an order-$N$ tensor $\Phi$ of samplable amplitudes, shown here for a 2D square lattice. A full CPS ansatz is formed by applying sequences of correlator operators, as shown here for a string-bond and an entangled plaquette cases. (c) An NQS based approach is equivalent to the application of $M$ GHZ-type correlator operators.}
\label{fig_nqs_projectors}
\end{figure}

Richer classes of references states can be derived from Hartree-Fock-Bogoliubov states of spin-1/2 fermions including Fermi sea, spin/charge density wave and Bardeen-Cooper-Schrieffer (BCS) states. In contrast to the product state $\ket{\Psi_+}$ these type of reference states are entangled and indeed can logarithmically violate the area-law~\cite{Eisert2010}. They are therefore a much more powerful starting point for describing strongly-correlated systems. In the BCS case a pair-product (or geminal) state of $N/2$ singlet pairs are defined as
\begin{eqnarray}
\ket{\Psi_{\rm BCS}} = \prod_{\sigma}\left(\sum_{i=1}^N\sum_{j=1}^N \alpha_{ij}\hat{c}^\dagger_{i\uparrow}\hat{c}^\dagger_{j\downarrow}\right)^{N/2}\vac,
\end{eqnarray}
where $\alpha_{ij}$ is a symmetric pair wave function and $\hat{c}^\dagger_{j\sigma}$ is the canonical creation operator for a fermion of spin $\sigma = \{\uparrow,\downarrow\}$ at site $j$. To connect to the qubit systems considered thus far a Gutzwiller projection~\cite{Gutzwiller1963}
\begin{equation}
\mathbbm{P}_g = \exp\left(-g\sum_{j} \hat{n}_{j\uparrow}\hat{n}_{j\downarrow}\right) = \prod_{j=1}^N \left[1 - (1-e^{-g})\hat{n}_{j\uparrow}\hat{n}_{j\downarrow}\right]
\end{equation}
is applied where the parameter $g>0$ is taken to the limit $g \rightarrow \infty$ so it fully projects out double occupancies on every site. At half-filling electronic configurations are spanned by qubit configuration basis states $\ket{\bf v}$, with $v_j = 0 =\,\, \uparrow$ and $v_j = 1 = \,\,\downarrow$, expressed in terms of fermions as
\begin{equation*}
\ket{{\bf v}} = \hat{c}^\dagger_{1,v_1} \hat{c}^\dagger_{2,v_2} \dots \hat{c}^\dagger_{N,v_N}\vac.
\end{equation*}
The fully projected half-filled BCS state gives a fermionic representation of an RVB spin state $\ket{\Psi_{\rm RVB}} = \mathbbm{P}_{g=\infty}\ket{\Psi_{\rm BCS}}$~\cite{Anderson1987}. The nearest-neighbour RVB spin state on a 2D square lattice considered in \secr{sec:rvb_state} has a fermionic representation with a $s + {\rm i}d$ pairing wave function, where $\alpha_{ij} = 1$ for $i$ and $j$ neighbouring along the $x$-axis of the lattice, $\alpha_{ij} = {\rm i}$ for $i$ and $j$ neighbouring along the $y$-axis of the lattice, and zero otherwise~\cite{Zhou2017}. 

Crucially the amplitudes $\braket{{\bf v}}{\Psi_{\rm RVB}}$ can be exactly and efficiently computed. Given the $N/2$ locations of 0's in $\ket{\bf v}$ are $i_1,\dots,i_{N/2}$, and those for the 1's are $j_1,\dots,j_{N/2}$, then~\cite{Gros1989}
\begin{equation*}
\fl \quad\quad
\braket{{\bf v}}{\Psi_{\rm RVB}} = {\rm sgn}(i_1,\dots,i_{N/2},j_1,\dots,j_{N/2}) \,{\rm det}\left[{\bf A}(i_1,\dots,i_{N/2},j_1,\dots,j_{N/2})\right].
\end{equation*}
Here ${\rm sgn}(i_1,\dots,i_{N/2},j_1,\dots,j_{N/2})$ is the sign of the permutation required to put the list of sites into numerical order, and $\bf A$ is an $N/2 \times N/2$ matrix defined as
\begin{equation}
{\bf A}(i_1,\dots,i_{N/2},j_1,\dots,j_{N/2}) = 
\left(
\begin{array}{ccc}
\alpha_{i_1j_1}  & \cdots  & \alpha_{i_1j_{N/2}}  \\
\vdots  & \ddots  & \vdots  \\
\alpha_{i_{N/2}j_1}  & \cdots  & \alpha_{i_{N/2}j_{N/2}}   
\end{array}
\right).
\end{equation}
Evaluating the determinant of a matrix with a size proportional to $N$ within $\braket{{\bf v}}{\Psi_{\rm RVB}}$ can be performed in $O(N^3)$ time. Analogous results hold for other projected mean-field fermionic states~\cite{Gros1989}. 

Armed with such non-trivial sampleable reference states $\ket{\Phi}$ dispenses with the need to have an explicit CPS/NQS construction for them like we considered in \secr{sec:rvb_state}. This allows the variational parameters within the correlator operators to instead be directed at modifying the physical properties of these states, such as correlation lengths, critical exponents and order parameters, that might otherwise be difficult to describe. For CPS correlator operators based on two-site correlators their action is entirely equivalent to well-known Jastrow factors~\cite{Jastrow1955} commonly used in variational Monte Carlo~\cite{Foulkes2001}, such as the spin-spin factor 
\begin{equation}
\mathcal{J}_s = \exp\left(-\sum_{i,j=1}^N u_{ij} \hat{\sigma}^z_{i}\hat{\sigma}^z_{j}\right),
\end{equation}
parameterised by scalars $u_{ij}$. Correlator product states are a natural framework for generalising Jastrow factors to higher-body correlators~\cite{Changlani2009}. A direct extension would be to use correlators spanning larger clusters of sites, however to scale to extensively sized correlators, while retaining efficient and exact sampleability, would instead require constructions such as string-bond or NQS based correlators. This is shown in \fir{fig_nqs_projectors}(b). While PEPS based correlator operators cannot be exactly and efficiently sampled, by employing approximate contraction methods they can also be used, as was proposed and analysed recently~\cite{Sikora2015}. 

From this perspective the elevation of NQS to correlator operators applied to suitable reference states, as shown in \fir{fig_nqs_projectors}(c), is particularly appealing. A key advantage of the NQS form is that the GHZ-type correlators are geometrically unbiased allowing the variational minimisation to tailor locality, multi-body and range of the correlators. Such properties of the RBM form have proven extremely useful in the context of machine learning features. Very recent work by Nomura {\em et al}~\cite{Nomura2017} implementing this approach on the Heisenberg and Hubbard models has shown substantially improved accuracy beyond conventional NQS and VMC approaches. Ongoing work using the TNT library~\cite{Alassam2017,TNTwebsite} is testing the performance of this NQS extension, with both higher dimensional hidden units and more layers, on a variety of paradigmatic spin and fermion models~\cite{Pei2018}.

\section{Conclusions}\label{sec:conclusions}
In this paper we have used the framework of tensor networks, and in particular the algebraic properties of the COPY tensor, to unify NQS with the broader class of CPS. This approach has illustrated how tensor network diagrammatics is a powerful tool for reasoning about quantum states beyond the conventional TNT ansatzes like MPS, TTN, PEPS and MERA. Using this we have revealed a number of simple observations about NQS, namely that they are based on an extensive correlator with a GHZ form related to the CPD tensor factorisation and that they naturally complement the MPS correlators used in string-bond states. Using the connection to CPS we presented a number of exact NQS representations of non-trivial states including, weighted-graph states, the Laughlin state, toric code states, and the RVB state. These examples showed how adaptable and expressive the NQS representation is, and suggest how further enhancements are gained by using higher-dimensional hidden units and/or a second hidden layer in the RBM. The major outlook of this work is the elevation of NQS to correlator operators allowing this novel approach to be applied to a much wider range of non-trivial reference states.  

\section*{Acknowledgements}
SRC gratefully acknowledges support from the EPSRC under grant No. EP/P025110/1. Also SRC would like to thank Dieter Jaksch, Jonathan Coulthard, Michael Lubasch, Michael Pei, Sam Pearce and Matthew Cook for helpful discussions.

\newpage

\appendix

\section{Tabulating hidden unit correlators}  \label{app:coupling_mats}
It is convenient to represent a hidden unit correlator $\Upsilon^{(i)}_{v_1v_2\cdots v_N}$ within an NQS by tabulating the constituent $2 \times 2$ coupling matrices as
\begin{equation}
 \fl \qquad
 \left[ \begin{array}{cc}
1 & e^{\frac{a_1}{M}}  \\
e^{\frac{b_i}{N}}  & e^{W_{i1} + \frac{a_1}{M} + \frac{b_i}{N}}  
\end{array}
\right]; \left[
\begin{array}{cc}
1 & e^{\frac{a_2}{M}}  \\
e^{\frac{b_i}{N}}  & e^{W_{i2} + \frac{a_2}{M} + \frac{b_i}{N}}  
\end{array}
\right]; \cdots ;\left[
\begin{array}{cc}
1 & e^{\frac{a_N}{M}}  \\
e^{\frac{b_i}{N}}  & e^{W_{iN} + \frac{a_N}{M} + \frac{b_i}{N}}  
\end{array}
\right]. \label{eq:correlator_tabulation}
\end{equation}
For any configuration $\bf v$ the value of $v_j = \{0,1\}$ indicates the relevant column of the $j$-th matrix, and we read off $\Upsilon^{(i)}_{v_1v_2\cdots v_N}$ by summing the product of elements along each row. For example $\Upsilon^{(i)}_{00\cdots0} = 1 + e^{b_i}$, $\Upsilon^{(i)}_{10\cdots0} = e^{\frac{a_1}{M}} + e^{b_i}e^{W_{i1} + \frac{a_1}{M}}$, and so on. The full NQS is then an amplitude-wise multiplication of each hidden unit correlator. We will find this tabulated form useful in the following appendices.

\section{NQS as a universal quantum state approximator} \label{app:approximator}
It was established by the machine learning community that RBMs can approximate arbitrarily well any probability distribution $p({\bf v})$ over ${\bf v} \in \{0,1\}^N$ given a sufficient number of hidden units. Specifically, if $p({\bf v})$ has support on $k$ configurations then $k+1$ hidden units can describe it exactly \cite{LeRoux2008}. The formal implication is that given an exponentially large number of hidden units an RBM can describe exactly any distribution $p({\bf v})$, and is therefore a universal approximator.

Here we provide a simpler proof of this result exploiting the coupling matrix tabulation. We refine the result by showing that to describe $k$ non-zero complex amplitudes $\Psi({\bf v})$ actually only requires $k$ hidden units. Suppose we want to describe a quantum state
\begin{equation}
\ket{\Psi} = \sum_{j=1}^k \Psi_j \kets{{\bf v}^{(j)}},
\end{equation}
comprising of a superposition of $k$ distinct configuration states $\kets{{\bf v}^{(j)}}$ and corresponding complex amplitudes $\Psi_j$. Then the we use $k$ hidden units with the $j$-th having the following coupling matrices 
\begin{equation}
\left[
\begin{array}{cc}
s  & s   \\
\eta_j(1-v^{(j)}_1)  & \eta_jv^{(j)}_1  
\end{array}
\right]; \cdots ;\left[
\begin{array}{cc}
s & s   \\
\eta_j (1-v^{(j)}_N)  & \eta_j v^{(j)}_N   
\end{array}
\right]. \label{eq:cmat_universal}
\end{equation}
Here $0<s<1$ is a suppression factor, and the coefficient in the matrices is
\begin{equation}
\eta_j = \frac{\Psi_j^{1/N}}{s^{k-1}}.
\end{equation}
In isolation the $j$-th hidden unit correlator is equivalent to a quantum state
\begin{equation}
\ket{\psi_j} = s^N\sum_{\bf x} \ket{\bf x} + \frac{\Psi_j}{s^{N(k-1)}}\kets{{\bf v}^{(j)}},
\end{equation}
superposing the uniform state with the $j$-th configuration state. The full NQS is built from all $k$ hidden units and its amplitudes are found by multiplying those of the states $\ket{\psi_j}$ together as
\begin{equation}
\fl \quad
\ket{\Psi_{\rm NQS}} = \prod_{j=1}^k\left(s^N + \frac{\Psi_j}{s^{N(k-1)}}\delta_{{\bf x},{\bf v}^{(j)}}\right) \ket{\bf x}  = s^{Nk} \sum_{\bf x} \ket{\bf x} + \sum_{j=1}^k \Psi_j\ket{{\bf v}_j}.
\end{equation} 
Here $\delta_{{\bf x},{\bf y}} = \delta_{x_1,y_1}\cdots\delta_{x_N,y_N}$ and since each ${\bf v}^{(j)}$ has disjoint amplitudes cross-terms vanish. We then take the limit $s \rightarrow 0$ to remove the uniform state from the superposition, although the exponential dependence with $N$ and $k$ means that any $s<1$ quickly achieves this effect in practice. 

\section{Transforming a CPD into an NQS} \label{app:cpd_to_nqs}
Here we show that a CPD can be expressed exclusively terms of binary COPY tensors forming an NQS geometry with two layers of hidden units. The CPD of an order-$N$ tensor is built around an order-$N$ $r$-dimensional COPY tensor, where $r$ is the tensor rank. The $N$ rectangular $r \times 2$ coupling are given by
\begin{equation}
\left[
\begin{array}{cc}
c^{(1)}_{10} & c^{(1)}_{11}  \\
c^{(2)}_{10} & c^{(2)}_{11} \\
\vdots & \vdots \\
c^{(r)}_{10} & c^{(r)}_{11} 
\end{array}
\right]; \left[
\begin{array}{cc}
c^{(1)}_{20} & c^{(2)}_{21}  \\
c^{(2)}_{20} & c^{(2)}_{21} \\
\vdots & \vdots \\
c^{(r)}_{20} & c^{(2)}_{21} 
\end{array}
\right]; \cdots; \left[\begin{array}{cc}
c^{(1)}_{N0} & c^{(1)}_{N1}  \\
c^{(2)}_{N0} & c^{(2)}_{N1} \\
\vdots & \vdots \\
c^{(r)}_{N0} & c^{(r)}_{N1}  
\end{array}
\right]. \label{eq:cmats_cpd}
\end{equation}
As a quantum state the CPD is equivalent a sum of $r$ product states
\begin{equation}
\ket{\Psi_{\rm CPD}} = \sum_{j=1}^r \ket{\phi_j}, \quad {\rm with} \quad \ket{\phi_j} = \otimes_{i=1}^N \left(c^{(j)}_{i0}\ket{0} + c^{(j)}_{i1}\ket{1}\right).
\end{equation}
A strategy for decomposing this state into an NQS essentially follows from \ref{app:approximator}. However, unlike configuration states $\kets{{\bf v}^{(j)}}$ the states $\ket{\phi_j}$ will in general have non-zero amplitudes for all configurations and so are not disjoint. Consequently cross-terms appear if the universal approximator construction is used. 

A way around this is to introduce $A = \lceil \log_2(r)\rceil$ ancilla physical qubits. For each state $\ket{\phi_j}$ we associate a unique configuration state over these ancilla, $\kets{{\bf a}^{(j)}}$. For $r$ hidden units we then define the coupling matrices of the $j$-th as
\begin{equation}
 \fl 
\left[
\begin{array}{cc}
s & s  \\
\eta^{(j)}_{10}  & \eta^{(j)}_{11}  
\end{array}
\right]; \cdots; \left[
\begin{array}{cc}
s & s  \\
\eta^{(j)}_{N0}  & \eta^{(j)}_{N1}  
\end{array}
\right]; \left[
\begin{array}{cc}
1 & 1 \\
(1-a^{(j)}_{1})  & a^{(j)}_{1} 
\end{array}
\right]; \cdots; \left[
\begin{array}{cc}
1 & 1  \\
(1-a^{(j)}_{A})  & a^{(j)}_{A} 
\end{array}
\right], \label{eq:cmats_cpd_binary}
\end{equation}
where the coefficients are
\begin{equation}
\eta^{(j)}_{i x_i} = \frac{c^{(j)}_{i x_i}}{s^{r-1}}.
\end{equation}
The full quantum state described by the NQS is then
\begin{eqnarray}
\ket{\Psi_{\rm NQS}} &=& \sum_{{\bf x},{\bf a}}\prod_{j=1}^r \left(s^N + \frac{1}{s^{N(r-1)}}\prod_{i=1}^N c^{(j)}_{ix_i} \delta_{{\bf a},{\bf a}^{(j)}}\right) \ket{\bf x}\ket{\bf a}, \\
&=& s^{Nr}\sum_{{\bf x},{\bf a}} \ket{\bf x}\ket{\bf a} + \left(\sum_{{\bf x}}\prod_{i=1}^N c^{(j)}_{ix_i}\ket{\bf x}\right)\kets{{\bf a}^{(j)}}, \\
&=& s^{Nr}\sum_{{\bf x},{\bf a}} \ket{\bf x}\ket{\bf a}  + \sum_{j=1}^r \ket{\phi_j}\kets{{\bf a}^{(j)}},
\end{eqnarray} 
owing to the disjoint amplitudes over the system + ancilla. As with the universal approximator we remove the uniform state by taking the limit $s\rightarrow 0$. Finally we reduce $\ket{\Psi_{\rm NQS}}$ to $\ket{\Psi_{\rm CPD}}$ by projecting out each ancilla qubit in the $\ket{+}$ state. The effect of this on the tensor network is to introduce a second layer of hidden units, as shown in \fir{fig_cpd_hidden_layer} for the projection of a single qubit. Doing this for all the ancilla qubits results in an NQS with a two-layer RBM geometry composed of $r$ units in the first layer and $\lceil \log_2(r)\rceil$ units in the second, as shown earlier in \fir{fig_cpd_to_nqs}.

\begin{figure}[ht]
\begin{center}
\includegraphics[scale=0.5]{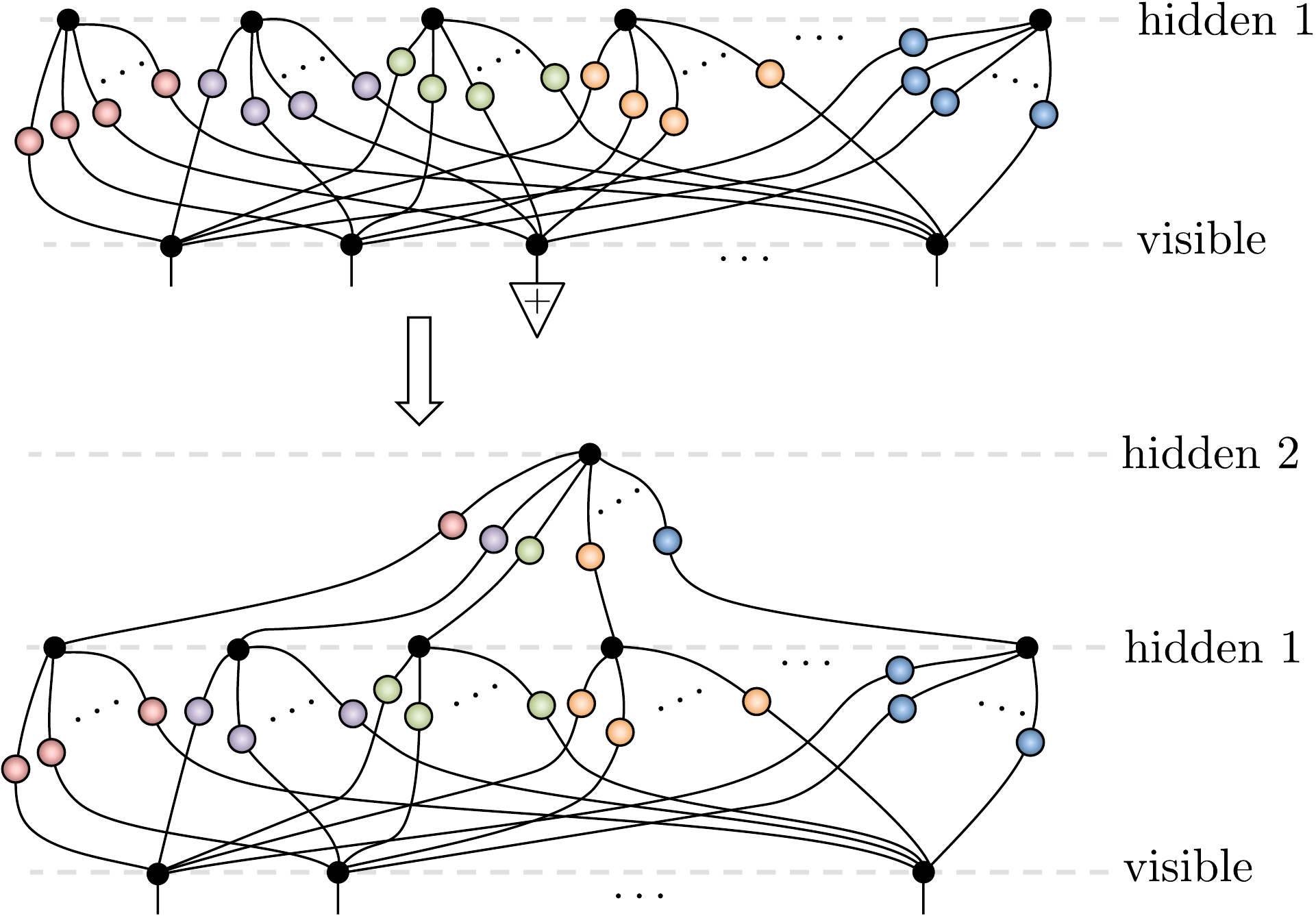}
\end{center}
\caption{Take an NQS with a single layer RBM geometry and project out one of the physical spins (the third in this case) in the visible layer in the $\ket{+}$ state. The resulting visible unit then becomes a hidden unit in a second hidden layer.}
\label{fig_cpd_hidden_layer}
\end{figure}

\section{Explicit construction of uniform number state} \label{app:uniform_number}
The NQS universal approximator constructions works by adding configurations upon with each hidden unit correlator multiplication. Here we illustrate how NQS can also work by cancellation. To represent $\ket{\Psi_n} $ we will introduce hidden unit correlators with uniform coupling matrices that project out all number sectors except $n$. We already have one such correlator useful for this task, namely the XOR tensor given by coupling matrices
\begin{equation}
{\bf C}_{\rm odd} = \left(\frac{1}{2}\right)^{1/N}\left[
\begin{array}{cr}
1 & -1  \\
1  & 1  
\end{array}
\right].
\end{equation}
From this we see that the bottom row gives $\sum_{\bf v} \ket{\bf v}$ while the top row gives $\sum_{\bf v} (-1)^{\mathcal{P}({\bf v})} \ket{\bf v}$. The sum of these two states therefore causes the cancellation of all $\ket{\bf v}$ in odd numbered sectors where $\mathcal{P}({\bf v})=1$. The XOR tensor is therefore an RBM correlator $\Upsilon_{\rm odd}$ describing a uniform superposition of even numbered sectors. This can be flipped around to give the RBM correlator $\Upsilon_{\rm even}$, with all the even numbered sectors cancelling out instead, by using the coupling matrices
\begin{equation}
{\bf C}_{\rm even} = \left(\frac{1}{2}\right)^{1/N}\left[
\begin{array}{cc}
(-1)^{1/N} & -(-1)^{1/N}  \\
1  & 1  
\end{array}
\right],
\end{equation}
where the top row is negated overall. Another simple case is an RBM correlator $\Upsilon_{\rm triv}$ that cancels out the trivial $n=0$ and $n=N$ sectors. This is achieved by coupling matrices
\begin{equation}
{\bf C}_{\rm triv} = \frac{1}{s^{1/N}}\left[
\begin{array}{cc}
(-1)^{1/N} & 1  \\
1  & (-1)^{1/N}  
\end{array}
\right].
\end{equation}
The amplitudes on all other sectors is now not uniform so the scale factor $s = (-1)^{N-n} + (-1)^{n}$ ensures all states in sector $n$ have unit amplitude. 

To project down to just sector $n$ we need to build RBM correlators $\Upsilon_{m\Rightarrow 0}$ that cancel out any given number sector $m>0$. The following form of coupling matrices is helpful
\begin{equation}
{\bf C}_{m\Rightarrow 0} = \frac{1}{s^{1/N}}\left[
\begin{array}{cr}
a & b  \\
1  & 1  
\end{array}
\right].
\end{equation}
In this case the top line gives $\sum_{\bf v} a^{N - \mathcal{S}({\bf v})} b^{\mathcal{S}({\bf v})}\ket{\bf v}$. By choosing
\begin{equation}
b = \frac{(-1)^{1/m}}{a^{N/m - 1}},
\end{equation}
all $\ket{\bf v}$ with $\mathcal{S}({\bf v}) = m$ cancel with the bottom row. So long as $a < 1$, all other terms remain non-zero. By choosing $s = 1 + a^{N- n} b^n$ all states in the number sector $n$ have unit amplitude.

To build an NQS representation of $\ket{\Psi_n}$ we therefore compose an appropriate set of these hidden unit correlators whose union of non-zero (unit) amplitudes is the number sector $n$. In fact $M = \lceil N/2 \rceil$ hidden units is sufficient with the following hidden unit correlators
\begin{eqnarray*}
n \textrm{ (even)} \quad & \ket{\Psi_n} \rightarrow & \Upsilon_{\rm triv}\Upsilon_{\rm odd}\prod_{m\textrm{ (even)}, m \neq n} \Upsilon_{m\Rightarrow 0}, \\ 
n \textrm{ (odd)} \quad & \ket{\Psi_n} \rightarrow & \Upsilon_{\rm triv}\Upsilon_{\rm even}\prod_{m\textrm{ (odd)}, m \neq n} \Upsilon_{m\Rightarrow 0}.
\end{eqnarray*}
While simple and efficient this construction is not always optimal. For example, the W-state can be shown to have an NQS representation using only $M = \lceil \log_2(N)\rceil$ hidden units. However, there is little utility in improving this construction since, as shown in \secr{sec:extensions}, it is unnecessary in practice to explicitly describe a uniform number state within an NQS.

\begin{figure}[ht]
\begin{center}
\includegraphics[scale=0.5]{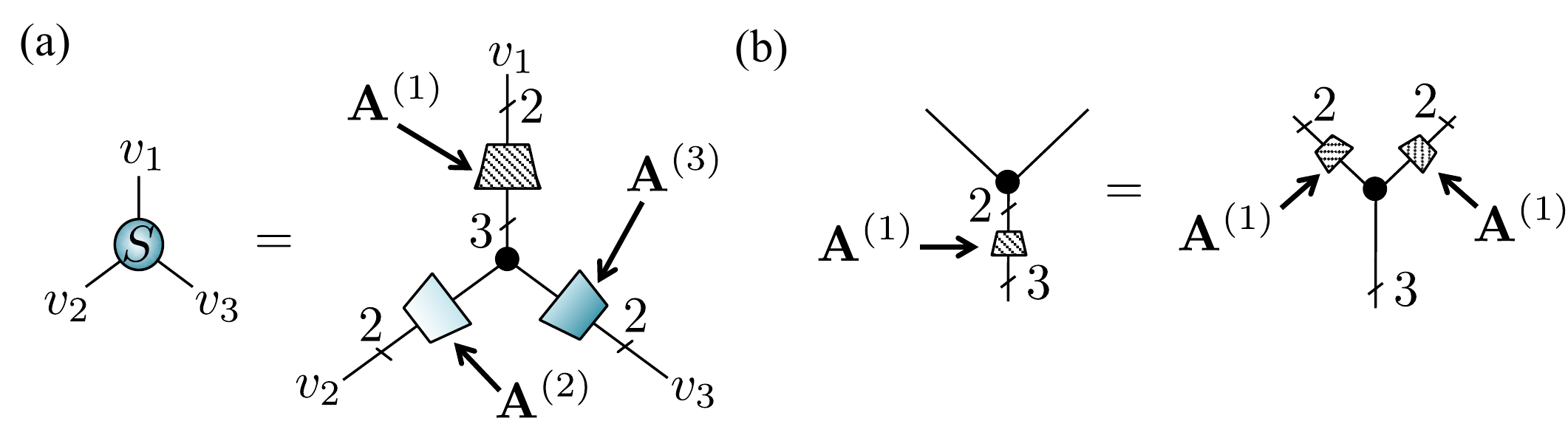}
\end{center}
\caption{(a) The CPD of the $S$ tensor with leg dimensions and component matrices ${\bf A}^{(j)}$ labelled. (b) When contracting the component matrix ${\bf A}^{(1)}$ with an order-3 2 dimensional COPY tensor it can be pulled through converting the COPY tensor to a 3 dimensional version.}
\label{fig_singlet_tensor}
\end{figure}

\section{Singlet tensor} \label{sec_singlet_tensor}
Following its definition given in \fir{fig_w_singlet}(b) the $S$ tensor can be readily seen to have a rank 3 CPD with $3 \times 2$ component matrices
\begin{equation}
{\bf A}^{(1)} = \left[
\begin{array}{cc}
0 & 1 \\
0 & 1 \\
1 & 0 
\end{array}
\right];\, {\bf A}^{(2)} = \left[
\begin{array}{cc}
1 & 0 \\
0 & 1 \\
1 & 1
\end{array}
\right]; \,{\bf A}^{(3)} = \left[
\begin{array}{rc}
0 & 1 \\
-1 & 0 \\
1 & 1
\end{array}
\right]. \label{eq:correlator_stensor}
\end{equation}
Together these give
\begin{equation}
S_{v_1v_2v_3} = \sum_{\alpha=1}^3 A^{(1)}_{\alpha v_1} A^{(2)}_{\alpha v_2} A^{(3)}_{\alpha v_3},
\end{equation}
as depicted in \fir{fig_singlet_tensor}(a). The component matrix ${\bf A}^{(1)}$ obeys a special property that when contracted with a 2 dimensional COPY tensor it can be pulled through and converts the COPY into a 3 dimensional variant, as shown in \fir{fig_singlet_tensor}(b). This was used earlier in \secr{sec:rvb_state} and \fir{fig_rvb_state}(c) to rewire the RVB state into a deep-RBM type NQS.

\section*{References}
\bibliographystyle{unsrt}
\bibliography{cps_nqs_paper}

\begin{thebibliography}{10}

\bibitem{Lauchli2012}
A~M L\"auchli, J~Sudan, and E~S S\o{}rensen.
\newblock {Ground-state energy and spin gap of spin-$\frac{1}{2}$
  Kagom\'e-Heisenberg antiferromagnetic clusters: Large-scale exact
  diagonalization results}.
\newblock {\em Phys. Rev. B}, 83:212401, 2011.

\bibitem{Eisert2010}
J~Eisert, M~Cramer, and M~B Plenio.
\newblock {Colloquium: Area laws for the entanglement entropy}.
\newblock {\em Rev. Mod. Phys.}, 82:277, 2010.

\bibitem{Snijders2012}
C~Snijders, U.~Matzat, and U-D Reips.
\newblock {'Big Data': Big gaps of knowledge in the field of Internet}.
\newblock {\em Int. J. Internet Science}, 7:1, 2012.

\bibitem{LeCun2015}
Y~LeCun, Y~Bengio, and G~Hinton.
\newblock Deep learning.
\newblock {\em Nature}, 521:436, 2015.

\bibitem{Goodfellow2016}
I~Goodfellow, Y~Bengio, and A~Courville.
\newblock {\em {Deep Learning}}.
\newblock MIT Press, 2016.
\newblock \url{http://www.deeplearningbook.org}.

\bibitem{Haykin2008}
S~S Haykin.
\newblock {\em {Neural networks and learning machines, 3rd ed.}}
\newblock Pearson, USA, 2008.
\newblock \url{http://www.pearsonhighered.com/haykin/}.

\bibitem{Beny2013}
C~B{\'e}ny.
\newblock {Deep learning and the renormalization group}.
\newblock {\em arXiv:1301.3124}.

\bibitem{Mehta2014}
P~Mehta and D~J Schwab.
\newblock {An exact mapping between the Variational Renormalization Group and
  Deep Learning}.
\newblock {\em arXiv:1410.3831}.

\bibitem{Carrasquilla2017}
J~Carrasquilla and R~G Melko.
\newblock {Machine learning phases of matter}.
\newblock {\em Nat. Phys.}, 13:43, 2017.

\bibitem{vanNieuwenburg2017}
E~P~L van Nieuwenburg, Y-H Liu, and S~D Huber.
\newblock {Learning phase transitions by confusion}.
\newblock {\em Nat. Phys.}, 13:435, 2017.

\bibitem{Broecker2017}
P~Broecker, J~Carrasquilla, R~G Melko, and S~Trebst.
\newblock {Machine learning quantum phases of matter beyond the fermion sign
  problem}.
\newblock {\em Scientific Reports}, 7:8823, 2017.

\bibitem{Wang2016}
L~Wang.
\newblock {Discovering phase transitions with unsupervised learning}.
\newblock {\em Phys. Rev. B}, 94:195105, 2016.

\bibitem{Chng2017}
K~Ch'ng, J~Carrasquilla, R~G Melko, and E~Khatami.
\newblock {Machine Learning Phases of Strongly Correlated Fermions}.
\newblock {\em Phys. Rev. X}, 7:031038, 2017.

\bibitem{Arsenault2015}
L-F Arsenault, O~A von Lilienfeld, and A~J Millis.
\newblock {Machine Learning for Many-Body Physics: Efficient Solution of
  Dynamical Mean-Field Theory}.
\newblock {\em arXiv:1506.08858}.

\bibitem{Torlai2016}
G~Torlai and R~G Melko.
\newblock {Learning thermodynamics with Boltzmann machines}.
\newblock {\em Phys. Rev. B}, 94:165134, 2016.

\bibitem{Huang2017}
L~Huang and L~Wang.
\newblock {Accelerated Monte Carlo simulations with restricted Boltzmann
  machines}.
\newblock {\em Phys. Rev. B}, 95:035105, 2017.

\bibitem{Liu2017}
J~Liu, Y~Qi, Z~Y Meng, and L~Fu.
\newblock {Self-learning Monte Carlo method}.
\newblock {\em Phys. Rev. B}, 95:041101, 2017.

\bibitem{Tubman2016}
N~M Tubman.
\newblock {Measuring quantum entanglement, machine learning and wave function
  tomography: Bridging theory and experiment with the quantum gas microscope}.
\newblock {\em arXiv:1609.08142}.

\bibitem{Torai2017}
G~Torlai, G~Mazzola, J~Carrasquilla, M~Troyer, R~G Melko, and G~Carleo.
\newblock {Many-body quantum state tomography with neural networks}.
\newblock {\em arXiv:1703.05334}.

\bibitem{Carleo2017}
G~Carleo and M~Troyer.
\newblock {Solving the quantum many-body problem with artificial neural
  networks}.
\newblock {\em Science}, 355:602, 2017.

\bibitem{Nomura2017}
Y~Nomura, A~Darmawan, Y~Yamaji, and M~Imada.
\newblock {Restricted-Boltzman Machine Learning for Solving Strongly Correlated
  Quantum Systems}.
\newblock {\em arXiv:1709.06475}.

\bibitem{Deng2016}
D-L Deng, X~Li, and S~Das Sarma.
\newblock {Exact Machine Learning Topological States}.
\newblock {\em arXiv:1609.09060}.

\bibitem{Gao2017}
X~Gao and L-M Duan.
\newblock {Efficient Representation of Quantum Many-body States with Deep
  Neural Networks}.
\newblock {\em arXiv:1701.05039}.

\bibitem{Deng2017}
D-L Deng, L~Xiaopeng, and S.~Das~Sarma.
\newblock Quantum entanglement in neural network states.
\newblock {\em Phys. Rev. X}, 7:021021, 2017.

\bibitem{Chen2017}
J~Chen, S~Cheng, H~Xie, L~Wang, and T~Xiang.
\newblock {On the Equivalence of Restricted Boltzmann Machines and Tensor
  Network States}.
\newblock {\em arXiv:1701.04831}.

\bibitem{Verstraete2008}
F~Verstraete, V~Murg, and J~I Cirac.
\newblock {Matrix product states, projected entangled pair states, and
  variational renormalization group methods for quantum spin systems}.
\newblock {\em Adv. Phys.}, 57:143, 2008.

\bibitem{Cirac2009}
J~I Cirac and F~Verstraete.
\newblock Renormalization and tensor product states in spin chains and
  lattices.
\newblock {\em J. Phys. A}, 42:504004, 2009.

\bibitem{Orus2014}
R~Orus.
\newblock {A practical introduction to Tensor Networks: Matrix Product States
  and Projected Entangled Pair States}.
\newblock {\em Ann. Phys.}, 349:117, 2014.

\bibitem{Changlani2009}
H~J Changlani, J~M Kinder, C~J Umrigar, and G~K-L Chan.
\newblock {Approximating strongly correlated wave functions with correlator
  product states}.
\newblock {\em Phys. Rev. B}, 80:245116, 2009.

\bibitem{Jastrow1955}
R~Jastrow.
\newblock {Many-Body Problem with Strong Forces}.
\newblock {\em Phys. Rev.}, 98:1479, 1955.

\bibitem{Alassam2017}
S~Al-Assam, S~R Clark, and D~Jaksch.
\newblock {The Tensor Network Theory library}.
\newblock {\em J. Stat. Mech.}, 2017:093102, 2017.

\bibitem{Schollwock2011}
U~Schollw\"{o}ck.
\newblock {The density-matrix renormalization group in the age of matrix
  product states}.
\newblock {\em Ann. Phys.}, 326:96, 2011.

\bibitem{Shi2006}
Y-Y Shi, L-M Duan, and G~Vidal.
\newblock {Classical simulation of quantum many-body systems with a tree tensor
  network}.
\newblock {\em Phys. Rev. A}, 74:022320, 2006.

\bibitem{Murg2010}
V~Murg, F~Verstraete, \"O Legeza, and R~M Noack.
\newblock {Simulating strongly correlated quantum systems with tree tensor
  networks}.
\newblock {\em Phys. Rev. B}, 82:205105, 2010.

\bibitem{Evenbly2014}
G~Evenbly and G~Vidal.
\newblock {Quantum Criticality with the Multi-scale Entanglement
  Renormalization Ansatz}.
\newblock {\em Chapter 4 in "Strongly Correlated Systems. Numerical Methods",
  edited by A. Avella and F. Mancini (Springer Series in Solid-State
  Sciences)}, pages 99--130, 2013.

\bibitem{Ran2017}
S-J Ran, E~Tirrito, C~Peng, X~Chen, G~Su, and M~Lewenstein.
\newblock {Review of Tensor Network Contraction Approaches}.
\newblock {\em arXiv:1708.09213}.

\bibitem{Vidal2003}
G~Vidal.
\newblock {Efficient classical simulation of slightly entangled quantum
  computations.}
\newblock {\em Phys. Rev. Lett.}, 91:147902, 2003.

\bibitem{Clark2004}
S~R Clark and D~Jaksch.
\newblock {Dynamics of the superfluid to Mott-insulator transition in one
  dimension}.
\newblock {\em Phys. Rev. A}, 70:043612, 2004.

\bibitem{Bruderer2010}
M~Bruderer, T~H Johnson, S~R Clark, D~Jaksch, A~Posazhennikova, and W~Belzig.
\newblock {Phonon resonances in atomic currents through Bose-Fermi mixtures in
  optical lattices}.
\newblock {\em Phys. Rev. A}, 82:043617, 2010.

\bibitem{Coulthard2017}
J~R Coulthard, S~R Clark, S~Al-Assam, A~Cavalleri, and D~Jaksch.
\newblock {Enhancement of superexchange pairing in the periodically driven
  Hubbard model}.
\newblock {\em Phys. Rev. B}, 96:085104, 2017.

\bibitem{Mendoza2015}
J~J Mendoza-Arenas, S~R Clark, and D~Jaksch.
\newblock Coexistence of energy diffusion and local thermalization in
  nonequilibrium $xxz$ spin chains with integrability breaking.
\newblock {\em Phys. Rev. E}, 91:042129, 2015.

\bibitem{Mendoza2016}
J~J Mendoza-Arenas, S~R Clark, S~Felicetti, G~Romero, E~Solano, D~G Angelakis,
  and D~Jaksch.
\newblock {Beyond mean-field bistability in driven-dissipative lattices:
  Bunching-antibunching transition and quantum simulation}.
\newblock {\em Phys. Rev. A}, 93:023821, 2016.

\bibitem{Znidaric2017}
M~Znidaric, J~J Mendoza-Arenas, S~R Clark, and J~Goold.
\newblock Dephasing enhanced spin transport in the ergodic phase of a many-body
  localizable system.
\newblock {\em Annalen der Physik}, 529:201600298, 2017.

\bibitem{Johnson2010}
T~H Johnson, S~R Clark, and D~Jaksch.
\newblock Dynamical simulations of classical stochastic systems using matrix
  product states.
\newblock {\em Phys. Rev. E}, 82:036702, 2010.

\bibitem{Johnson2015}
T~H Johnson, T~J Elliott, S~R Clark, and D~Jaksch.
\newblock {Capturing Exponential Variance Using Polynomial Resources: Applying
  Tensor Networks to Nonequilibrium Stochastic Processes}.
\newblock {\em Phys. Rev. Lett.}, 114:090602, 2015.

\bibitem{Foulkes2001}
W~M~C Foulkes, L~Mitas, R~J Needs, and G~Rajagopal.
\newblock {Quantum Monte Carlo simulations of solids}.
\newblock {\em Rev. Mod. Phys.}, 73:33, 2001.

\bibitem{Gubernatis2016}
J~Gubernatis, N~Kawashima, and P~Werner.
\newblock {\em {Quantum Monte Carlo Methods: Algorithms for Lattice Models}}.
\newblock Cambridge University Press, 2016.
\newblock
  \url{http://www.cambridge.org/catalogue/catalogue.asp?isbn=9781107006423}.

\bibitem{Lou2007}
J~Lou and A~W Sandvik.
\newblock Variational ground states of two-dimensional antiferromagnets in the
  valence bond basis.
\newblock {\em Phys. Rev. B}, 76:104432, 2007.

\bibitem{Nightingale2001}
M~P Nightingale and V~Melik-Alaverdian.
\newblock {Optimization of Ground- and Excited-State Wave Functions and van der
  Waals Clusters}.
\newblock {\em Phys. Rev. Lett.}, 87:043401, 2001.

\bibitem{Toulouse2007}
J~Toulouse and C~J Umrigar.
\newblock {Optimization of quantum Monte Carlo wave functions by energy
  minimization}.
\newblock {\em J. Chem. Phys.}, 126:084102, 2007.

\bibitem{Sorella2001}
S~Sorella.
\newblock {Generalized Lanczos algorithm for variational quantum Monte Carlo}.
\newblock {\em Phys. Rev. B}, 64:024512, 2001.

\bibitem{Marti2010}
K~H Marti, B~Bauer, M~Reiher, M~Troyer, and F~Verstraete.
\newblock Complete-graph tensor network states: a new fermionic wave function
  ansatz for molecules.
\newblock {\em New J. Phys.}, 12:103008, 2010.

\bibitem{Mezzacapo2009}
F~Mezzacapo, N~Schuch, M~Boninsegni, and J~I Cirac.
\newblock {Ground-state properties of quantum many-body systems:
  entangled-plaquette states and variational Monte Carlo}.
\newblock {\em New J. Phys.}, 11:083026, 2009.

\bibitem{Neuscamman2011}
E~Neuscamman, H~Changlani, J~Kinder, and G~K-L Chan.
\newblock {Nonstochastic algorithms for Jastrow-Slater and correlator product
  state wave functions}.
\newblock {\em Phys. Rev. B}, 84:205132, 2011.

\bibitem{Schuch2008}
N~Schuch, M~M Wolf, F~Verstraete, and J~I Cirac.
\newblock {Simulation of Quantum Many-Body Systems with Strings of Operators
  and Monte Carlo Tensor Contractions}.
\newblock {\em Phys. Rev. Lett.}, 100:040501, 2008.

\bibitem{Biamonte2011}
J~D Biamonte, S~R Clark, and D~Jaksch.
\newblock {Categorical Tensor Network States}.
\newblock {\em AIP Adv.}, 1:042172, 2011.

\bibitem{Denny2012}
S~J Denny, J~D Biamonte, D~Jaksch, and S~R Clark.
\newblock {Algebraically contractible topological tensor network states}.
\newblock {\em J. Phys. A}, 45:015309, 2012.

\bibitem{Alassam2011}
S~Al-Assam, S~R Clark, C~J Foot, and D~Jaksch.
\newblock {Capturing long range correlations in two-dimensional quantum lattice
  systems using correlator product states}.
\newblock {\em Phys. Rev. B}, 84:205108, 2011.

\bibitem{Huse1988}
D~A Huse and V~Elser.
\newblock {Simple Variational Wave Functions for Two-Dimensional Heisenberg
  Spin-1/2 Antiferromagnets}.
\newblock {\em Phys. Rev. Lett.}, 60:2531, 1988.

\bibitem{Fischer2012}
A~Fischer and C~Igel.
\newblock {An Introduction to Restricted Boltzmann Machines}.
\newblock {\em Progress in Pattern Recognition, Image Analysis, Computer
  Vision, and Applications: 17th Iberoamerican Congress, CIARP 2012, Buenos
  Aires, Argentina, September 3-6, 2012. Proceedings, eds. Alvarez, Luis and
  Mejail, Marta and Gomez, Luis and Jacobo, Julio}, 176:14, 2012.

\bibitem{Hinton2000}
G~E Hinton.
\newblock {Training products of experts by minimizing contrastive divergence}.
\newblock {\em Technical Report GCNU, Gatsby Unit, University College London},
  pages 2000--004, 2000.

\bibitem{MNISTwebsite}
Y~LeCun, C~Cortes, and C~J~C Burges.
\newblock {\em {MNIST database}}.
\newblock \url{http://yann.lecun.com/exdb/mnist/}.

\bibitem{LeRoux2008}
N~Le~Roux and Y~Bengio.
\newblock Representational power of restricted boltzmann machines and deep
  belief networks.
\newblock {\em Neural Computation}, 20(6):1631--1649, 2008.

\bibitem{Trefethen1997}
L~N Trefethen and D~Bau.
\newblock {\em {Numerical Linear Algebra}}.
\newblock SIAM, 1997.
\newblock
  \url{http://www.cambridge.org/catalogue/catalogue.asp?isbn=9781107006423}.

\bibitem{Hackbusch2016}
W~Hackbusch.
\newblock {\em {Tensor Spaces and Numerical Tensor Calculus}}.
\newblock Springer, 2012.
\newblock \url{http://www.springer.com/gp/book/9783642280269}.

\bibitem{Kolda2009}
T~G Kolda and B~W Bader.
\newblock {Tensor Decompositions and Applications}.
\newblock {\em SIAM Review}, 51:455, 2009.

\bibitem{Raussendorf2001}
R~Raussendorf and H~J Briegel.
\newblock {A One-Way Quantum Computer}.
\newblock {\em Phys. Rev. Lett.}, 86:5188, 2001.

\bibitem{Clark2005}
S~R Clark, C~Moura Alves, and D~Jaksch.
\newblock {Efficient generation of graph states for quantum computation}.
\newblock {\em New J. Phys.}, 7:124, 2005.

\bibitem{Hein2004}
M~Hein, J~Eisert, and H~J Briegel.
\newblock {Multiparty entanglement in graph states}.
\newblock {\em Phys. Rev. A}, 69:062311, 2004.

\bibitem{Bondy2011}
A~Bondy and U.S.R. Murty.
\newblock {\em {Graph Theory}}.
\newblock Graduate Texts in Mathematics. Springer London, 2011.

\bibitem{Anders2006}
S~Anders, M~B Plenio, W~D\"ur, F~Verstraete, and H-J Briegel.
\newblock {Ground-State Approximation for Strongly Interacting Spin Systems in
  Arbitrary Spatial Dimension}.
\newblock {\em Phys. Rev. Lett.}, 97:107206, 2006.

\bibitem{Laughlin1983}
R~B Laughlin.
\newblock {Anomalous Quantum Hall Effect: An Incompressible Quantum Fluid with
  Fractionally Charged Excitations}.
\newblock {\em Phys. Rev. Lett.}, 50:1395, 1983.

\bibitem{Kitaev2003}
A~Y Kitaev.
\newblock {Fault-tolerant quantum computation by anyons}.
\newblock {\em Ann. Phys.}, 303:2, 2003.

\bibitem{Verstraete2006}
F~Verstraete, M~M Wolf, D~Perez-Garcia, and J~I Cirac.
\newblock {Criticality, the Area Law, and the Computational Power of Projected
  Entangled Pair States}.
\newblock {\em Phys. Rev. Lett.}, 96:220601, 2006.

\bibitem{Batchelor1996}
M~T Batchelor, H~W~J Bl{\"o}te, B~Nienhuis, and C~M Yung.
\newblock {Critical behaviour of the fully packed loop model on the square
  lattice}.
\newblock {\em J. Phys A}, 29:L399, 1996.

\bibitem{Fisher1961}
M~E Fisher.
\newblock {Statistical Mechanics of Dimers on a Plane Lattice}.
\newblock {\em Phys. Rev.}, 124:1664, 1961.

\bibitem{Rokhsar1988}
D~S Rokhsar and S~A Kivelson.
\newblock {Superconductivity and the Quantum Hard-Core Dimer Gas}.
\newblock {\em Phys. Rev. Lett.}, 61:2376, 1988.

\bibitem{Anderson1973}
P~W Anderson.
\newblock Resonating valence bonds: A new kind of insulator?
\newblock {\em Mater. Res. Bull.}, 8:153, 1973.

\bibitem{Anderson1987}
P~W Anderson.
\newblock {The Resonating Valence Bond State in La$_2$CuO$_4$ and
  Superconductivity}.
\newblock {\em Science}, 235:1196, 1987.

\bibitem{Savary2017}
L~Savary and L~Balents.
\newblock {Quantum spin liquids: a review}.
\newblock {\em Rep. Prog. Phys.}, 80:016502, 2017.

\bibitem{Zhou2017}
Y~Zhou, K~Kanoda, and T-K Ng.
\newblock {Quantum spin liquid states}.
\newblock {\em Rev. Mod. Phys.}, 89:025003, 2017.

\bibitem{Liang1988}
S~Liang, B~Doucot, and P~W Anderson.
\newblock {Some New Variational Resonating-Valence-Bond-Type Wave Functions for
  the Spin-1/2 Antiferromagnetic Heisenberg Model on a Square Lattice}.
\newblock {\em Phys. Rev. Lett.}, 61:365, 1988.

\bibitem{Schuch2012}
N~Schuch, D~Poilblanc, J~I Cirac, and D~P\'erez-Garc\'{\i}a.
\newblock {Resonating valence bond states in the PEPS formalism}.
\newblock {\em Phys. Rev. B}, 86:115108, 2012.

\bibitem{Gutzwiller1963}
M~C Gutzwiller.
\newblock {Effect of Correlation on the Ferromagnetism of Transition Metals}.
\newblock {\em Phys. Rev. Lett.}, 10:159, 1963.

\bibitem{Gros1989}
C~Gros.
\newblock {Physics of projected wavefunctions}.
\newblock {\em Ann. Phys.}, 189:53, 1989.

\bibitem{Sikora2015}
O~Sikora, H-W Chang, C-P Chou, F~Pollmann, and Y-J Kao.
\newblock {Variational Monte Carlo simulations using tensor-product projected
  states}.
\newblock {\em Phys. Rev. B}, 91:165113, 2015.

\bibitem{TNTwebsite}
S~Al-Assam, S~R Clark, and D~Jaksch.
\newblock {\em {The Tensor Network Theory Library}}.
\newblock \url{http://www.tensornetworktheory.org}.

\bibitem{Pei2018}
M~Pei, S~Pearce, and S~R Clark.
\newblock {Neural-network quantum state projection approach}.
\newblock {\em in preparation}, 2017.

\end{thebibliography}
    
\end{document}